\documentclass[11pt,reqno]{amsart}
\pdfoutput=1
\usepackage[hidelinks]{hyperref}
\usepackage{packages}

\newcommand{\e}{e}

\theoremstyle{plain}
\newtheorem*{Thm*}{Main Theorem}

\newtheorem{innercustomthm}{Theorem}
\newenvironment{customthm}[1]
  {\renewcommand\theinnercustomthm{#1}\innercustomthm}
  {\endinnercustomthm}

\begin{document}

  \title[Quantum Yang-Mills Theory in Two Dimensions]{Quantum Yang-Mills Theory in Two Dimensions:\\Exact versus Perturbative}
  \author{Timothy Nguyen}
\date{\today}

\begin{abstract}
   The standard Feynman diagrammatic approach to quantum field theories assumes that perturbation theory approximates the full quantum theory at small coupling even when a mathematically rigorous construction of the latter is absent. On the other hand, two-dimensional Yang-Mills theory is a rare (if not the only) example of a nonabelian (pure) gauge theory whose full quantum theory has a rigorous construction. Indeed, the theory can be formulated via a lattice approximation, from which Wilson loop expecation values in the continuum limit can be described in terms of heat kernels on the gauge group. It is therefore fundamental to investigate how the exact answer for 2D Yang-Mills compares with that of the continuum perturbative approach, which a priori are unrelated. In this paper, we provide a mathematically rigorous formulation of the perturbative quantization of 2D Yang-Mills, and we consider perturbative Wilson loop expectation values on $\R^2$ and $S^2$ in Coulomb gauge, holomorphic gauge, and axial gauge (on $\R^2$). We show the following equivalences and nonequivalences between these gauges:
   (i) Coulomb and holomorphic gauge are equivalent and are independent of the choice of gauge-fixing metric; (ii) both are inequivalent with axial-gauge. Additionally, we show that the asymptotics of exact lattice Wilson loop expectations  on $S^2$ agree with perturbatively computed expectations in holomorphic gauge for simple closed curves to all orders. However, as a consequence of (ii), this result is necessarily false on $\R^2$. Our work therefore presents fundamental progress in the analysis of how continuum perturbation theory succeeds or fails in capturing the asymptotics of the continuum limit of the lattice theory.
\end{abstract}

\maketitle

\vspace{-.25in}

\tableofcontents

\section{Introduction}

Yang-Mills theory provides a theoretical framework for describing the physics of elementary particles and has profoundly impacted the study of partial differential equations and differential topology. The classical (Euclidean) Yang-Mills action can be written as
\begin{equation}
  S_{YM}(A) = \frac{1}{2e^2}\int_\Sigma \left<F_A \wedge * F_A\right>, \label{intro:action}
\end{equation}
where $F_A$ is the gauge-field strength, $\left<\cdot,\cdot\right>$ is an ad-invariant inner product on the Lie algebra of the gauge group, $e$ is a coupling constant, and $\Sigma$ is the underlying space assumed to be a smooth, orientable Riemannian manifold. In the quantum theory, this classical action is inserted into a (formal) path integral,  from which one can compute various physical quantities in terms of a Feynman diagrammatic expansion. The process of evaluating and understanding such Feynman diagrams is what leads to many of the basic features of quantum Yang-Mills theory, such as perturbative renormalizability \cite{tHV, Col, Cos} and asymptotic freedom \cite{GW, Pol}.

On the other hand, quantum Yang-Mills theory can also be formulated on a lattice, whereby one obtains a mathematically rigorous, nonperturbative approach that avoids the formal aspects of the continuum theory outlined above. Indeed, working on a lattice ensures that all quantities can be computed in terms of well-defined finite-dimensional integrals. Here, one has to introduce a suitable discretization of the action (\ref{intro:action}) the details of which we will return to later. But as one is ultimately interested in a theory that extends all the way down to the relevant microscopic scales, one would like to take a continuum limit in which the lattice spacing becomes finer and finer.

Supposing this to be achieved, we obtain two independent, a priori distinct constructions of quantum Yang-Mills theory. While quite different, both the perturbative methods of Feynman diagrams and the numerical simulations of lattice methods have yielded spectacular agreement with experimental data in various settings \cite{Mak, PS, Rothe}. Naturally then, one should consider how these two methods compare at the level of precise mathematics. Specifically, since one regards the continuum formulation as perturbative and the lattice formulation as nonperturbative, one expects in the limit of small coupling that the two formulations should somehow converge. For emphasis, we state this as the following\\

\noindent \textbf{Question: }\textit{For Yang-Mills theory, what is the relationship between the perturbative results obtained in the continuum formulation and the nonperturbative results obtained from the continuum limit of the lattice formulation, as the coupling constant is sent to zero?\\}

This paper is an investigation into this basic question, for which we are unaware of any prior rigorous work by the mathematical community.

A priori, our question is well-posed only if we know how to take the continuum limit of the lattice formulation. This is a very difficult problem in dimensions three and four, for which there exists old work by Balaban \cite{Bal85,Bal87} that is unfortunately not easily accessible. However, we are in the fortunate situation that in two dimensions, quantum Yang-Mills theory has a well-known and elegant lattice continuum limit due to Migdal \cite{Mig}, which was subsequently generalized to surfaces by Witten \cite{Wit2D} and then systematically developed by Levy \cite{Levy}. The aim of this paper is to compare this continuum limit with perturbative two-dimensional Yang-Mills theory.

Our approach is as follows. The quantities of interest to us are expectation values of Wilson loop observables, which are the basic gauge-invariant observables of any gauge theory. Given an oriented closed curve $\gamma$ and a conjugation-invariant function $f$ on the gauge group $G$ of our theory (without loss of generality, we always take $f$ to be trace in an irreducible representation of $G$), we obtain the Wilson loop observable $W_{f,\gamma}$ which takes a connection $A$, computes the holonomy of $A$ about $\gamma$, and applies $f$ to this group-valued element:
$$W_{f,\gamma}(A) = f(\mr{hol}_\gamma(A)).$$
We can compute the expectation value of $W_{f,\gamma}(A)$ exactly using the continuum of the lattice approach or perturbatively using the methods of Feynman diagrams:
\begin{align*}
  \left<W_{f,\gamma}\right>_{\Sigma} &:= \textrm{(exact) expectation value} \\
  \left<W_{f,\gamma}\right>_{\Sigma,\, pert} &:= \textrm{perturbative expectation value}.
\end{align*}
The exact expectation value is defined mathematically precisely in Section 2; for the perturbative expectation, defined in Section \ref{Sec:Pert}, a few clarifying remarks are in order. We will primarily be considering the special cases $\Sigma = S^2$ or $\R^2$. These cases are natural for several reasons. First, their topology is such that there is a unique minimal Yang-Mills connection modulo gauge-equivalence, namely the trivial connection\footnote{We will be focusing only on topologically trivial bundles. This is not a real restriction, see Remark \ref{Rem:topology}.}. For $\Sigma$ of higher genus, the presence of a continuous moduli of flat connections makes the perturbation theory more involved. Thus, for $\Sigma = S^2$ or $\R^2$, the perturbative expectation $\left<W_{f,\gamma}\right>_{\Sigma, pert}$ involves a Feynman diagrammatic expansion about only the trivial connection. Secondly, in the case of $S^2$, having a compact underlying space conveniently eliminates  infrared divergences. In fact, there will be instances in which we are forced to regard $\R^2$ as the limit of $S^2$ when the area of the latter is sent to infinity. We refer to this limiting procedure as ``decompactification''.

Next, note that in two dimensions, the Hodge star operator $*$ appearing in the integral (\ref{intro:action}) is specified entirely in terms of an area form $d\sigma$ on $\Sigma$ (and not on a full metric tensor). It follows that the coupling constant $\lambda_0 = e^2$ has dimensions of inverse area. (In other words, scaling the area form by $\ell^2$ and the coupling constant $\lambda_0$ by $\ell^{-2}$ preserves the action.) Thus, for $\Sigma$ compact, we define the dimensionless coupling constant
\begin{equation}
  \lambda = 
             \lambda_0|\Sigma|
            \label{eq:lambda}
\end{equation}
where $|\cdot|$ denotes area with respect to $d\sigma$. Thus, $\left<W_{f,\gamma}\right>_\Sigma$ is a function of $\lambda$ while $\left<W_{f,\gamma}\right>_{\Sigma, pert}$ is a formal power series in $\lambda$. For $\Sigma = \R^2$, since we always take the area form to be the standard flat area form, we work directly with the parameter $\lam_0$.

Finally and quite crucially, the perturbative expectation value $\left<W_{f,\gamma}\right>_{\Sigma, pert}$ requires the choice of a suitable gauge-fixing procedure. We consider several such choices. The most natural choice of gauge to consider is \textit{Coulomb gauge} (also known as Landau gauge). Here, one chooses an auxiliary metric $g = g_{ij}$ and imposes the gauge-fixing condition $d^*A = 0$ to eliminate longitudinal modes. The geometric nature of Coulomb gauge makes it applicable for arbitrary $\Sigma$. For $\R^2$, we will regard Coulomb gauge expectations on $\R^2$ as the decompactification limit of Coulomb gauge expectations on $S^2$, since Coulomb gauge on $\R^2$ has infrared diverges.  Next, for $\Sigma = S^2$ or $\R^2$, we also have available another choice of ``gauge", namely \textit{holomorphic gauge}. This gauge is referred to as (Euclidean) light-cone gauge in the physics literature.  Here, writing $A$ as $A_z dz + A_{\bar z}d\bar z$ in terms of holomorphic and anti-holomorphic components (with respect to a chosen auxiliary conformal structure), holomorphic gauge imposes the condition $A_{\bar z} = 0$. While the interpretation of this condition as a gauge-fixing condition requires some additional analysis \cite{Ngu2016}, Feynman diagrams can be meaningfully generated from this ansatz notwithstanding. Finally, on $\Sigma = \R^2$, we can also consider the especially simplifying \textit{axial-gauge}, in which we eliminate the component of a connection along a fixed direction. As we will clarify later, our useage of the word axial-gauge is an abbreviation for what we call \textit{stochastic axial-gauge}.

Thus, we consider in this paper 
$$\left<W_{f,\gamma}\right>_{\Sigma,\, pert} = \left<W_{f,\gamma}\right>_{\Sigma,C}, \;\left<W_{f,\gamma}\right>_{\Sigma,\, hol},\;\textrm{or } \left<W_{f,\gamma}\right>_{\Sigma,\, ax}$$
the perturbative expectation on $\Sigma$ corresponding to Coulomb gauge, holomorphic gauge, or axial-gauge, respectively, with the first two of these requiring the choice of an auxiliary metric compatible with the given area form. These gauges are all defined  mathematically precisely in Section \ref{Sec:Pert}. Moreover, as is standard in perturbative quantum field theory, the presence of ultraviolet divergences requires the use of a regularization scheme. For Coulomb gauge, we choose a heat-kernel regulator, which is especially adapted to the underlying space being curved (unlike standard dimensional/momentum-cutoff regularization methods on flat space). For the remaining holomorphic and axial gauges, they do not require a regulator since the Feynman diagrams they generate are finite. (Upon closer inspection however, these latter gauges were obtained through an appropriate regularization scheme applied at a more fundamental starting point. We elucidate this point in a moment.)

Our results can be summarized as follows, which combined with the results from \cite{NguSYM}, we depict pictorially in Figure 1. First we consider the case of $S^2$:

\begin{customthm}{1}\label{MainThm1}
  Consider Yang-Mills theory on $(S^2, d\sigma)$ with arbitrary compact gauge group $G$. Pick any compatible metric for use as a gauge-fixing metric.
  \begin{enumerate}
    \item  $\left<W_{f,\gamma}\right>_{S^2,C}$, defined using a heat-kernel regularization scheme, is finite without any need for counterterms and is independent of the choice of gauge-fixing metric. Moreover, $\left<W_{f,\gamma}\right>_{S^2,C}$ is invariant under area-preserving diffeomorphisms, i.e. those which preserve the areas of the regions complementary to $\gm$.
    \item We have
    \begin{equation}
     \left<W_{f,\gm}\right>_{S^2,hol}=\left<W_{f,\gm}\right>_{S^2,C}.
    \end{equation}
    In particular, $\left<W_{f,\gamma}\right>_{S^2,hol}$ is also independent of the choice of gauge-fixing metric and invariant under area-preserving diffeomorphisms.
    \item Let $\gamma$ be a simple closed curve. Then
    \begin{equation}
 	\lim_{\lam\to 0}\left<W_{f,\gamma}\right>_{S^2} \sim \left<W_{f,\gamma}\right>_{S^2, hol}. \label{eq:main}
    \end{equation}
    Here, $\sim$ means that the left-hand side of (\ref{eq:main}) has an asymptotic series given by the right-hand side\footnote{That is, $\lim_{\lam\to0}\psi(\lambda) \sim \sum_{n \geq 0} c_n \lambda^n$ if $\psi(\lambda) - \sum_{n=0}^Nc_n\lambda^n = o(\lambda^N)$ as $\lambda \to 0$ for every $N$. We write $\lim_{\lambda \to 0}$ to emphasize that we are considering small $\lambda$ asymptotics.}. Moreover, $\left<W_{f,\gamma}\right>_{S^2, hol}$ is given explicitly by (the asymptotic series for) the Gaussian integral over $\g$ given by formula (\ref{eq:ThmExact1}). This series differs from $\left<W_{f,\gamma}\right>_{S^2}$ by exponentially small ``instanton" corrections, see Remark \ref{Rem:Inst}.
  \end{enumerate}
\end{customthm}

For (i), that no counterterms are needed even as the regulator is removed is shown in Theorem \ref{ThmFinite} via explicit computations. The proof of invariance under area-preserving diffeomorphisms is shown in Theorem \ref{Thm:Area}. Statement (ii), which proves the equivalence of Coloumb and holomorphic gauge (to all orders in perturbation theory), is surprising from a purely mathematical point of view since the constructions are very different. From a practical point of view, since holomorphic gauge is much more computationally feasible than Coulomb gauge, (ii) represents a dramatic simplification. Statement (iii) provides an explicit computation of the asymptotics of the exact Wilson loop expectation of a simple closed curve to all orders in the coupling. Moreover, it shows explicitly how the exact answer and the perturbative answer differ through asymptotically zero instanton corrections. There has been previous work on (iii) in the physics literature \cite{BasGriVia99, BNV, GP}, most notably \cite{GP}, which establishes (\ref{eq:main}) for circular loops. Our proof of (iii) is an immediate consequence of the work of \cite{GP} and (ii). Nevertheless, the result (\ref{eq:main}) is somewhat mysterious, see Section \ref{Sec:Discussion} for some discussion.

Next, we consider the case of $\R^2$.

\begin{customthm}{2}\label{MainThm2}
 Consider Yang-Mills theory on $\R^2$ with the standard flat area form and with arbitrary compact gauge group $G$.
 \begin{enumerate}
  \item The decompactification limit
  \begin{equation}
   \left<W_{f,\gm}\right>_{\R^2,C} := \lim_{S^2 \to \R^2}\left<W_{f,\gm}\right>_{S^2,C}
  \end{equation}
  exists.
  \item We have
   \begin{equation}\left<W_{f,\gm}\right>_{\R^2,hol} = \lim_{S^2 \to \R^2}\left<W_{f,\gm}\right>_{S^2,hol} \label{eq:decompactify_hol}
  \end{equation}
and the equivalence of holomorphic gauge with Coulomb gauge:
  \begin{equation}\left<W_{f,\gm}\right>_{\R^2,hol} = \left<W_{f,\gm}\right>_{\R^2,C}.
  \end{equation}
  \item Holomorphic gauge does not capture the asymptotics of exact Wilson loop expectations:
 \begin{align}
      \lim_{\lambda \rightarrow 0}\left<W_{f,\gamma}\right>_{\R^2} \not\sim \left<W_{f,\gamma}\right>_{\R^2, hol}. \label{eq:main_false}
    \end{align}
   As a consequence, we have
   \begin{equation}
    \left<W_{f,\gm}\right>_{\R^2,ax} \neq \left<W_{f,\gm}\right>_{\R^2,hol} = \left<W_{f,\gm}\right>_{\R^2,C}.
   \end{equation}
 \end{enumerate}
\end{customthm}

Parts (i) and (ii) of Theorem 2 follow from Theorem 1 by ensuring that a decompactification limit exists. For (ii), a similar equivalence, with Coulomb gauge on $\R^2$ replaced by Feynman gauge, has been previously checked to second order in special cases \cite{BasNar96, BasGriVia99}. That case (iii) differ in Theorems 1 and 2 is remarkable. One the one hand, it contradicts the tenant that different choices of gauge-fixing should not affect the evaluation of perturbative Wilson loop expectations (barring anomalies). On the other hand, such a discrepancy is consistent with the fact that different regularization schemes for a quantum field theory can lead to different results. Indeed, axial-gauge, as we have defined it in Section \ref{Sec:Axial}, implicitly uses a ``stochastic regulator'' \cite{NguSYM}, whereas holomorphic gauge uses the Wu-Mandelstam-Liebrandt (WML) regulator \cite{SK}. As explained in \cite{Ngu2016}, the WML regulator leads to a family of ``generalized axial-gauges'', all of which are equivalent and of which one of them is holomorphic gauge. The purpose of these stochastic and WML regulators is to resolve the fact that $\frac{1}{k_1^2}$, the reciprocal of the Fourier transform of $-\pd_1^2$ naturally arising from the Yang-Mills action (\ref{eq:YM-axial-gauge}) in axial-gauge, does not define a distribution. Another physical interpretation of
(\ref{eq:main_false}), in light of (\ref{eq:decompactify_hol}), is that the decompactification limit of perturbative expansion on $S^2$ ``remembers instantons'' \cite{BasGriVia99}.

Parts (i) and (ii) above readily extend to expectations of products of Wilson loop observables. In light of (iii) above, we formulate the following conjecture:

\begin{Conjecture}\label{Conj:main}
 On $S^2$, for general closed curves $\gm$, we have
    \begin{align}
      \lim_{\lambda \rightarrow 0}\left<W_{f,\gamma}\right>_{S^2} \sim \left<W_{f,\gamma}\right>_{S^2, hol}. \label{eq:conj_scc}
    \end{align}
\end{Conjecture}

Altogether, our results shed light on the central tenant of quantum field theory which asserts that the Feynman diagrammatic perturbative expansion yields an asymptotic series for the exact expectation of observables:
\begin{align}
  \lim_{\lambda \rightarrow 0}\left<O\right> \sim \left<O\right>_{pert}. \label{intro:expect}
\end{align}
Despite enormous efforts to place quantum field theory on firm mathematical foundations, ascertaining the validity or violation of the fundamental consistency condition (\ref{intro:expect}) in the context of gauge-theories appears to have been overlooked by the mathematical community.

\begin{figure}[h]
\includegraphics[scale=0.3]{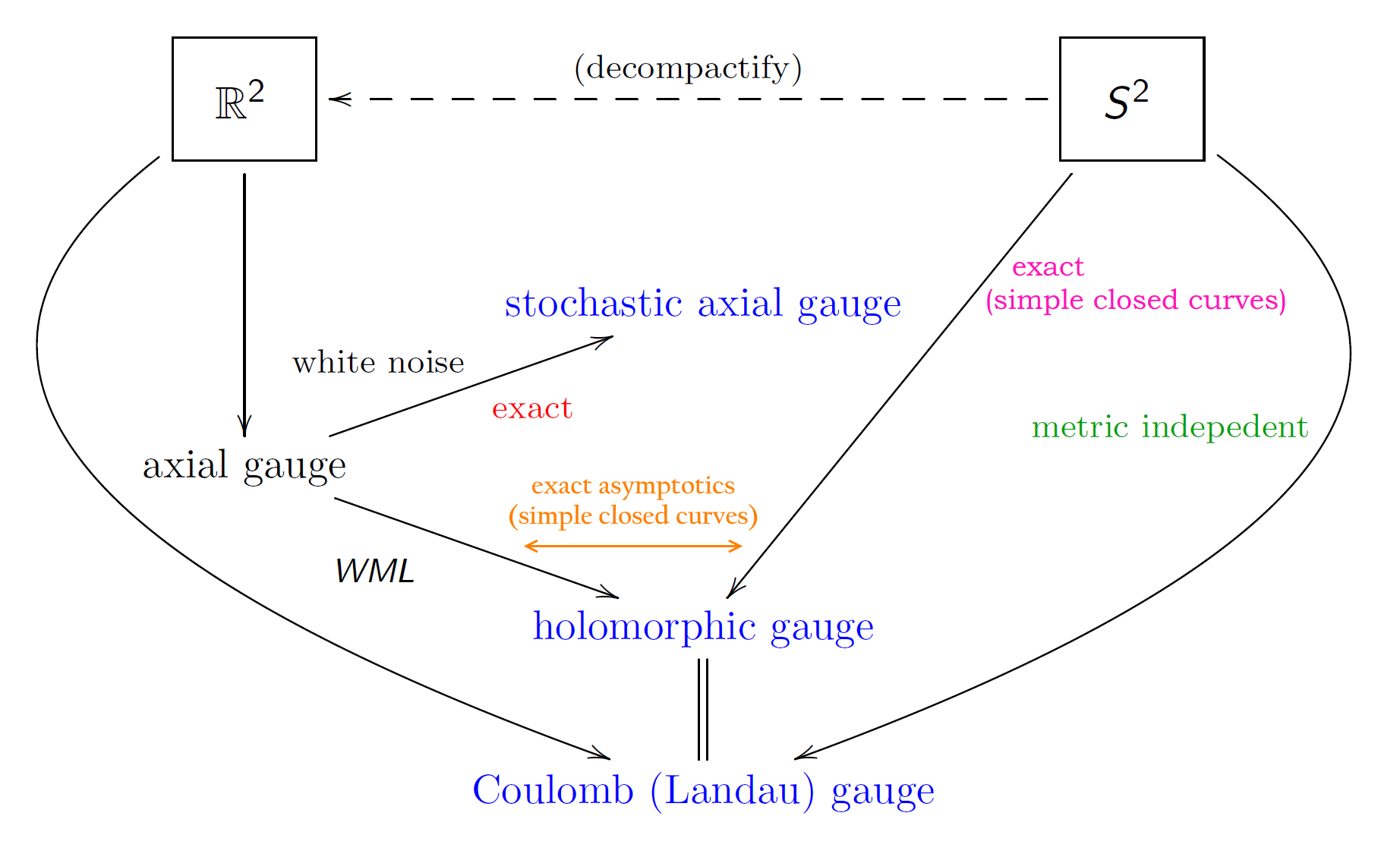}
\textsc{Figure 1.} Roadmap of equivalences and results.\\[3ex]
\end{figure}

Figure 1 illustrates how the nexus of our main results above along with our related work \cite{Ngu2016, NguSYM} fit together. Results (i) and (ii) on $S^2$ and $\R^2$ are given by the outermost arrows and the equality between holomorphic and Coulomb gauge. Theorem 2(iii) is given by the two different arrows emanating from axial gauge, yielding the inequivalent stochastic axial gauge and holomorphic gauge. Theorem 1(iii) yields the statement about exact asymptotics for simple closed curves on $S^2$. Moreover, the explicit formula provided by Theorem 1(iii), as well as the decompactification limit of this formula, provides the double arrow for the explicit asymptotics. Finally, the exact arrow corresponding to stochastic axial-gauge is a consequence of the work of \cite{NguSYM}, showing that
\begin{equation}\left<W_{f,\gm}\right>_{\R^2} = \left<W_{f,\gm}\right>_{\R^2, ax},
\end{equation}
so that a fortiori $\left<W_{f,\gm}\right>_{\R^2} \sim \left<W_{f,\gm}\right>_{\R^2, ax}$.\\

Our paper is organized as follows. In Section 2, we discuss the lattice formulation of 2D Yang-Mills theory and describe how its continuum limit yields a rigorous construction of a Yang-Mills measure. The most important outcome of this is that one obtains concrete formulas for Wilson loop expectation values. In Section 3, we describe the entirely different methods of perturbative quantization of continuum Yang-Mills theory. Here we give a rapid, self-contained setup of the formal perturbation theory in all our gauges, of which the most technical is Coulomb gauge. For the latter, we use the most direct procedure available: the Faddeev-Popov method. While many physics treatments of the Faddeev-Popov method use formal arguments to assert that it leads to a gauge-invariant construction (i.e. independence of the choice of gauge-fixing metric), proving that this is so mathematically rigorously takes a fair amount of sophistication.

In order to establish our gauge-invariance results (i) and (ii), we use the Batalin-Vilkovisky method of quantization, which is powerful enough to handle the situation in which there are zero ghost modes (which we do find ourselves in since we quantize about a trivial connection). This is performed Section \ref{SecGI} and \ref{Sec:C=hol}, with Section \ref{Sec:Finite} providing necessary auxiliary computations. These sections constitute the main technical achievements of this paper, most noticeable the interpolating analysis we perform
between holomorphic and Coulomb gauge, for which we are unaware of any prior work. Finally in Section \ref{Sec:Asymp}, we use analytic and Lie theoretic tools to relate the asymptotics of exact Wilson loop expectation values to perturbative calculations. We finish with a discussion of our results and future directions. Since the methods of perturbative quantum field theory are not well-known to most mathematicians, our appendix provides background on Wick's Theorem so as to make this paper as self-contained as possible.\\

\noindent\textit{Acknowledgments.} The author would like to thank Vasily Pestun for many helpful discussions and for pointing out a crucial error in an earlier version of this paper. Pestun also referred the author to several references in the physics literature which the author had missed. The author would also like to acknowledge helpful discussions with Greg Moore and with Tom Parker concerning heat kernel methods.

\section{2D Yang-Mills Measure}\label{Sec:Measure}

Euclidean quantum Yang-Mills theory, in any dimension, can be given a rigorous formulation on a lattice. As there is no canonical choice for how to discretize a continuum theory, many formulations are possible. In two dimensions it is most convenient to work with the one due to Migdal \cite{Mig}, which was later refined and extended by many others \cite{Dri}, \cite{Wit2D}, \cite{Levy}. This formulation is invariant with respect to lattice subdivision, so that the continuum limit of  taking the lattice spacing to zero is in some sense already inherent.

We recall this formulation (following \cite{Levy}) in the general setting of when the underlying space is a closed, connected surface $\Sigma$ endowed with an area form $\lambda_0d\sigma$, where $d\sigma$ is some reference area form and $\lambda_0$ a coupling constant. Let $G$ be a compact Lie group and $\Gamma$ be a triangulation of $\Sigma$, that is, a finite set of edges (mappings of intervals $[0,1]$ into $\Sigma$, injective on the interiors) joined at their vertices such that their complement consists of a disjoint union of faces homeomorphic to disks. Assign an arbitrary orientation to each edge. To these oriented edges $e$ of $\Gamma$, we assign group-valued elements $g_e \in G$, which form the basic variables of the discretized Yang-Mills theory. To each face $F$, we can compute its area $|F|$ (with respect to $d\sigma$) and assign an orientation to the boundary $\pd F$ of $F$, thereby inducing an orientation on all the edges which comprise it. Write $\pd F = \pm e_n \cdots \pm e_1$ as a concatenation of edges occurring in their natural cyclic order (well-defined up to cyclic permutation), where one has $\pm$ according to whether the preassigned orientation of the edge $e_i$ is the same or opposite of that induced from $\pd F$. Define
$$g_{\pd F} = g_{e_n}^{\pm 1}\cdots g_{e_1}^{\pm 1},$$
the product of the group valued elements associated to $\pd F$ with the appropriate corresponding powers. It is well-defined up to cyclic permutation of the factors and an overall inversion.

Pick any bi-invariant metric on $G$. This is obtained from an ad-invariant metric on its Lie algebra $\g$, which we denote by $\left<\cdot,\cdot\right>$. One obtains an associated Laplace-Beltrami operator $\Delta$ on $G$. The heat kernel for $\Delta$ is given by the function $K_t(g)$, $t > 0$, which satisfies
\begin{equation}
  (e^{-t\Delta/2}f)(h) = \int_G K_t(hg^{-1})f(g)dg \label{def:heatkernel}
\end{equation}
for all smooth functions $f$ on $G$.  Here $dg$ denotes normalized Haar measure on $G$. The function $K_t(g)$ satisfies
\begin{equation}
  K_t(g^{-1}) = K_t(g), \qquad K_t(gh) = K_t(hg), \qquad \textrm{for all } g,h \in G. \label{Kprop}
\end{equation}

Assign the weight $K_{\lambda_0|F|}(g_{\pd F})$ to $F$. Properties (\ref{Kprop}) show that this weight is independent of the orientation of $\pd F$ and the cyclic ordering of the edges in $\pd F$.  The Yang-Mills measure associated to $\Gamma$ is the measure
\begin{align}
  d\mu_{\Gamma, \lambda_0d\sigma} = \Big(\prod_{F \in F(\Gamma)} K_{\lambda_0|F|}(g_{\pd F})\Big)\prod_{e \in E(\Gamma)} dg_e, \label{def:dmu}
\end{align}
on $G^{|E(\Gamma)|}$, where $E(\Gamma)$ and $F(\Gamma)$ are the set of edges and faces of $\Gamma$, respectively. The Yang-Mills partition function for $(\Sigma, \lambda_0d\sigma)$ is then
\begin{equation}
  Z_{\Sigma, \lambda_0d\sigma} = \int_{G^{|E(\Gamma)|}} d\mu_{\Gamma, \lambda_0d\sigma}. \label{eq:Z}
\end{equation}
Our lattice action is such that the partition function is independent of the choice of triangulation $\Gamma$. This is a simple consequence of the fact that the heat kernel obeys the convolution property
\begin{align}
  \int_G K_{t_1}(g_1g^{-1})K_{t_2}(gg_2)dg = K_{t_1+t_2}(g_1g_2) \label{eq:heatconv}
\end{align}
so that (\ref{eq:Z}) is invariant under subdivision. See \cite{Wit2D} for further details.

We can make formula (\ref{eq:Z}) more explicit. A surface $\Sigma$ of genus $h$ can be represented as a $2h$-gon with sides appropriately identified. Applying the above formula using the (degenerate) triangulation whose only face is such a $2h$-gon, we obtain
$$Z_{\Sigma, \lambda_0d\sigma} = \int_{G^{2h}} K_{\lambda_0|\Sigma|}(a_1b_1a_1^{-1}b_1^{-1}\cdots a_hb_ha_h^{-1}b_h^{-1})\prod da_i \prod db_i.$$

In a quantum field theory, we are interested not only in the partition function but also in the expectation values of (gauge-invariant) observables. For continuum gauge theories, we have the Wilson loop observables $W_{f,\gm}(A)$, which form a rich set of observables since they form a dense collection of functions on the space of connections modulo gauge. This is a simple consequence of the following fact:

\begin{Lemma}
  Let $A_1$ and $A_2$ be two connections on a principal $G$-bundle $P$ over a connected base manifold $M$. Suppose that their holonomies about every loop based at some given point agree. Then $A_1$ and $A_2$ are gauge-equivalent.
\end{Lemma}

\Proof Denote the given basepoint in question by $p$. Given any path $\gamma$ in $M$, let $P_i(\gamma)$ denote parallel transport from $\gamma(0)$ and $\gamma(1)$ using $A_i$. Our hypotheses imply that given any path $\gamma$ joining $p$ to any other point $q \in M$, the automorphism $P_2(\gamma)P_1(\gamma)^{-1}$ of $P_q$, the fiber of $P$ over $q$, is independent of $\gamma$. Indeed, this follows from  $P_1(\tilde\gamma)^{-1}P_1(\gamma) = P_2(\tilde\gamma)^{-1}P_2(\gamma)$ for any other path $\tilde \gamma$ joining $p$ to $q$. Letting $q$ vary, this yields for us a well-defined bundle automorphism $g$. This automorphism is the desired gauge transformation intertwining $A_1$ and $A_2$, since $g^*A_1$ and $A_2$ define equal parallel transport operators.\End

In the lattice formulation, we would like to compute the expectation $\left<W_{f,\gamma}\right>_\Sigma$ with respect to the lattice Yang-Mills measure induced by some triangulation $\Gamma$ on $\Sigma$, with $W_{f,\gamma}$ suitably defined. The lattice formulation makes it clear how to express such an expectation value in terms of a closed formula. We assume $\gamma$ is \textit{regular}, namely, that it is a finite union of piecewise smoothly embedded curves. Since $\gamma$ is regular, we can consider its image as an oriented finite graph $\hat\gamma$ on $\Sigma$ (the maximal components on which $\gamma$ is injective constitute the edges of $\hat\Gamma$). Embed $\hat\gamma$ in some triangulation $\Gamma$ of $\Sigma$. Write $\hat\gamma = \pm e_k \cdots \pm e_1$ as a sequence of edges in the order that they occur in the parametrization of $\gamma$, with the $\pm$ according to whether the given orientation of $e_i$ agrees with the one induced from $\gamma$. We then obtain $g_\gamma = g_{e_k}^{\pm 1}\cdots g_{e_1}^{\pm 1}$ as above.

\begin{Definition}
We have
  \begin{equation}
  \left<W_{f,\gamma}\right>_\Sigma = \frac{1}{Z_{\Sigma, \lambda_0d\sigma}}\int_{G^{|E(\Gamma)|}} f(g_\gamma) d\mu_{\Gamma, \lambda_0d\sigma}. \label{def:WLE}
\end{equation}
\end{Definition}
The subdivision invariance property of our lattice formulation implies that (\ref{def:WLE}) does not depend on the choice of $\Gamma$ containing $\hat\gamma$.

Strictly speaking, in the lattice formulation of gauge-theories, one chooses a very fine triangulation of $\Sigma$ and Wilson loops with $\gamma$ adapted to the triangulation. The above formula, which allows arbitrary regular $\gamma$ (and then extends to arbitrary continuous $\gamma$) takes into account the continuum limit, since as the triangulation gets finer and finer, one can approximate arbitrary curves. The end result is that (\ref{def:WLE}) provides us with an operational definition of the Yang-Mills measure. A more refined treatment \cite[Definition 2.10.4]{Levy} shows that such a Yang-Mills measure is a measure on $\mc{F}(L\Sigma, G)/\mc{G}$, the quotient space of functions $\mc{F}(L\Sigma, G)$ from $L\Sigma$ (the based loops on $\Sigma$) to $G$, modulo the action of the group $\mc{G}$ of gauge transformations. (Given $f \in \mc{F}(L\Sigma, G)$ and $g \in \mc{G}$, we have $(g\cdot f)(\gamma) = g(\gamma(0)) f(\gamma) g(\gamma(0))^{-1}$.) Leaving out many details, such a measure is obtained by constructing a probability space in which one has a $G$-valued random variable $H_\gamma$, for every $\gamma \in L\Sigma$, defined as follows. Given $\gamma$ (assumed to be regular without loss of generality), the law of $H_\gamma$ is obtained by conditioning the Yang-Mills measure $d\mu_{\Gamma}$ on $G^{|E(\Gamma)|}$, where $\Gamma \supset \hat\gamma$; namely, one conditions on the element $g_\gamma$ determined by $\gamma$. Such a law determines a $G$-valued process $\{H_\gamma\}_{\gamma \in L\Sigma}$ and thus a measure on $G^{L\Sigma} = \mc{F}(L\Sigma, G)$, which due to its invariance under $\mc{G}$, descends to a measure on $\mc{F}(L\Sigma, G)/\mc{G}$.

One can informally regard a measure on $\mc{F}(L\Sigma, G)/\mc{G}$ as a measure on the space $\A/\G$ of connections $\A$ modulo gauge transformations. Indeed, we have an inclusion $\A/\G \to \mc{F}(L\Sigma, G)/\mc{G}$, given by mapping a connection $A$ to the function $\mr{hol}_{(\cdot)}(A)$, which takes a based loop $\gamma$ to $\mr{hol}_\gamma(A)$.  This makes sense for $A$ sufficiently regular; if $A$ is continuous say, then the resulting holonomy function will depend continuously on $\gamma$. Thus, $\mc{F}(L\Sigma, G)/\mc{G}$ represents ``generalized connections". One can also interpret the Yang-Mills measure in terms of white-noise measures, with Wilson loop observables being given by stochastic parallel transport \cite{Dri, Sen}.

We only mention these measure-theoretic interpretations so as to note that Yang-Mills theory in two dimensions has a rigorous construction in accords with the demands of constructive quantum field theory \cite{GJ}. For our present purposes however, we are only concerned with the formula (\ref{def:WLE}), which gives the exact expectation value of a Wilson loop observable.

We will be mainly concerned with the case $\Sigma = S^2$ for reasons explained in the introduction. To that end, let us specialize (\ref{def:WLE}) to $\Sigma = S^2$ and the case of a simple closed curve, which we will analyze in Section \ref{Sec:Asymp}.

\begin{Corollary}
  Let $\gamma$ be a simple closed curve on $S^2$, with $R_1$ and $R_2$ the connected components of $S^2\setminus\gamma$. Then
    \begin{align}
      \left<W_{f,\gamma}\right>_\Sigma = \frac{\int_G f(g)K_{\lambda_0|R_1|}(g)K_{\lambda_0|R_2|}(g)dg}{K_{\lambda_0|S^2|}(1)} \label{exact1}
    \end{align}
\end{Corollary}

\Proof The graph of $\gamma$ consists of a single edge which we label by $g$. We now apply (\ref{def:WLE}) with $\Gamma = \hat\gamma$ and make use of properties (\ref{Kprop}) and (\ref{eq:heatconv}).\End

\begin{Remark}\label{eq:R2}
 The above analysis carries over to $\R^2$ (equipped with the standard area form). For unbounded regions of $\R^2$, we use $K_\infty(g) \equiv 1$. Using the fact that $\int_G K_t(g)dg = 1$ for all $t$, the partition function $Z_{\R^2,\lambda_0d^2x}$ simply becomes unity. Since the area of $\R^2$ is not normalizable, the effective coupling constant for Yang-Mills theory on $\R^2$ is simply $\lambda = \lambda_0$. (We can think of $\lambda$ as $\lambda_0$ times the area of the unit square, which is one.)
\end{Remark}

\begin{Remark}\label{Rem:topology}
 The Yang-Mills measure we describe in fact consists of an average of all possible topological bundle types over $\Sigma$ \cite{Levy2}. In \cite{Levy2}, to a graph $G$ over $\Sigma$ and a bundle type over $\Sigma$, a more refined construction associates to such datum a Yang-Mills measure on a configuration space that covers $G^{E(\Gamma)}$. However, when $\lambda \to 0$, the dominant contribution to the Yang-Mills measure comes from the trivial bundle. On the continuum side, this arises from connections near the trivial connection having holonomies close to $1$. On the lattice side, this arises from $K_t(g)$ concentrating near $g = 1$ for $t$ small. In this way, nontrivial bundles will make asymptotically zero contributions due to exponentially small factors arising from the topology (i.e. instantons). Thus, for our analysis at small coupling, it suffices to work with the averaged Yang-Mills measure described above. For $G$ simply-connected, all $G$-bundles over $\Sigma$ are trivial.
\end{Remark}

We will compute the asymptotics of (\ref{exact1}) as $\lambda = \lambda_0|S^2| \to 0$ in Section \ref{Sec:Asymp} to give an explicit example of the relation between exact asymptotics and perturbation theory.

\section{Perturbation Theory}\label{Sec:Pert}

In the previous section, we discussed the full quantum Yang-Mills theory, in which one obtains expectation values for all possible Wilson loops $W_{f,\gamma}$ from a well-defined Yang-Mills measure. This gives a complete, rigorous construction of the quantum Yang-Mills theory insofar as it provides a mathematical realization of the formal expression 
\begin{equation}
  \left<W_{f,\gamma}\right>_\Sigma = \frac{\int dA\, W_{f,\gamma}(A) e^{-S_{YM}(A)}}{\int dA\, e^{-S_{YM}(A)}}, \label{formalexp}
\end{equation}
which supposes the existence of a suitable Yang-Mills measure $dA\, e^{-S_{YM}(A)}$ on the space of connections (modulo gauge). Here, our basic field $A$ is a connection on the trivial $G$-bundle over $\Sigma$ so that the space of connections $\A$ can be identified with $\Omega^1(\Sigma; \g)$ and the group of gauge transformations can be identified with $G$-valued functions on $\Sigma$.

In this section, we discuss the \textit{perturbative quantization} of Yang-Mills theory, whereby one computes expectation values not with regard to a true measure but through a perturbative expansion in Feynman diagrams about a minimal configuration of the classical action. In other words, we proceed by way of the standard paradigm of quantum field theory since its earliest inception: regard the right-hand side of (\ref{formalexp}) not as a number but as a notational device for generating a formal power series in the coupling constant $\lambda$.

The methods by which one generates such a formal power series, and the manner in which one establishes its resulting properties, are described with mixed approaches and differing degrees of rigor and generality in the physics and mathematics literature. For gauge theories, whatever approach one adopts, one ultimately needs to choose a gauge-fixing condition and show that the resulting outcome, i.e., the series expansion obtained from (\ref{formalexp}), is independent of the choice of gauge. For the case at hand, a degree of sophistication is required since we work on curved space, in which case the majority of treatments which quantize Yang-Mills theory on flat space do not readily apply. Indeed, flat space techniques such as working in momentum space and using dimensional regularization \cite{Col, PS} are not available to us.

We thus find it instructive to describe our quantization procedure from first principles, albeit in a succinct manner. That being so, \textit{this section is written using both rigorous mathematics and the physically motivated ideas from which they are derived.} We trust that the reader finds this two-track narrative enlightening rather than confusing.

We have four main tasks. First, we describe mathematically precisely how to generate the perturbative Feynman diagrammatic expansion of 2D Yang-Mills theory. Here, we present a variety of constructions. First, we present the standard Faddeev-Popov procedure to apply Coulomb (Landau) gauge-fixing. Due to its generality and natural geometric underpinnings, we regard Coulomb gauge as the most fundamental choice of gauge. We then describe the holomorphic and axial gauges, which require underlying topological assumptions to implement, but take an especially simple form in two dimensions.

Second, we show that the Coulomb gauge expectation of a Wilson loop observable is independent of the auxiliary Riemannian metric chosen to perform Coulomb gauge-fixing. This second step requires greater sophistication than that involved in the first step, whereby we use a blend of ideas from \cite{AS1, AS2, CM, Cos} to establish gauge-invariance in Section \ref{SecGI}. In broad strokes, we apply the Batalin-Vilkovisky (BV) formalism in the form developed by \cite{Cos}, which provides a powerful algebraic framework by which to analyze the gauge-dependence of a quantization scheme. The BV formalism is used to show that the family of gauges obtained from interpolation between two metrics all yield equal perturbative Wilson loop expectation values. As necessary step of this procedure is showing that the BV and Faddeev-Popov formulations of Coloumb gauge are equivalent, see Lemma \ref{Lemma:BV=FP}. Moreover, the metric-indepence of Coulomb gauge is also what gives rise to the invariance of perturbative Wilson loop expectations under area-preserving diffeomorphisms, see Theorem \ref{Thm:Area}.

Third, we establish the equivalence between Coulomb gauge and holomorphic gauge, i.e. that Wilson loop expectations in these gauges agree to all orders in perturbation theory. This also involves using the Batalin-Vilkovisky formalism, where it is applied to a family of gauges interpolating between Coulomb gauge and holomorphic gauge. To the best of our knowledge, this family of gauges has not been previously considered in the literature.

Our fourth and final task in this section supplies key computations that go into the tasks above. Namely, we show that Yang-Mills theory is finite in Coulomb gauge and in the family of gauges we construct interpolating between Coulomb and holomorphic gauge. That is, no counterterms are needed as the heat-kernel regularization parameter is sent to zero.

As a guide to the reader, the Sections \ref{SecGI} and \ref{Sec:C=hol} using the BV formalism are the most abstract and conceptually demanding. For those mainly interested in explicit computations, Sections \ref{SecGI}--\ref{Sec:C=hol} may be treated as a theoretical black box, with Section \ref{Sec:Finite} supplying more concrete details.

\subsection{Definitions of Gauges}

In order to compute perturbative Wilson loop expectations, we need to choose a suitable gauge-fixing condition. This is because the path integrals in (\ref{formalexp}) should only be over the space of physically distinct configurations, i.e., those which are gauge-inequivalent. A \textit{gauge-fixing condition} is thus the choice of a local-slice for the action of the gauge-group\footnote{Or possibly with respect to based gauge transformations, i.e. those that are fixed to be the identity at a point. Based gauge transformations act freely on the space of connections and its coset space with respect to all gauge transformations is simply a copy of $G$. Since the latter is finite-dimensional, this residual gauge-freedom is unproblematic.}, i.e. a submanifold transverse to the action of the gauge group (within the vicinity of the trivial connection, the connection about which we perform perturbation theory). We describe the different (families of) gauge-fixing conditions we employ, namely the Coulomb, holomorphic, and axial gauges, and the manner in which they determine a corresponding perturbative Wilson loop expectation through a Feynman diagrammatic expansion.

\subsubsection{Coulomb gauge (via Faddeev-Popov quantization)}

Consider any compact Riemann surface $\Sigma$ equipped with an area form $d\sigma$.

\begin{Definition}
  A Riemannian metric $g = g_{ij}$ on $(\Sigma, d\sigma)$ is \textit{compatible} if its induced area form agrees with $d\sigma$.
\end{Definition}


A compatible metric $g_{ij}$ along with the inner product $\left<\cdot,\cdot\right>$ on $\g$ yields for us an inner product $\int_\Sigma \left<\alpha \wedge *\beta\right>$ on $\Omega^\bullet(\Sigma; \g)$ and a corresponding adjoint operator $d^*: \Omega^\bullet(\Sigma; \g) \to \Omega^{\bullet - 1}(\Sigma; \g)$ of the exterior derivative $d$. We consider the following \textit{Coulomb gauge-fixing condition}
\begin{equation}
d^*A = 0 \label{eq:Coulomb_gauge}
\end{equation}
and let
\begin{equation}
  \mc{A}_C = \{A : d^*A = 0\}. \label{eq:Cslice}
\end{equation}
Since gauge transformations act via $A \mapsto gAg^{-1} + gdg^{-1}$, the above gauge-fixing condition eliminates all infinitesimal gauge degrees of freedom (which span the space $\im d$) within a neighborhood of the trivial connection. We call $g_{ij}$ a choice of \textit{gauge-fixing metric}.


Our path integral
$$\frac{1}{Z}\int dA\, W_{f,\gamma}(A) e^{-S_{YM}(A)}$$
is to be replaced with the gauge-fixed path integral
\begin{equation}
  \frac{1}{Z}\int_{\mc{A}_C} dA\, \mr{det}(d^*d_A) W_{f,\gamma}(A) e^{-S_{YM}(A)}. \label{eq:FPint1}
\end{equation}
The determinant factor $\mr{det}(d^*d_A)$ is the Faddeev-Popov determinant that weights gauge orbits appropriately\footnote{For a rigorous treatment in the finite dimensional setting, see e.g. \cite{Ngu}.}. This determinant can be evaluated via the introduction of anticommuting fields, or ghosts. This is because there is a well-defined theory for fermionic integration in finite dimensions that produces this determinant factor (see Appendix \ref{Sec:Wick}), and we can extrapolate from this an analogous procedure in the infinite dimensional case. We proceed as follows:

Introduce a pair of ghost fields $\omega, \bar\omega$, which are each $\g$-valued functions on $\Sigma$. To keep them separate, we denote the space of $\omega$ and $\bar\omega$ by $\Omega^0(\Sigma; \g)$ and $\overline{\Omega}^0(\Sigma; \g)$, respectively. Define 
$$\Omega^0_\bot(\Sigma;\g) = \{\omega \in \Omega^0(\Sigma;\g) : \int_\Sigma \omega\, d\sigma = 0\}$$
and similarly for $\overline{\Omega}^0_\bot(\Sigma; \g)$. Let
$$\frak{C} = \Omega^0_\bot(\Sigma; \g) \oplus \overline{\Omega}^0_\bot(\Sigma; \g) \oplus \A_C,$$
consisting of the total space of gauge-fixed connections and ghosts that are orthogonal to constants. The latter condition is to eliminate the kernel of $d: \Omega^0(\Sigma; \g) \to \Omega^1(\Sigma; \g)$, which arises from the Lie algebra of the constant gauge transformations (which act trivially on the trivial connection).

We now replace (\ref{eq:FPint1}) with
\begin{equation}
  \frac{1}{Z}\int_{\frak{C}} d\bar\omega d\omega dA\, W_{f,\gamma}(A) e^{-S(A, \bar\omega,\omega)} \label{eq:FPint2}
\end{equation}
where\footnote{It is not necessary to multiply $\int (\bar\omega, d^*d_A\omega)$ by $\frac{1}{\lambda_0^2}$, but this ensures that all terms in the perturbative expansion are weighted equally in the coupling constant.}
$$S = \frac{1}{2\lambda_0}\int \left<F_A \wedge *F_A\right> + \frac{1}{\lambda_0}\int \left<\bar\omega, d^*d_A\omega\right> d\sigma.$$
The integration over $\omega, \bar\omega$ formally produces the determinant factor $\mr{det}(d^*d_A)$ via Lemma \ref{LemmaDet}.

It is with (\ref{eq:FPint2}) that we can perform a Feynman diagrammatic expansion. This is done as follows.
Group the extended Yang-Mills action into a quadratic kinetic part and the remaining higher order interaction part, which one regards as a perturbation of the former. Here, we use $F_A = dA + \frac{1}{2}[A,A]$ and $d_A = d + [A,\cdot]$. In this way, we have
\begin{align}
  e^{-S} = e^{-S_{kin}}e^{I}
\end{align}
where
\begin{align}
S_{kin} &= \frac{1}{2\lambda_0}\int \left<A \wedge *\, d^*dA\right> + \frac{1}{\lambda_0}\int \left<\bar\omega, d^*d\omega\right> d\sigma \label{eq:FPSkin} \\
I &=  -\frac{1}{2\lambda_0}\int \left<[A,A] \wedge * dA\right> - \frac{1}{8\lambda_0}\int \left<[A,A] \wedge *[A,A]\right> - \frac{1}{\lambda_0}\int \left<\bar\omega, d^*[A,\omega]\right>d\sigma \label{eq:FPI}
\end{align}

The next step is to write
\begin{align}
  \frac{1}{Z}\int_{\frak{C}} d\bar\omega d\omega dA\, W_{f,\gamma}(A) e^{-S(A, \bar\omega,\omega)} = \frac{1}{Z}\int_{\frak{C}} d\bar\omega d\omega dA\, e^{-S_{kin}}W_{f,\gamma}(A)e^{I} \label{eq:pathintexpand}
\end{align}
and then expand $e^I$ as a formal series in powers of $\lambda_0=e^2$. These terms, multiplied with $W_{f,\gamma}$, each give multilinear functionals on the space of fields. One then ``integrates" each of these terms against the ``Gaussian measure"  $\frac{1}{Z}d\bar\omega d\omega dA e^{-S_{kin}}$ defined on $\frak{C}$, thereby producing a formal power series in $\lambda$. In reality, what one is really doing is performing the algebraic operation given by Wick's Theorem. This operation is described in Appendix \ref{Sec:Wick}. We describe how this generalizes to the quantum field theoretic setting at hand.

To invoke the appropriate analog of Lemma \ref{LemmaWickP}, we need to describe the propagator $P$ as well as the appropriate expansion of $W_{f,\gamma}e^I$ as a Taylor series (i.e. an infinite sum over polynomial functions). The Taylor expansion of $e^I$ is automatic from the definition of the exponential function and we obtain a formal series in powers of $\lambda$. For $W_{f,\gamma}$, we obtain a Taylor expansion via the representation of $\mr{hol}_\gamma(A)$ in terms of path ordered exponentials. Namely,
\begin{align*}
  \mr{hol}_\gamma(A) &= \mc{P}\exp\left(-\int_\gamma A\right) \\
  &= 1 + \sum_{n=1}^\infty (-1)^n\int_{1 \geq t_n \geq \cdots \geq t_1 \geq 0} A(t_n)\cdots A(t_1)
\end{align*}
where $A(t) = A_\mu(\gm(t))\gm^\mu(t)$. Note that this presentation assumes we have chosen an embedding $G$ into the group of unitary matrices $U(V)$ on a vector space $V$, so that elements of $\g \subset \mr{End}(V)$ can be multiplied.

Without loss of generality, we can take $f = \tr_V$ to be trace in an irreducible representation $\rho: G \to \mr{End}(V)$, since the linear span of such functions is dense in the space of conjugation-invariant functions on $G$. This allows us to expand $W_{f,\gamma}(A)$ as a Taylor series in $A$:
\begin{equation}
  W_{f,\gm}(A) = \tr_V(1) + \sum_{n=1}^\infty(-1)^n\int_{1 \geq t_n \geq \cdots \geq t_1 \geq 0} \tr_V\Big(\rho(A(t_n))\cdots \rho(A(t_1))\Big). \label{eq:W_expansion}
\end{equation}
Here, we also write $\rho: \g \to \mr{End}(V)$ to denote the induced Lie algebra homomorphism. The above representation uses the fact that parallel transport is equivariant with respect to group homomorphisms:$$\rho(\mr{hol}_\gm(A)) = \mr{hol}_\gm(\rho(A)).$$
Altogether, this describes the expansion of the integrand $W_{f,\gamma}(A)e^I$ as a Taylor series in the field variables $A, \omega, \bar\omega$.

Next, the gauge-fixed path integral (\ref{eq:FPint2}) determines for us a \textit{propagator} $P$, which is a Green's operator determined by the kinetic operator occurring in $S_{kin}$. Specifically, we have the orthogonal decomposition
\begin{align*}
  \mc{A} &= \im\! *\!d \oplus \ker d \\
  \Omega^0(\Sigma; \g) &= \Omega^0_\bot(\Sigma;\g) \oplus \R \\
  \overline{\Omega}^0(\Sigma; \g) &= \overline{\Omega}^0_\bot(\Sigma;\g) \oplus \R.
\end{align*}
The decomposition for $\mc{A}$ depends on our compatible metric, while the other two depend only on $d\sigma$. The kinetic action $S_{kin}$ is formed out of the Laplace-Beltrami operator $\Delta = \Delta_g$ restricted to $\frak{C}$. The Green's operator we are interested in is the operator
\begin{equation}
P = P_C :=  \Delta^{-1}|_{\im *d \oplus \Omega^0_\bot(\Sigma;\g) \oplus \overline{\Omega}^0_\bot(\Sigma;\g)}
\end{equation}
which extends to the zero operator on the orthogonal complement. Here, the $C$ denotes Coulomb gauge (with respect to a chosen metric). We have
\begin{equation}P_C = P^{bos} \oplus P^{fer},\label{eq:FPprop}
\end{equation}
 according to the restriction of $P_C$ to bosonic $A$ and fermionic $\omega, \bar\omega$ fields.

More explicitly, the decomposition (\ref{eq:FPprop}) is given as follows. The inner product on $\Omega^1(\Sigma; \g)$ allows us to identify the operator $P^{bos}$ with its integral kernel\footnote{In what follows, all tensor products are completed, see footnote \ref{fnote:complete}.} \begin{align}
  P^{bos}_{\mu,\nu}(x,y) \otimes {\mr{id}_\g} \in \big(\im\!*\!d \otimes \im\!*\!d\big) \otimes \big(\g \otimes \g\big) \subset \big(\Omega^1(\Sigma) \otimes \Omega^1(\Sigma)\big) \otimes \big(\g \otimes \g\big). \label{eq:Pboskernel}
\end{align}
Here, we have separated the integral kernel of $P^{bos}$ as the part $P^{bos}_{\mu,\nu}(x,y)$ which acts as $\Delta^{-1}$ on scalar-valued differential forms and the identity operator on $\g$. We can write the latter an element $e_a \otimes e_a$ of $\g \otimes \g$ using the inner product on $\g$, where $e_a$ is an orthonormal basis of $\g$. If we wish to incorporate the Lie-algebra dependence into the notation, we write $P^{bos, ab}_{\mu,\nu}$.

\begin{Notation}
 Hereafter, the use of variables $x$ and $y$ in an integral kernel expression such as $P(x,y)$ will always denote outgoing and incoming ``dummy'' variables. Thus, in the above, the operator associated to the integral kernel $P^{bos}(x,y)$ is given by
 \begin{align}\alpha \mapsto P^{bos}(\al) &= \int_{\Sigma_y} P^{bos,ab}(x,y) \wedge *\alpha^b(y)\\
 &= \int_{\Sigma_y} P^{bos}(x,y) \wedge *\alpha(y).
 \end{align}
(Here, we have suppressed differential form indices, as we typically do for all differential-form objects; in the first line, we made the Lie algebra dependence explicit, and in the second line, the identity operation on the Lie algebra is suppressed.) This notation is to distinguish an operator $P$ (acting on differential forms) from its corresponding integral kernel $P(x,y)$ (with respect to a specified pairing, either the $L^2$-pairing or some version of the wedge pairing which we consider later, tensored with the inner product pairing on $\g$). However, to avoid overly cumbersome notation, instead of $\pd_{P(x,y)}$, which denotes contraction with the integral kernel $P(x,y)$ (see the appendix) we will instead write $\pd_P$. By slight abuse of language, we will refer to both the operator $P$ and its integral kernel $P(x,y)$ as being a propagator. At times, we may also drop the dependence of $P(x,y)$ on its Lie algebra part, since it always the identity tensor. 
\end{Notation}

The integral kernel for $P^{bos}$ yields a contraction operator $\pd_{P^{bos}}$ satisfying 
\begin{align}
  \pd_{P^{bos}}A^a_\mu(x)A^b_\nu(x) = P^{bos}_{\mu,\nu}(x,y)\delta^{ab}. \label{eq:2pt_fun}
\end{align}
In other words, (\ref{eq:2pt_fun}) expresses the bosonic two-point function for Yang-Mills theory in Coulomb gauge.

The fermionic propagator $P^{fer}$ is obtained from restricting $P_C$  to $\Omega^0_\bot(\Sigma;\g) \oplus \overline{\Omega}^0_\bot(\Sigma;\g)$ suitably interpreted. Namely, we have the skew-symmetric pairing \begin{align*}
\left<\bar\omega,\omega\right>_{fer} = -\left<\omega,\bar\omega\right> = \int \left<\bar\omega,\omega\right>d\sigma
\end{align*}
with which we can use to express the integral kernel of $P^{fer}$ as an element of $\Lambda^2(\Omega^0_\bot(\Sigma;\g) \oplus \overline{\Omega}^0_\bot(\Sigma;\g))$. Thus, letting $G^{(0)}(x,y)$ denote the Green's function for $\Delta^{-1}$ on $\Omega^0_\bot(\Sigma; \g)$, i.e.
$$(\Delta^{-1}\al)(x) = \int G^{(0)}(x,y)\alpha(y)d\sigma(y), \qquad \al \in \Omega^0_\bot(\Sigma; \g),$$
then $P^{fer}$ satisfies
\begin{align*}
  \pd_{P^{fer}}\omega^a(x)\bar\omega^b(y) = G^{(0)}(x,y)\delta^{ab} \\
  \pd_{P^{fer}}\bar\omega^a(x)\omega^b(y) = -G^{(0)}(x,y)\delta^{ab}.
\end{align*}
Note that the above expresses the fermionic sign rule in which a minus sign is picked up by switching  the order of $\bar\omega$ and $\omega$. Altogether, the above defines $P^{fer}(x,y)$ uniquely as an element of $$\Lambda^2(\Omega^0_\bot(\Sigma;\g) \oplus \overline{\Omega}^0_\bot(\Sigma;\g)) \subset \left(\overline{\Omega}^0_\bot(\Sigma;\g) \otimes \Omega^0_\bot(\Sigma;\g)\right) \oplus \left(\Omega^0_\bot(\Sigma;\g) \otimes \overline{\Omega}^0_\bot(\Sigma;\g)\right).$$

We now suppose $\Sigma = S^2$. This way, $\im\! *\!d = \mc{A}_C$ and $\Delta$ restricted to $\mc{A}_C$ has no zero modes. This allows us to conclude that the Feynman diagrammatic expansion of (\ref{eq:pathintexpand}), following Lemma \ref{LemmaWickP}, is formally given by the expression\footnote{If there were zero modes, the definition of $\left<W_{f,\gamma}\right>_{pert}$ should be modified to have a residual integration over these modes after performing the expansion (\ref{eq:naiveexp}).}
\begin{equation}
\left<W_{f,\gamma}\right>_{\Sigma, C} ``=" e^{\lambda_0\pd_{P_C}}\left(W_{f,\gamma}e^{I}\right)\Big|_{conn, 0}. \label{eq:naiveexp}
\end{equation}
Here, the subscripts ``$conn$" and ``$0$" refer to the fact that we only wish to consider those Feynman diagrams which consist of a single component connected\footnote{In path integral notation, the normalization factor $\frac{1}{Z}$ in (\ref{eq:pathintexpand}) eliminates disconnected components of Feynman diagrams.} to $W_{f,\gamma}$ and which have no external edges, respectively.

The formal definition (\ref{eq:naiveexp}) fails a priori because the resulting Feynman integrals we obtain are divergent. Thus, we need to choose a suitable regularization procedure, i.e. a way of mollifying the integral kernel $P_C(x,y)$ to a smooth one $P_{C,\eps}(x,y)$, $\eps > 0$, with $P_{C,\eps} \to P_C$ as $\eps \to 0$. Our regularization procedure is via the heat kernel method, which regulates $P_C$ via
\begin{align}
  P_{C,\eps} = \int_\eps^\infty e^{-t\Delta}dt\bigg|_{\frak{C}}. \label{eq:heatreg}
\end{align}
Note that as $\eps \to 0$, we recover $\Delta^{-1}|_{\frak{C}}$, which is most easily seen by diagonalizing $\Delta$ and working on individual eigenspaces.

The integral kernel of (\ref{eq:heatreg}) is smooth for all $\eps > 0$. Thus, we can replace $P_C$ with $P_{C,\eps}$ in (\ref{eq:naiveexp}) and obtain a well-defined formal power series in $\lambda$. Having chosen a regularization procedure as above, we also need to perform renormalization, i.e. \textit{counterterms} need to be introduced. These are additional $\eps$-dependent (and $\lambda$ dependent) interactions $I^{CT}_\eps$ one adds to $I$. One is supposed to arrange the $I^{CT}_\eps$ so that
$$\lim_{\eps \to 0} e^{\lambda_0\pd_{P_{C,\eps}}}\left(W_{f,\gamma}e^{I+I^{CT}_\eps}\right)\Big|_{conn, 0}$$
exists as a formal power series in $\lambda$. (In general, additional counterterms may also be needed to renormalize observables.) A very nice feature of two-dimensional Yang-Mills theory is that in fact no counterterms are needed as $\eps \to 0$. This is proven in Theorem \ref{ThmFinite}. Thus, we can form the following definition:

\begin{Definition} Fix a compatible metric on $\Sigma = S^2$. The perturbative expectation value of a Wilson loop $W_{f,\gamma}$ in  Coulomb gauge (\ref{eq:Cslice}) is the formal series in $\lambda=\lambda_0|\Sigma|$ defined by
  \begin{equation}
\left<W_{f,\gm}\right>_{\Sigma,C} = \lim_{\eps \to 0} e^{\lambda_0\pd_{P_{C,\eps}}}\left(W_{f,\gamma}e^{I}\right)\Big|_{conn, 0}. \label{eq:defpertexp}\end{equation}
\end{Definition}
This yields for us a mathematically rigorous definition of the perturbative expectation value of a Wilson loop in Coulomb gauge, with the preceding discussion revealing its physical origins.\\

On $\R^2$, defining Coulomb gauge in a way that ensures that Wilson loop expectations are well-defined requires extra care because of the noncompactness of $\R^2$. That is, we have the problem of infrared diverges, which manifests itself by way of $\frac{1}{|\xi|^2}$ (the Fourier transform for the Green's function of the Laplacian away from $\xi = 0$) not being integrable near the origin. In particular, the limit
$$\lim_{L \to \infty} \int_\eps^L e^{-t\Delta}dt$$
does not exist, since we can think of $L$ as an infrared regulator.

We proceed by a roundabout path: we define Coulomb gauge expectations of Wilson loops on $\R^2$ as a decompactification limit of such expectations on $S^2$:

\begin{Definition}\label{Def:C_R2} Equip $\R^2$ with the canonical flat area form. The perturbative expectation value of a Wilson loop $W_{f,\gamma}$ in Coulomb gauge (\ref{eq:Cslice}) on $\R^2$ is the formal series in $\lambda_0$ defined as follows. Consider the decompactification limit $(S^2, d\sigma_i) \to \R^2$ in the sense that we regard $\gamma \subset \R^2 = S^2\setminus\{\infty\}$ and let $d\sigma_i$ be a sequence of round area forms that converge to the area form on $\R^2$. Then we define
\begin{equation}
 \left<W_{f,\gm}\right>_{\R^2,C} = \lim_{S^2 \to \R^2}\left<W_{f,\gm}\right>_{S^2,C},
\end{equation}
where the limit on the right-hand side is with respect to any sequence of metrics compatible with the sequence of area forms $d\sigma_i$.
\end{Definition}

That this definition is well-defined will follow once we show that Wilson loop expectations on $S^2$ are independent of the choice of compatible metric, that Coulomb gauge is equivalent to holomorphic gauge, and that the above limit exists if Coulomb gauge is replaced with holomorphic gauge (see Proposition \ref{Prop:well_defined}). We merely record the definition here for convenience and will make use of it in Section \ref{Sec:Asymp}.

\subsubsection{Holomorphic Gauge}

Computations in Coulomb gauge are difficult to perform due to the presence of many complicated Feynman diagrams arising from the interactions $I$. This leads us to conside the more computationally tractable holomorphic gauge in which there are no interactions, i.e., the only terms which contribute to Feynman diagrams are those arising from $W_{f,\gm}$ in (\ref{eq:W_expansion}). The interpretation of holomorphic gauge as a gauge-fixing condition is subtle, see \cite{Ngu2016}. Nevertheless, holomorphic gauge is defined as follows.

Given a fixed area form on $\Sigma$, the choice of a compatible metric is equivalent to the choice of a conformal structure.
So for $\Sigma = S^2$, pick a compatible metric and consider the resulting complex structure it induces. We can use it complexify the space of connections $\A = \Omega^1(\Sigma; \g)$ to $\A_c = \Omega^1(\Sigma; \g_c)$ where $\g_c$ is the complexification of $\g$. The Yang-Mills action extends complex-linearly to connections belonging to $\A_c$ (by extending the inner product on $\g$ complex-linearly). We say $A$ is in \textit{holomorphic gauge} if it is a differential form of type $(1,0)$, i.e. \begin{equation}
 A \in \Omega^{1,0}(\Sigma; \g_c). \label{eq:hol_gauge}                                                                                                                                                                                                                                                                                                                                                                                                                                                                                                                                                                          \end{equation}
The Yang-Mills action in holomorphic gauge becomes
\begin{equation}
YM(A) = \int \left<A \wedge \bar\pd * \bar\pd A\right>, \qquad A \in \Omega^{1,0}(\Sigma; \g_c). \label{eq:holgauge}
\end{equation}
Indeed, the quadratic terms in the curvature $F_A$ vanish in holomorphic gauge.

On $S^2$, since there are no nontrivial holomorphic $1$-forms, the pairing (\ref{eq:holgauge}) is nondegenerate. Hence, the operator $\bar\pd*\bar\pd: \Omega^{1,0}(\Sigma;\g_c) \to  \Omega^{0,1}(\Sigma;\g_c)$ is invertible and it has an integral kernel, with respect to the wedge pairing, belonging to $\Omega^{1,0}(\Sigma) \otimes \Omega^{1,0}(\Sigma)$. This
yields a corresponding \textit{holomorphic gauge propagator} 
$$P_{hol} \in \Omega^{1,0}(\Sigma; \g_c) \otimes \Omega^{1,0}(\Sigma; \g_c)$$
by tensoring with the identity tensor in $\g \otimes \g$. Explicitly, if $z$ and $w$ are local holomorphic coordinates on $\bC = S^2\setminus\{0\}$ with respect to the standard conformal structure on $S^2$, we have
\begin{equation}
P_{hol}(z,w) = \left(\frac{1}{\pi}\frac{1}{(1+|z|^2)(1+|w|^2)}dz \frac{\bar z - \bar w}{z - w}dw\right)e_a\otimes e_a. \label{eq:holP}
\end{equation}
Note that because the holomorphic gauge propagator is uniformly bounded, there is no difficulty in defining integrals of the $P_{hol}$ supported along a regular curve $\gm$.

\begin{Definition}
Let $\Sigma = S^2$. Fix a compatible metric on $S^2$, which determines a holomorphic gauge propagator. The perturbative expectation value of a Wilson loop $W_{f,\gm}$ in holomorphic gauge (\ref{eq:hol_gauge}) is the formal series in $\lambda = \lambda_0|\Sigma|$ defined by
\begin{equation}\left<W_{f,\gm}\right>_{hol,S^2} = \left<W_{f,\gm}\right>_{hol} := e^{\lambda_0\pd_{P_{hol}}}W_{f,\gm}\bigg|_0, \label{eq:holexp}
\end{equation}
where $P_{hol}$ is given by (\ref{eq:holP}). (Note that because there are no interactions, all Feynman diagrams are automatically connected.)
\end{Definition}

On $\R^2 = \bC$, the holomorphic gauge propagator, which is the unique homogeneous Green's function for $\bar\pd * \bar \pd$, can also be explicitly computed and it is given by 
\begin{equation}
P_{hol}(z,w) = \left(\frac{1}{4\pi}dz \frac{\bar z - \bar w}{z - w}dw\right)e_a \otimes e_a. \label{eq:holP_R2}
\end{equation}
Thus, we have
\begin{Definition}
Equip $\R^2$ with the canonical flat area form and flat metric. The perturbative expectation value of a Wilson loop $W_{f,\gm}$ in holomorphic gauge (\ref{eq:hol_gauge}) on $\R^2$ is the formal series in $\lambda_0$ defined by
\begin{equation}\left<W_{f,\gm}\right>_{hol,\R^2} := e^{\lambda_0\pd_{P_{hol}}}W_{f,\gm}\bigg|_0, \label{eq:holexpR2}
\end{equation}
where $P_{hol}$ is given by (\ref{eq:holP_R2}).
\end{Definition}

\subsubsection{Axial Gauge}\label{Sec:Axial}
On $\R^2$, since we have a global coordinate system $(x^0,x^1$), any connection $A = A_0dx^0 + A_1dx^1$ can be gauge-transformed into one in which
\begin{align}
A_1 & \equiv 0. \label{eq:ax-gauge}
\end{align}
For $A$ satisfying (\ref{eq:ax-gauge}), the Yang-Mills action becomes
\begin{equation}
\frac{1}{2\lam_0}\int_{\R^2} (\pd_1A_0)^2 d^2x \label{eq:YM-axial-gauge}
\end{equation} and the propagator in axial-gauge is determined from the appropriate Green's function for $\pd_1^2$. This axial-gauge propagator is given by
\begin{equation}
P_{pax}(x,y) = -\frac{1}{2}|x^1 - y^1|\delta(x^0 - y^0)e_a\otimes e_a, \label{eq:Pax}
\end{equation}
where the subscript $pax$ stands for partial axial-gauge\footnote{There are actually two different axial gauges, complete axial-gauge and partial axial-gauge. The latter has a simpler propagator and is the one most familiar, and so we use choose this one in (\ref{Def:Wax}), even though partial axial-gauge is not a true gauge in the sense that there are still infinitely many gauge-degrees of freedom remaining (the $x^1$-independent gauge-transformations). It is shown in \cite{NguSYM} that the complete axial-gauge and partial axial-gauge yield equivalent Wilson loop expectations.}.

\begin{Definition}\label{Def:Wax}
 On $\R^2$, the perturbative Wilson loop expectation in axial-gauge (\ref{eq:ax-gauge}) is the formal series in $\lam_0$ given by
 \begin{equation}
  \left<W_{f,\gm}\right>_{\R^2,ax} = e^{\lam_0\pd_{P_{pax}}}W_{f,\gm}\bigg|_0. \label{eq:defWax}
 \end{equation}
\end{Definition}

The experienced reader will recognize that even though $P_{pax}$ is singular, (\ref{eq:defWax}) yields a meaningful ansatz for generating integrals. For a more careful treatment of Definition \ref{Def:Wax}, see \cite{NguSYM}.

\subsection{Metric-independence of Coulomb gauge}\label{SecGI}

In this section, we prove the following theorem:

\begin{Theorem}\label{ThmGI}
Let $\Sigma = S^2$. Then
  $\left<W_{f,\gamma}\right>_{\Sigma, C}$ is independent of the choice of gauge-fixing metric.
\end{Theorem}

We prove this theorem using the Batalin-Vilkovisky (BV) formalism. The power of the BV formalism is that it captures the notion of gauge-invariance in an algebraic manner that is well-adapted for perturbative quantization. For additional background and insights regarding this formalism, we refer the reader to \cite{Cos, Sch, Wein}. We will dive directly into the formalism, which is an adaptation of the approach of \cite{Cos}.

For our present purposes, a fundamental aspect of the BV formalism consists of being able to find a chain complex on which the propagator, viewed as an operator, becomes a chain homotopy between the identity map and the projection onto the zero modes of our theory. Let us unravel this rather involved statement. Consider the following chain complex

\begin{equation}\begin{CD}
\mr{degree:} & \quad -1 & & 0 & & 1 & & 2 \\
  &    \quad \Omega^0(\Sigma,\g) @>{-d}>> \Omega^1(\Sigma,\g) @>{d*d}>> \Omega^{1,\dag}(\Sigma,\g) @>{-d}>> \Omega^2(\Sigma,\g) \\
  & \quad X & & A & & A^\dag & & X^\dag  \vspace{.15in}\\
\end{CD}\label{eq:BVcomplex}
\end{equation}
consisting of the ghost field $X$, gauge field $A$, antifield $A^\dag$, and antighost $X^\dag$. (We use $X$ instead of $\omega$ for our ghosts now, since in the BV formalism they appear in the action in a different form.) The space $\Omega^{1,\dag}(\Sigma,\g)$ is just a separate copy of $\Omega^1(\Sigma; \g)$ to keep track of the field $A^\dag$. We call this chain complex $\cE$, which is our total space of (all) fields, and its components have degree listed as above. We have the graded component decomposition
\begin{equation}
 \cE = \oplus_{k=-1}^2 \cE^{[k]}
\end{equation}
as given by (\ref{eq:BVcomplex}). An operator $T: \cE \to \cE$ has \textit{degree $d$} if $T = \oplus T^{[k]}$ with
$$T^{[k]}: \cE^{[k-d]} \to \cE^{[k]}.$$
Given a (degree-homogeneous) element $a$ of $\cE$, we denote its degree by $|a|$ and its component in $\cE^{[k]}$ by $a^{[k]}$.

\begin{Definition}
  A gauge-fixing operator $Q^\dag: \cE \to \cE$ is an operator of degree $-1$ that satisfies (i) $(Q^\dag)^2 = 0$; (ii) $[Q,Q^\dag]$ is a generalized Laplace-type operator\footnote{This means the operator is of the form $\nabla^*\nabla + F$, where $\nabla$ is a covariant derivative, $\nabla^*$ its adjoint, and $F$ a bundle endomorphism. Also, in what follows we must interpret the commutator $[\cdot,\cdot]$ in the graded sense as explained in the appendix.}. 
\end{Definition}

A gauge-fixing operator $Q^\dag$ yields for us a Hodge-like decomposition $\cE = \im Q \oplus \im Q^\dag \oplus \ker [Q,Q^\dag]$. More importantly, it allows us to define a Feynman diagrammatic expansion. In our situation, we are concerned with the following gauge-fixing operators:

\begin{Definition}
  Given a compatible metric $g=g_{ij}$ on $\Sigma=S^2$, define the (Coulomb) gauge-fixing operator $Q^\dag_g$ on $\cE$ via
  \begin{equation}\begin{CD}
    \Omega^0(\Sigma,\g) @<{-d^*}<< \Omega^1(\Sigma,\g) @<{-(\Pi_{\im *d}})*<< \Omega^{1,\dag}(\Sigma,\g) @<{-d^*}<< \Omega^2(\Sigma,\g).
  \end{CD}\label{eq:gfcomplex}
  \end{equation}
  Here, $\Pi_{\im *d}$ denotes the orthogonal projection onto $\im *d \subset \Omega^1(\Sigma,\g)$ with respect to $g_{ij}$.
\end{Definition}

Observe that $[Q,Q^\dag_g] = \Delta_g$ is the Laplce-Beltrami operator acting on all of $\cE$. On $S^2$,
$$\mc{H} := \ker \Delta_g$$
is spanned by the constant functions and the fixed area form $d\sigma$ on $S^2$. Thus, $\mc{H}$ is independent of the choice of compatible metric $g$. We can define the pseudoinverse $[Q,Q^\dag_g]^{-1}$, which is zero on $\ker [Q,Q^\dag_g]$ and the inverse of $[Q,Q^\dag_g]$ on the orthogonal complement $\cE_\bot$ to $\mc{H}$. The complement $\cE_\bot$ and the orthogonal projection onto it are independent of compatible $g$, since in this case, we have
\begin{equation}
 \cE_\bot = \{a \in \cE : \int a^{[0]} d\sigma = \int (*a^{[2]})d\sigma = 0\} \label{eq:cEbot}
\end{equation}

A gauge-fixing operator allows us to construct a propagator. The propagator we get from $Q^\dag_g$, which we call the BV propagator, differs from the propagator we obtained in the Faddeev-Popov procedure in the previous section. However, BV quantization leads to the same set of Feynman integrals as Faddeev-Popov quantization, as we will see shortly.

\begin{Definition}
  Given the gauge-fixing operator $Q^\dag_g$, we obtain the corresponding \textit{BV propagator} $P_g = Q^\dag_g[Q,Q^\dag_g]^{-1}$. It is a degree $-1$ operator on $\cE$ which satisfies
  \begin{equation}
    [Q, P_g] = \mr{id} - \Pi \label{homotopy}
  \end{equation}
  where $\Pi$ is the orthogonal projection onto $\mc{H}$. We denote the components of $P_g: \cE^{[k]} \to \cE^{[k-1]}$ by $P_g^{[k]}$. We call $P_g^{[0]}$ the bosonic and $P_g^{[\pm 1]}$ the bosonic and fermionic parts, respectively.
\end{Definition}

Equation (\ref{homotopy}) is the statement that $P_g$ is a chain homotopy from $\mr{id}$ and $\Pi$. To provide some heuristic insight into the significance of this fact, consider the following. The right-hand side of (\ref{homotopy}) is $g$-independent. Thus,
\begin{equation}
  [Q, d_{met}P_g] = 0 \label{eq:Qclosed}
\end{equation}
where $d_{met}$ is the exterior derivative on the space of compatible metrics on $S^2$ (such a space is a smooth, connected subvariety inside the space of all metrics). Equation (\ref{eq:Qclosed}) is the statement that $d_{met} P_g$ is closed as an element of $\mr{Hom}_\R(\cE_\bot)$, the space of $\R$ linear maps on $\cE_\bot$, endowed with the differential $[Q,\cdot]$ naturally induced from $\cE$. On the other hand, since $Q$ is acyclic on $\cE^\bot$, it follows that $[Q,\cdot]$ is acyclic on $\mr{Hom}_\R(\cE_\bot)$. Thus, we have
\begin{equation}
  d_{met} P_g \in \im [Q,\cdot]. \label{Qexact}
\end{equation}
Observe that $Q$ arises from infinitesimal gauge-transformations (and the linearized equations of motion). Thus, one can interpret equation (\ref{Qexact}) as stating that the propagator $P_g$, under changes of the metric $g$, changes by gauge degrees of freedom. Such an identity is a manifestation of gauge-invariance. After a detailed analysis,  (\ref{Qexact}) and the gauge-invariance of the underlying classical theory ultimately lead to gauge-invariance of the quantum theory.

The remainder of this section makes the above remarks precise. We need to do the following:
\begin{enumerate}
  \item Convert the BV propagator $P_g$, defined as an operator, into an integral kernel $P_g(x,y)$ so as to be placed on the edges of Feynman diagrams;
  \item Describe the BV action so as to obtain the vertices to be used in Feynman diagrams;
  \item Use the appropriate analog of (\ref{Qexact}) and the underlying gauge invariance of classical Yang-Mills theory and Wilson loop observables to establish gauge-invariance of Wilson loop expectation values. (This exploits the fact that no counterterms, in particular those that might have spoiled gauge-invariance, are needed for quantization.)
\end{enumerate}

We have streamlined our approach in this manner because it then explains what would otherwise be many mysterious sign rules in what follows. All such choices of signs can be viewed as being carefully crafted so as to ensure (i)--(iii) above hold. In what follows, we carry out steps (i)--(iii) in a somewhat abstract, but coordinate-independent manner. Explicit computations to help make our approach more explicit are carried out in Appendix \ref{Sec:AppBV}.\\

\noindent \underline{Step (i):} \\

To convert an operator to an integral kernel, we need a suitable pairing that induces a convolution operator. In other words, given a linear operator $K: \cE \to \cE$ and a bilinear pairing $\left<\cdot,\cdot\right>_\R: \cE \otimes \cE \to \R$, we want to express $K$ as an integral kernel $K(x,y) \in \cE \otimes \cE$ in the following way. For $a \in \cE$, we want $K(a)$ to be given by applying $1 \otimes \left<\cdot,\cdot\right>_\R$, with the appropriate signs to $K(x,y) \otimes a \in \cE \otimes \cE \otimes \cE$, which is to say that $K(x,y)$ is the integral kernel of $K$ with respect to our chosen bilinear pairing $\left<\cdot,\cdot\right>_\R$.

The pairing $\left<\cdot,\cdot\right>_\R$ and correct sign rule is determined by the following constraint. We want
\begin{equation}
([Q,K])(x,y) = Q(K(x,y)).
\end{equation}
Here $Q$ acts on $\cE \otimes \cE$ in the natural way as a derivation:
\begin{equation}
Q(a \otimes b) = Qa \otimes b + (-1)^{|a|}a \otimes Qb.
\end{equation}

\begin{Definition}
  Define the pairing $\left<\cdot,\cdot\right>_{BV}: \cE \otimes \cE \to \Omega^2(\Sigma)$ by
  \begin{align}
  \begin{split}
    \left<A, A^\dag\right>_{BV} &= \left<A \wedge A^\dag\right>\\
    \left<A^\dag, A\right>_{BV} &= \left<A^\dag \wedge A\right> \\
    \left<X, X^\dag\right>_{BV} &= -\left<X \wedge X^\dag\right> \\
    \left<X^\dag, X\right>_{BV} &= \left<X^\dag \wedge X\right>
   \end{split}\label{eq:BVpairing_dens}
  \end{align}
  where $\left<\cdot,\cdot\right>$ is the inner product on $\g$. We then obtain the pairing
  \begin{equation}
    \left<\cdot,\cdot\right>_\R = \int_\Sigma\left<\cdot,\cdot\right>_{BV}: \cE \otimes \cE \to \R. \label{eq:BVpairing}
  \end{equation}
  Both these pairings are of degree $-1$, i.e. $\left<a,b\right>_\R \neq 0$ only for $|a| + |b| = 1$, and are skew-symmetric. By slight abuse of language, we use the term \textit{BV-pairing} to denote either (\ref{eq:BVpairing_dens}) or (\ref{eq:BVpairing}), with which pairing we have in mind clear from the context.
\end{Definition}

In particular, $\left<\cdot,\cdot\right>_\R$ is an odd symplectic form, i.e., it is skew-symmetric, nondegenerate, and of odd degree. From $\left<\cdot,\cdot\right>_\R$, we obtain a corresponding convolution operator on $\cE$:

\begin{Definition}\label{def:conv}
  Given $K(x,y) \in \cE \otimes \cE$, define the BV convolution operator $K(x,y)\circledast: \cE \to \cE$ as follows. On simple tensors $K(x,y) = K_1(x) \otimes K_2(y) \in \cE \otimes \cE$, we have
  \begin{equation}
  K(x,y)\circledast a = (-1)^{|K_2|}K_1\left<K_2,a\right>_\R \label{BVconv}
  \end{equation}
  This determines $\circledast$ on general $K(x,y)$ from bilinearity and completion. We say that $K(x,y)$ is the \textit{BV integral kernel} of the corresponding operator $ K(x,y)\circledast$.
\end{Definition}

Note that if $|K(x,y)| = d$, then $K\circledast$ has degree $d-1$. The sign rules in the definition of the BV pairing and in the definition (\ref{BVconv}) are carefully chosen so that the following are true:

\begin{Lemma}We have that
\begin{enumerate}
  \item $Q$ is skew-adjoint with respect to $\left<\cdot,\cdot\right>_\R$, i.e.
  $$\left<a, Qb\right>_\R = -(-1)^{|a|}\left<Qa,b\right>_\R$$
  \item For all $K(x,y) \in \cE \otimes \cE$, we have $[Q,K(x,y)\circledast] = \big(Q(K(x,y))\big)\circledast$.
\end{enumerate}
\end{Lemma}

\Proof These are both straightforward computations. We check (ii) on simple tensors $K(x,y) = K_1(x) \otimes K_2(y)$:
\begin{align*}
  [Q,K(x,y)\circledast](a) &= Q(-1)^{|K_2|}K_1\left<K_2,a\right>_\R - (-1)^{|K|+1}(-1)^{|K_2|}K_1\left<K_2,Qa\right>_\R \\
  \big(Q(K(x,y))\big)\circledast a &= (-1)^{|K_2|}QK_1\left<K_2,a\right>_\R + (-1)^{|K_1|}(-1)^{|K_2|+1}K_1\left<QK_2,a\right>_\R
\end{align*}
We use the skew-adjointness property (i) to equate the above two lines.\End

Given our propagator operator $P_g = Q^\dag_g[Q_g,Q^\dag_g]^{-1}$, we obtain the corresponding propagator integral kernel $P_g(x,y) \in \cE \otimes \cE$ via
\begin{align}
  P_g = P_g(x,y) \circledast. \label{eq:BVprop}
\end{align}
The regulated propagator $P_{g,\eps}$ is obtained from regulating $[Q,Q^\dag_g]^{-1}$ in $P_g$:
\begin{align}
  P_{g,\eps} &= Q^\dag_g \int_\eps^\infty e^{-t[Q,Q^\dag_g]}dt \label{eq:BVprop_eps}\\
  &= Q^\dag_g \int_\eps^\infty e^{-t\Delta_g}dt, \label{eq:BVprop_eps2}
\end{align}
from which we obtain the regulated integral kernel $P_{g,\eps}(x,y)$ satisfying
\begin{equation}
  P_{g,\eps} = P_{g,\eps}(x,y)\circledast
\end{equation}

\begin{Remark}
 The integral kernels for propagators in the Faddeev-Popov setting were with respect to the $L^2$-pairing (with respect to the gauge-fixing metric $g$). Thus, while the bosonic BV and Faddeev-Popov progator operators differ, their integral kernels, as obtained from the BV-pairing and $L^2$-pairing respectively, agree (see Lemma \ref{Lemma:BV=FP}). While the $L^2$-pairing is more natural from an analytic-geometric standpoint, for the purpose of establishing metric-independence, the BV-pairing is more convenient because it is intrinsic to the underlying space and not metric-dependent.
\end{Remark}

\noindent \underline{Step (ii):}\\

In the BV formalism, we have an extended BV action $S$ consisting of the ordinary (bosonic) action $S_{bos}$, a ghost action $S_{gh}$ which encodes infinitesimal gauge-symmetries, and a Chevalley-Eilenberg action $S_{CE}$, which accounts for the Lie-algebraic structure of the infinitesimal gauge-transformations. These are functions on the space $\cE$ in the sense of Definition \ref{DefFun}, i.e., they are elements of $\mr{Sym}(\cE^*)$. Explicitly,
\begin{align}
  S &= S_{bos} + S_{gh} + S_{CE} \nonumber\\
  S_{bos}(A) &= S_{YM}(A) \nonumber\\
  S_{gh}(A^\dag, A, X) &= \frac{1}{\lam_0}\left<A^\dag, -d_AX\right>_\R  \nonumber\\
  S_{CE}(X^\dag, X_1,X_2) &= -\frac{1}{\lam_0}\left<X^\dag, [X_1,X_2]\right>_\R. \label{S_CE}
\end{align}

The quadratic part of $S$ yields a kinetic term
\begin{align}
  S_{kin} = \frac{1}{2\lam_0}\int\left<A, d*dA\right> - \frac{1}{\lam_0}\int\left<A^\dag, dX\right>
\end{align}
while the negative of the cubic and quartic parts of $S$ yield for us the interaction terms:
\begin{align}
  I = -\frac{1}{2\lam_0}\int \left<[A,A] \wedge *dA\right> - \frac{1}{8\lam_0}\int \left<[A,A] \wedge *[A,A]\right> + \frac{1}{\lam_0}\int \left<A^\dag \wedge [A,X]\right> -S_{CE}. \label{eq:BVI}
\end{align}
Define $I_{bos}$, $I_{gh}$, and $I_{CE}$ to be the first two, the third, and last terms of $I$, respectively. Thus, we have
\begin{align}
 S &= S_{kin} - I\\
 &= S_{kin} - I_{bos} - I_{gh} - I_{CE}
\end{align}

The basis for the extended action is as follows. The BV pairing induces a BV bracket, which allows us to convert action functionals to vector fields. Conceptually, it is the (odd) Poisson bracket corresponding the BV pairing.

\begin{Definition}
  Let $F$ be a local functional, i.e., one given by the integral of a polydifferential function of the fields. Then there is a unique local vector field $\delta^{BV}F$ such that
  $$\pd_v F = (-1)^{|v|}\int_\Sigma \left<v, \delta^{BV}F\right>_{BV}$$
  for all $v \in \cE$.
  The \textit{BV bracket} $\{F,G\}$ between a local functional $F$ and an arbitrary functional $G$ is given by
  $$\{F,G\} = \pd_{\delta^{BV}F}G.$$
  where one must interpret $\pd_{\delta^{BV}F}$ in the sense of (\ref{eq:pdvf}).
\end{Definition}

The BV action, which has degree zero, satisfies the following \textit{master equation}\footnote{Note that because $\{\cdot,\cdot\}$ is an odd bracket, it satisfies $\{F,G\} = (-1)^{|F||G|}\{G,F\}$ for $F$ and $G$ local so that the master equation is not vacuous.}:
\begin{equation}
  \{S,S\} = 0. \label{CME}
\end{equation}
We can expand this equation as follows. First, we have
\begin{equation}
 \{S_{kin},\cdot\} = \frac{1}{\lam_0}Q,
\end{equation}
regarded as a derivation on the space $\Sym(\cE^*)$ (the action of $Q$ on $\cE^*$ itself is the one induced from $Q$ on $\cE$ via pullback). Let
\begin{equation}
 \bar I = \lam_0 I.
\end{equation}
Then (\ref{CME}) can be written as
\begin{align}
  -Q\bar I + \frac{1}{2}\{\bar I, \bar I\} = 0.\label{ME}
\end{align}

If we decompose this equation by ghost number via (\ref{eq:CME1}--\ref{eq:CME3}), equation (\ref{eq:CME1}) expresses gauge-invariance of the classical action $S_{bos}$, (\ref{eq:CME2}) expresses that infinitesimal gauge transformations act as a Lie algebra, and (\ref{eq:CME3}) expresses the Jacobi identity.
Thus the master equation encodes symmetries and their algebraic consistency relations into a single equation. Expressing all such relations in a compact manner facilitates the analysis of symmetries, and in particular gauge-invariance, when quantizing. Next, we describe the Feynman diagrammatic expansion in the BV formalism, which involves applying Wick's Theorem to the BV propagator and the above BV interaction. It yields the same expansion as the Faddeev-Popov prcoedure, as the following lemma shows:

\begin{Lemma}\label{Lemma:BV=FP}
  Fix a compatible metric for Coulomb gauge-fixing. Consider the Faddeev-Popov propagator (\ref{eq:FPprop}) and interactions (\ref{eq:FPI}) and the Batalin-Vilkovisky propagator (\ref{eq:BVprop}) and interactions (\ref{eq:BVI}) which for notational clarity we denote by here by $P_{FP}$, $I_{FP}$, $P_{BV}$, and $I_{BV}$, respectively. Then
  \begin{equation}
    \lim_{\eps \to 0}e^{\lambda_0\pd_{P_{BV,\eps}}}e^{I_{BV}}\Big|_0 = \lim_{\eps \to 0}e^{\lambda_0\pd_{P_{FP,\eps}}e^{I_{FP}}}\Big|_0. \label{eq:compare}
  \end{equation}
As a consequence, we have
\begin{equation}
    \lim_{\eps \to 0}e^{\lambda_0\pd_{P_{BV,\eps}}}W_{f,\gamma}e^{I_{BV}}\Big|_{conn,0} = \lim_{\eps \to 0}e^{\lambda_0\pd_{P_{FP,\eps}}}W_{f,\gamma}e^{I_{FP}}\Big|_{conn,0}. \label{eq:compareW}
\end{equation}
\end{Lemma}

\Proof First, we check that the bosonic parts of the integral kernels of the propagators coincide. We have
\begin{align}
 P_{FP,\eps}^{bos} &= \Pi_{\im *d}\circ \int_\eps^\infty e^{-t\Delta}dt \label{eq:PFPregulated}\\
 P_{BV,\eps}^{[0]} &= -\Pi_{\im *d} \circ \int_\eps^\infty e^{-t\Delta}dt \circ *.
\end{align}
where in the second line, we used that the Hodge star $*$ commutes with the integral. Thus, for all $\beta \in \Omega^1(\Sigma)$, we have
\begin{align*}
 P_{FP,\eps}^{bos}(\beta) &= P_{BV,\eps}^{[0]}(*\beta)\\
 \int_{\Sigma_y} P_{FP,\eps}^{bos, ab}(x,y) \wedge *\beta^b(y) &= \int_{\Sigma_y} P_{BV,\eps}^{[0], ab}(x,y) \wedge (*\beta^b),
\end{align*}
where in the second-line, we used the $L^2$-pairing and $BV$-convolution pairing to convert to integral kernels. Since $\beta$ was arbitrary, we have
\begin{equation}
P^{bos, ab}_{FP,\eps}(x,y) = P^{[0],ab}_{BV,\eps}(x,y). \label{eq:P_FP=P_BV}
\end{equation}

So the only remaining issue consists in comparing the fermionic propagators and the interactions that have fermions. We can ignore $I_{CE}$ from the BV interactions because the propagator has no $X^\dag$ component, so that upon setting external leg variables equal to zero, those that depend on $X^\dag$ are annihilated. One can show that the fermionic propagators are related as follows. Note that

\begin{align*}
  P^{fer,ab}_{BV}(x,y) \in \Big(\cE^{[-1]} \oplus \cE^{[1]}\Big) \oplus \Big(\cE^{[1]} \oplus \cE^{[-1]}\Big)
\end{align*}
corresponding to $P_{BV,\eps}^{[-1]}:\cE^{[0]} \to \cE^{[-1]}$ and $P_{BV,\eps}^{[1]}: \cE^{[2]} \to \cE^{[1]}$, respectively. On the other hand, 
\begin{align}
  P^{fer,ab}_{FP}(x,y) \in  \Big(\overline{\Omega}^0(\Sigma; \g) \otimes \Omega^0(\Sigma; \g)\Big) \oplus \Big(\Omega^0(\Sigma; \g) \otimes \overline{\Omega}^0(\Sigma; \g)\Big) . \label{eq:Pfer2parts}
\end{align}
Next, observe that $-P^{fer,ab}_{BV,\eps}(x,y)$ is given by $(*d \otimes 1) \oplus (1 \otimes *d)$ applied to the two corresponding components of $P^{fer,ab}_{FP,\eps}(x,y)$ in (\ref{eq:Pfer2parts}), i.e. $*d$ acts on the $\overline{\Omega}^0(\Sigma; \g)$ factor. This follows, for instance, from the computation
\begin{align*}
  \int_{\Sigma_y} (1 \otimes *d_y)P^{fer,ab}_{FP,\eps}(x,y) \wedge A^b(y) &= -\int_{\Sigma_y} P^{fer,ab}_{FP,\eps}(x,y) \wedge * d^*A^b(y) \\
  &= -P^{fer}_{FP,\eps}(d^*A)\\
  &= P_{BV,\eps}^{[-1]}(A).
\end{align*}
and similarly for $P_{BV,\eps}^{[1]}$. It follows that $(1 \otimes *d)$ applied to the $\Omega^0(\Sigma;\g) \otimes \bar\Omega^0(\Sigma;\g)$ component of $P^{fer,ab}_{FP}(x,y)$ is equal to the $\Omega^0(\Sigma;\g) \otimes \Omega^{1,\dag}(\Sigma;\g)$ component of $-P^{fer,ab}_{BV}(x,y)$ due to the sign rule (\ref{BVconv}). On the other hand, if we identify $*d\bar\omega \in \im Q_{g}^\dag$ with $-A^\dag$ and $\omega$ with $X$, then
\begin{align*}
  I_{FP,gh} &= -\int \left<\bar\omega, d^*[A,\omega]\right>d\sigma\\
  &= \int \left<*d\bar\omega \wedge [A,\omega]\right> \\
  & {\rightarrow} -\int\left<A^\dag \wedge [A,X]\right>\\
  &= -I_{BV,gh}
\end{align*}
So $I_{FP,gh}$ (the third term of (\ref{eq:FPI})) and $I_{BV,gh}$ have opposite signs under this correspondence. Thus, in passing from Faddeev-Popov to BV fermonic Feynman integrals, the former correspond to BV integrals with interactions and propagators replaced with their negatives, thus resulting in no net difference. It now follows that (\ref{eq:compare}) and (\ref{eq:compareW}) hold.\End

\noindent \underline{Step (iii):}\\

We now turn to the heart of the proof of Theorem \ref{ThmGI}. We have a few algebraic preliminaries to establish:

\begin{Lemma}\label{LemmaAlg}
 Let $v \in \cE$. For any linear operator $D$ on $\cE$, which acts as a derivation on $\Sym(\cE)$ and dually as a derivation on $\Sym(\cE^*)$, we have
 \begin{enumerate}
   \item $\pd_{Dv} = [\pd_v,D]$
   \item $\pd_{DK} = [\pd_K,D]$
 \end{enumerate}
 as operators on $\Sym(\cE^*)$.
\end{Lemma}

\Proof (i) Both $\pd_{Dv}$ and $[\pd_v, D]$ are derivations of degree $(-1)^{|Dv|}$, so it suffices to check $\pd_{Dv} = [\pd_v,D]$ on $\cE^*$. The statement is then automatic.

(ii) Without loss of generality, let $K = K_1 \otimes K_2$. Using (i), then
\begin{align*}
  \pd_{DK} &= \pd_{(DK_1\otimes K_2 + (-1)^{|D||K_1|}K_1\otimes DK_2)} \\
  &= \frac{1}{2}\left(\pd_{K_2}\pd_{DK_1} + (-1)^{|D||K_1|}\pd_{DK_2}\pd_{K_1}\right)\\
  &= \frac{1}{2}\left(\pd_{K_2}[\pd_{K_1},D] + (-1)^{|D||K_1|}[\pd_{K_2},D]\pd_{K_1}\right) \\
  &= \frac{1}{2}[\pd_{K_2}\pd_{K_1},D]\\
  &= [\pd_K,D].\End
\end{align*}

Let $K_t = e^{-t\Delta}$ be the heat-operator associated to $\Delta = [Q,Q^\dag_g]$ at time $t$. Note that $K_\infty = \Pi$ is the orthogonal projection onto $\ker \Delta$. Because $K_t$ is a degree zero operator, its integral kernel is of degree $1$.

\begin{Definition}
  Define the BV Laplacian $\Delta_{BV}$ to be the ``divergence operator"
$$\Delta_{BV} := \pd_{K_0}$$
We also consider the regulated versions
$$\Delta_{BV,\eps} := \pd_{K_\eps}$$
These operators are of degree one.
\end{Definition}
Note that since $K_{0}$, being the integral kernel of the identity operator, is a $\delta$-function along the diagonal, $\Delta_{BV}(F)$ is not well-defined for arbitrary $F$. The BV Laplacian is intimately related to the master equation (\ref{CME}), for further reading see \cite{Cos}. In our condensed presentation, we obtain $\Delta_{BV}$ as a byproduct of the gauge-invariance analysis that is about to follow.

\begin{Lemma}\label{LemmaBracketID}
  If at least one of $F$ or $G$ is a local action functional, then
\begin{align}
  \{F,G\} = \lim_{\eps \to 0}\Big(\Delta_{BV,\eps}(FG) - \Delta_{BV,\eps}(F)G - (-1)^{|F|}F\Delta_{BV,\eps}(G)\Big). \label{eq:bracketid}
\end{align}
\end{Lemma}

\Proof This is a straightforward algebraic computation by noting that $\{F,G\}$ is the part of $\pd_{K_0}(FG)$ in which the $K_0 = \lim_{\eps \to 0}K_\eps$ contraction joins $F$ to $G$. This is what the right-hand side of (\ref{eq:bracketid}) expresses, since it subtracts from $\Delta_{BV,\eps}(FG)$ those contractions that involve only $F$ or $G$ alone.\End

Next, we need to consider how the propagator $P_g$ varies as the metric varies. Here, we follow the ideas of \cite[Proposition 10.7.2]
{Cos}. Let $g_\tau$ be a family of compatible metrics parametrized by $\tau$ belonging to the standard $n$-simplex $\Delta^n$ (for us, $n = 1$ suffices since any two compatible metrics can be joined by a $1$-parameter family of compatible metrics). Consider the complex
$$\C = \Omega^\bullet(\Delta^n) \otimes \cE,$$
the space of $\tau$-dependent elements of $\cE$ tensored with differential forms on $\Delta^n$. The differential on $\C$ is  $$d_\C = d_{dR} + Q$$
induced from the de-Rham $d_{dR}$ differential on $\Omega^\bullet(\Delta^n)$ and the differential $Q$ on $\cE$, i.e.
$$d_\C(\eta \otimes a) = d_{dR}\eta \otimes a + (-1)^{|\eta|}\eta \otimes Qa, \qquad \eta \in \Omega^\bullet(\Delta^n), a \in \cE.$$
The space $\cC$ allows us to differentiate $\tau$-dependent elements of $\cE$ within the setting of a chain-complex (and not just merely an ordinary derivative).

We consider all $\Omega^\bullet(\Delta^n)$-linear functionals on $\cE$, i.e.
$$\Sym_{\Omega^\bullet(\Delta^n)}(\C^*) = \Omega^\bullet(\Delta^n) \otimes \Sym(\cE^*).$$
That is, an element of $\Sym_{\Omega^\bullet(\Delta^n)}(\C^*)$
is a section of the bundle over $\Delta^n$ whose fiber over $\tau \in \Delta^n$ is a polydifferential functional of elements of $\cE$ valued in $\Lambda^\bullet(T^*_\tau\Delta^n)$. The space $\Sym_{\Omega^\bullet(\Delta^n)}(\C^*)$ accounts for $\tau$-dependent functionals on $\cE$ along with their exterior derivatives in the $\tau$-directions. The complex $\C$ is a graded module over $\Omega^\bullet(\Delta^n)$ while $\Sym_{\Omega^\bullet(\Delta^n)}(\C^*)$ is a graded algebra over $\Omega^\bullet(\Delta^n)$ in the natural way. In what follows, we extend our previous constructions to the appropriate families version, which is not entirely straightforward. All operators in question will be parametrized by $\tau \in \Delta^n$ and are $\Omega^\bullet(\Delta^n)$-linear operators on $\C$.

We have a family of gauge-fixing operators
$$Q^{\dag}_\tau := Q^{\dag}_{g_\tau}$$
given by a family of compatible metrics $g_\tau$, $\tau \in \Delta^n$. The degree $-1$ maps $Q^\dag_t$ on $\cE$ induce degree $-1$ maps on $\C$ in the natural way. We then obtain a family of Laplace-type operators
$$\tilde\D_\tau = [d_{dR} + Q, Q^\dag_\tau] = [Q, Q^\dag_\tau] + [d_{dR}, Q^\dag_\tau], \quad \tau \in \Delta^n$$
where the first term $[Q, Q^\dag_\tau]$ is a Laplacian and the second term $[d_{dR}, Q^\dag_\tau]$ is nilpoent (since wedging with a $1$-form in the $\Delta^n$-direction is a nilpotent operation). The $\tilde{\;}$ supserscript is to emphasize the fact that $\tilde D_\tau$ has components in de Rham degree greater than zero; it is not simply a $\tau$-dependent operator on $\cE$.

Because $Q_\tau^\dag$ varies along compatible metrics (or along our path connecting Coulomb to holomorphic gauge), we have $[d_{dR},Q_\tau^\dag]h = 0$ for all $h \in \H$. Since $[Q,Q^\dag_\tau]$ is invertible on the orthogonal complement $\cE_\bot$ to $\H$, we have
$$\ker \tilde\D_\tau = \Omega^\bullet(\Delta^n) \otimes \H, \quad \tau \in \Delta^n$$
is constant. Likewise, we have $\im \tilde\D_\tau \subset \Omega^\bullet(\Delta^n) \otimes \cE_\bot$. Hence we can define $\tilde\D_\tau^{-1} = [d_{dR} + Q, Q^\dag_\tau]^{-1}$ to be the inverse of $\tilde\D_\tau$ on $\Omega^\bullet(\Delta^n) \otimes \cE^\bot$. We obtain a family of propagators
$$\tilde P_\tau = Q^\dag_\tau[d_{dR} + Q, Q^\dag_\tau]^{-1}.$$
It can be regarded as a $\Omega^\bullet(\Delta^n)$-linear chain-homotopy on $\C$:
\begin{equation}
[d_{dR} + Q, \tilde P_\tau] = \mr{id} - \Pi, \label{eq:familiesCH}
\end{equation}
where, by slight abuse of notation, we identify $\Pi$  on $\cE$ with its $\Omega^\bullet(\Delta^n)$-linear extension to $\C$. As before, we can implement ``heat-kernel regularization'' for $\tilde P_\tau$:
$$\tilde P_{\tau,\eps} = Q_t^\dag \int_\eps^\infty e^{-t\tilde\D_\tau}dt, \qquad \eps \geq 0.$$
Here, the operator $e^{-t\tilde\D_\tau}$, for each fixed $\tau$, has to be interpreted as the strongly-continuous semigroup associated to $\tilde\D_\tau$, since $\tilde\D_\tau$ is no longer a Laplace-type differential operator (the term $[d_{dR},Q^\dag_\tau]$ contains a pseudodifferential term arising from the component $Q^{\dag,[0]}_\tau$). Since $[d_{dR},Q^\dag_\tau]$ is nilpotent, the spectrum of $\tilde\D_\tau$ equals the spectrum of $[Q,Q^\dag_\tau]$ and hence is nonnegative. Thus, the Hille-Yosida theorem implies that $\tilde\D_\tau$ generates a strongly continuous semigroup $e^{-t\tilde\D_\tau}$, such that for $t = 0$ we have the identity operator and for $t = \infty$ we have projection onto $\Omega^\bullet(\Delta^n) \otimes \H$.

Equation (\ref{eq:familiesCH}) becomes
\begin{equation}
 [d_{dR} + Q, \tilde P_{\tau,\eps}] = e^{-\eps \tilde\D_\tau} - \Pi, \label{eq:familiesCHeps}
\end{equation}
The integral kernel $\tilde P_{\tau,\eps}(x,y)$ of $\tilde P_{\tau,\eps}$ (with respect to BV convolution) is an element of $\Omega^\bullet(\Delta^n) \otimes \cE \otimes \cE$
and is of degree zero.

Define $\tilde K_{\tau,t} = e^{-t\tilde \D_\tau}$, for $t \in [0,\infty]$. Its integral kernel $\tilde K_t(x,y)$ is an element of $\Omega^\bullet(\Delta^n) \otimes \cE \otimes \cE$
that is of degree one. For $t = \infty$, then $\tilde K_{\tau,\infty} = \tilde K_\infty$ is independent of $\tau$ and equal to $\Pi$.

We have the families BV Laplacians
$$\tilde \Delta_{BV,\tau,t} = \pd_{\tilde K_{\tau,t}}.$$
Equation (\ref{eq:familiesCH}), translated from operators to integral kernels with respect to BV convolution, implies that
\begin{equation}
(d_{dR} + Q) \tilde P_{\tau,\eps}(x,y) = \tilde K_{\tau,\eps}(x,y) - \tilde K_{\infty}(x,y) \label{eq:familiesCHint}
\end{equation}
where $d_{dR} + Q$ acts on $\Omega^\bullet(\Delta^n) \otimes \cE \otimes \cE$ as a (graded) derivation.

Since $(Q^\dag_\tau)^2 = 0$ and $Q^\dag_\tau$ commutes with $\tilde \D_\tau$ for every $\tau$, we have
\begin{equation}
\tilde P^2_{\tau,\eps} = 0.
\end{equation}

\noindent \textbf{Proof of Theorem \ref{ThmGI}:} Let $O = W_{f,\gamma}$ be a Wilson loop observable. If $g_0$ and $g_1$ are two compatible metrics, join them by a path of compatible metrics. This yields for us a family of metrics indexed by $\tau$ along a $1$-simplex $\Delta^1$. We will show that the resulting family of Wilson loop expectations is constant along $\Delta^1$, by showing that $d_{met}\left<O\right> = 0$.

Let $=_k$ denote equality in de-Rham degree $k$. For notational clarity, we drop the explicit dependence on $\tau$. Also, we denote the de Rham differential on $\Omega^\bullet(\Delta^1)$ by $d_{met}$. Thus, we have
\begin{equation}
 \left<O\right> =_0 \lim_{\eps\to 0}e^{\lambda_0 \pd_{\tilde P_\eps}}Oe^{I}|_{conn,0}
\end{equation}
since in de Rham degree zero, $\tilde P_\eps$ is just the family of propagators $P_\eps$. We have
\begin{equation}
 d_{met}\left<O\right> =_1 \lam_0 \lim_{\eps \to 0}e^{\lambda_0 \pd_{\tilde P_\eps}}\pd_{[d_{met}\tilde P_\eps(x,y)]}Oe^{I}|_{conn,0},  \label{eq:dmet1}
\end{equation}
i.e. we sum over all possible Feynman diagrams in which we must also Wick contract with a derivative of the propagator $d_{met}\tilde P(x,y)$. From (\ref{eq:familiesCHint}), we have
\begin{equation}
d_{met}\tilde P_\eps(x,y) = -Q\tilde P_\eps(x,y) + \tilde K_\eps(x,y) + \tilde K_\infty(x,y). \label{eq:familiesCHint2}
\end{equation}
We can use Lemma \ref{LemmaAlg} to write
$$\pd_{Q\tilde P_\eps(x,y)} = \pd_{[Q,\tilde P_\eps]} = [\pd_{\tilde P_\eps},Q].$$
Moreover, it is also easy to verify the algebraic identity $$[e^{\lam_0\pd_{\tilde P_\eps}},Q] = e^{\lam_0\pd_{\tilde P_\eps}}\lam_0[\pd_{\tilde P_\eps},Q].$$
Finally, we also have $(QF)|_0 = 0$ for any functional $F$, since $Q$ annihilates constant functionals. Thus,
$$[e^{\lam_0\pd_{\tilde P_\eps}},Q]F\bigg|_0 = e^{\lam_0\pd_{\tilde P_\eps}}QF\bigg|_0.$$

So using the above three equations, replacing $I$ with $\bar I/\lambda_0$, then (\ref{eq:dmet1}) and (\ref{eq:familiesCHint2}) imply
\begin{align}
 d_{met}\left<O\right> &=_1 \lim_{\eps \to 0}e^{\lam_0\pd_{\tilde P}}\big(-Q + \lam_0\tilde\Delta_{BV,\eps} - \lam_0\tilde\Delta_{BV,\infty}\big)Oe^{\bar I/\lambda_0}\bigg|_{conn,0}\\
  \begin{split}
  &=_1 \lambda_0\lim_{\eps\to0}e^{\lambda_0 \pd_{\tilde P_\eps}}\Big(-QO + \{\bar I,O\} + \lam_0\tilde\Delta_{BV,\eps} O\Big)e^{\bar I/\lam_0}\bigg|_{conn,0} \\
  & \qquad + \lam_0^{-1}\lim_{\eps\to0}e^{\lambda_0 \pd_{\tilde P_\eps}}\Big(-Q\bar I + \frac{1}{2}\{\bar I, \bar I\} + \lam_0\tilde\Delta_{BV,\eps} \bar I\Big)e^{\bar I/\lam_0}\bigg|_{conn,0} \\
  & \qquad\qquad
-\lambda_0\lim_{\eps\to0}e^{\lambda_0 \pd_{\tilde P_\eps}}\tilde\Delta_{BV,\infty} \Big(O e^{\bar I/\lam_0}\Big)\bigg|_{conn,0}.\end{split}\label{eq:dmet_last3}
\end{align}
In the second line, we made repeated use of Lemma \ref{LemmaBracketID} to express $\tilde\Delta_\eps$ of a product in terms of $\tilde\Delta_\eps$ of the individual factors and the BV-bracket. We have $-QO + \{\bar I,O\} = -\lam_0\{S, O\} = 0$ since $O$ is gauge-invariant. We have $\tilde\Delta_{BV,\eps} O = 0$ since $O$ has no antifield components. Next, $-Q\bar I + \frac{1}{2}\{\bar I, \bar I\} = 0$ by the master equation (\ref{ME}). We have $\tilde\Delta_{BV,\eps} I = 0$ since $\tilde K_\eps$ is symmetric in Lie-algebra indices, $\eps > 0$, while $I$ is skew-symmetric in them.

The final term of (\ref{eq:dmet_last3}) vanishes by the following. The operation $\tilde\Delta_{BV,\infty} = \pd_{\tilde K_\infty}$ contracts $d\sigma$ into an $X^\dag$ entry and $1$ into an $X$ entry (and there is also the $\mr{id}_\g$ component one must contract). We have $\tilde\Delta_{BV,\infty} Oe^I = O(\frac{1}{2}\{I,I\}_\infty + \tilde\Delta_{BV,\infty} I)e^I$, where $\{\cdot,\cdot\}_\infty$ denotes the contraction of the two input functionals using $\pd_{\tilde K_\infty}$, with $\tilde K_\infty$ placed on distinct functionals (cf. Lemma \ref{LemmaBracketID}). The term $\tilde\Delta_{BV,\infty} I$ vanishes since $\tilde K_\infty$ is symmetric in Lie algebra indices. We thus have to consider
\begin{align}
  \frac{1}{2}\{I,I\}_\infty = \{I_{gh}, I_{CE}\}_\infty + \frac{1}{2}\{I_{CE},I_{CE}\}_\infty \label{eq:KinftyI}
\end{align}
The second term of (\ref{eq:KinftyI}) has an uncontracted $X^\dag$ argument, for which $\tilde P_\eps$ will not be able to contract (there is no degree zero element of $\Omega^\bullet(\Delta^1) \otimes \cE \otimes \cE$ that has a component in $\cE^{[2]}$). Thus the operation $|_0$, which makes external leg variables zero, will annihilate all diagrams with external $X^\dag$ legs. So we need only consider diagrams arising from the first term of (\ref{eq:KinftyI}). However, all such diagrams vanish using the condition $\tilde P_\eps^2 = 0$. Indeed, one has an external $A$ and $A^\dag$ leg in $\{I_{gh}, I_{CE}\}_\infty$, and the placement of propagators $\tilde P_\eps$ on these legs yields the integral kernel for $\tilde P^2_\eps$ (since $X = 1$ was contracted into the remaining slot of $I_{gh}$).

Altogether, we have shown that all terms of $d_{met}\left<O\right>_{C}$ vanish. This establishes gauge-invariance.\End

As a result of metric-independence, we can deduce that, like the exact expectation, the perturbative expectation of Wilson loops are invariant under area preserving diffeomorphisms.

\begin{Theorem}\label{Thm:Area}
  Let $\gamma$ be a regular curve on $S^2$. Then for any diffeomorphism $\Phi: S^2 \to S^2$ which preserves the areas of the regions complementary to $\gm$, we have
  $$\left<W_{f,\gamma}\right>_{S^2, C} = \left<W_{f,\Phi(\gamma)}\right>_{S^2, C}.$$
\end{Theorem}

\Proof First, we have the following straightforward covariance property: for any (continuously differentiable) diffeomorphism $\Psi: S^2 \to S^2$, we have
\begin{equation}\left<W_{f,\gamma}\right>_{S^2, C, d\sigma} = \left<W_{f,\Psi(\gamma)}\right>_{S^2, C, \Psi_*(d\sigma)},
\label{eq:cov}
\end{equation}
where the two sides above denote expectations with respect to $d\sigma$ and $\Psi_*(d\sigma)$, respectively. Indeed, the terms of $\left<W_{f,\gamma}\right>_{S^2, C, d\sigma}$ consist of (regulated) Feynman diagrams formed out of any compatible metric, and by pushing forward all these diagrams under $\Psi$, we obtain (regulated) Feynman diagrams computed with respect to a corresponding compatible metric for the area form $\Psi_*(d\sigma)$. Letting the regulator go to zero, we obtain the result.

In particular, equation 
(\ref{eq:cov}) holds with $\Psi = \Phi$. Our result follows by one further application of (\ref{eq:cov}) if we can find a diffeomorphism $\Psi$ such that $\Psi$ leaves $\Phi(\gm)$ invariant and $\Psi_*(\Phi_*(d\sigma)) = d\sigma$. This is possible as a consequence of \cite{BMPR}, which in particular, establishes the following. Given two volume forms $\omega_1$ and $\omega_2$ on a surface $M$ with corners, if they have equal volumes and agree pointswise at the boundary corners, then there exists a diffeomorphism $\Psi: M \to M$ such that $\Psi_*(\omega_1) = \omega_2$ and $\Psi$ is the identity map restricted to the boundary. We apply this result as follows. We have $S^2\setminus\Phi(\gm)$ is the union $\bigcup R_i$ of manifolds with corners, one for each connected component $R_i$ of $S^2\setminus\Phi(\gm)$. Let $d\sigma' = \Phi_*(d\sigma)$. Suppose $d\sigma'$ and $d\sigma$ agree at the points of self-intersection of $\Phi(\gm)$ (the source of corner points). By the aforementioned result, we can find diffeomorphisms $\Psi_i: R_i \to R_i$ such that 
\begin{equation}
(\Psi_i)_*(d\sigma') = d\sigma \label{eq:R_i}                                                                                                                                                                                                                                                                                                                                                                                                                                                                                                                                                                                                                                                                                                                                                                                                                                                                                                                                                                                                                                                                                                                                                                                                                                                                                                                                                                                                                                                                                                                                                                                                                                                                                                                                                                                                                                                                                                                                                                                                                                                                                                                                           \end{equation}
 on $R_i$ and then patch the $\Psi_i$ together since the $\Psi_i$ agree along common boundaries (contiguous $\Psi_i$ fix their common boundary and their first derivatives agree by virtue of (\ref{eq:R_i})). This yields our desired map $\Psi$. 

So suppose $d\sigma'$ and $d\sigma$ disagree at a self-intersection point $p$. Choose a small open disk $D$ centered at $p$ and work in a coordinate chart such that all components of $\gm$ passing through $D$ are diameters of $D$. Then via a compactly-supported nonlinear radial dilation about $p$, we can diffeomorphically map $D$ to itself in such a way that it preserves the image of $\gm$ and it maps $d\sigma'|_p$ to $d\sigma|_p$. Apply such a diffeomorphism to neighborhoods of every self-intersection point of $\Phi(\gamma)$ and extend it to the rest of $S^2$ by the identity map. Letting $\Xi$ denote the resulting diffeomorphism of $S^2$, we have that
\begin{align*}
\left<W_{f,\Phi(\gm)}\right>_{S^2, C, \Phi_*(d\sigma)} &= \left<W_{f,\,\Xi(\Phi(\gm))}\right>_{S^2, C,\, \Xi_*(\Phi_*(d\sigma))}\\
&= \left<W_{f,\Phi(\gm)}\right>_{S^2, C,\, \Xi_*(\Phi_*(d\sigma))}. 
\end{align*}
The first line applies the covariance property (\ref{eq:cov}) and the second line uses the fact that the Feynman integrals induced from a Wilson loop are independent of the parametrization of the underlying curve ($\Xi$, which preserves the image of $\Phi(\gm)$, is such that $\Xi(\Phi(\gm))$ is a reparametrization of $\Phi(\gm)$). We can now proceed as before, but with $d\sigma' = \Xi_*(\Phi_*(d\sigma))$, since now $d\sigma'$ and $d\sigma$ agree at the self intersection points of $\Phi(\gm)$.\End

\subsection{Coulomb gauge = Holomorphic gauge}\label{Sec:C=hol}

In this section, we use the Batalin-Vilkovisky formalism of the previous section to prove the equivalence of Coulomb gauge with holomorphic gauge on $\Sigma = S^2$. We have the following result:

\begin{Theorem}\label{Thm:C=Hol}
 Pick any compatible metric on $S^2$ determining a corresponding Coulomb gauge and holomorphic gauge. Then
 \begin{equation}
  \left<W_{f,\gm}\right>_{S^2,C} =  \left<W_{f,\gm}\right>_{S^2,hol}. \label{eq:WC=Whol}
 \end{equation}
 Consequently, $\left<W_{f,\gm}\right>_{hol}$ is invariant under area-preserving diffeomorphisms.
\end{Theorem}

\Proof As a consequence of Theorem \ref{Thm:Area}, we need only establish (\ref{eq:WC=Whol}). In the same way that holomorphic gauge needs to be interpreted in terms of a real integration cycle \cite{Ngu2016}, the proof of Theorem \ref{Thm:C=Hol} exploits this idea in the Batalin-Vilkovisky context. Namely, we complexify the BV complex (\ref{eq:BVcomplex}) and connect the Coulomb gauge and holomorphic gauge through a one-paramemter family of gauge-fixing operators, one which interpolates between the subspace $\im *d \subset \Omega^1(\Sigma,\g)$ and a totally real subspace of $\Omega^{1,0}(\Sigma;\g_c)$.

For the sake of brevity, we write $\Omega^1_c = \Omega^1(\Sigma; \g_c)$, and similarly for $\Omega^{1,0}$ and $\Omega^{0,1}$. We have the decompositions
\begin{align*}
\Omega^1_c &= \big(\bC \otimes \im d \big) \oplus \big(\bC \otimes \im *d\big)\\
 \Omega^1_c &= \Omega^{1,0} \oplus \Omega^{0,1}
\end{align*}
with $\Omega^{1,0}$ and $\Omega^{0,1}$ the $\mp i$ eigenspaces of $*$, respectively. Concretely, we have
\begin{align*}
 \Omega^{1,0} &= \{(df + *dg) + i(-dg + *df)\} \\
 \Omega^{0,1} &= \{(df + *dg) + i(dg - *df)\}
\end{align*}
given by the graphs of $\pm i*: \Omega^1(\Sigma;\g_c) \to i\Omega^1(\Sigma;\g_c)$, where $f$ and $g$ denote real-valued functions. Thus, we can define a totally real subspace of $\Omega^{1,0}$ by restricting the graph of $i*$ to $\im *d$:
\begin{align*}
\Omega^{1,0}_r &:= \{\alpha \in \Omega^{1,0} :
\alpha = *df - idf\}.
\end{align*}
We have
$$\Omega^{1,0} = \Omega^{1,0}_r \oplus i\Omega^{1,0}_r.$$

Next, we have the following complexification of the BV-complex:
\begin{equation}
\xymatrix{
\Omega^0_c \ar^{-d}[r] & \Omega^1_c \ar^{d*d}[r] & \Omega^1_c \ar^{-d}[r] & \Omega^2_c
} \label{eq:c-complex}
\end{equation}
The differential $Q$ and the Coulomb gauge-fixing operator $Q^\dag$, defined as in (\ref{eq:gfcomplex}) with respect to some fixed compatible metric, extend complex-linearly. Denote the complex (\ref{eq:c-complex}) by $\E_c$; it consists of the terms $\E_c^{[k]}$ supported in degree $k$, $-1 \leq k \leq 2$.

Define the one-parameter family of isomorphisms
$U_\tau: \E_c \to \E_c$, $0 \leq \tau \leq 1$, by
\begin{align*}
U_\tau|_{\E_c^{[k]}} = \begin{cases}
               \Pi_{\im d} + (1 + \tau i*)\Pi_{\im *d} & k = 0, 1\\
               \mr{id} & k \neq 1
              \end{cases}
\end{align*}
where the complementary orthogonal projections $\Pi_{\im d}$ and $\Pi_{\im *d}$   (with respect to the chosen compatible metric) are extended complex linearly to $\Omega^1_c$. One can check that $U_\tau$ commutes with the differential $Q$, so that the $U_\tau$ are chain isomorphisms. We have $U_0$ is the identity and $U := U_1$ maps $\im\!*\!d \subset \cE^{[k]}$ to $\Omega^{1,0}_r$, $k=1,2$.

Define
\begin{equation}
 Q_\tau^\dag = U_\tau Q^\dag U_\tau^{-1}. \label{eq:Q_t^dag}
\end{equation}
Then
$$[Q,Q_\tau^\dag] = U_\tau[Q,Q^\dag]U_\tau^{-1} = [Q,Q^\dag]$$
since $\Delta = [Q,Q^\dag]$ commutes with $U_\tau$. We have the $\tau$-dependent propagator
\begin{align}
P_\tau &= Q^\dag_\tau[Q,Q^\dag_\tau]^{-1}\\
&= Q^\dag_\tau[Q,Q^\dag]^{-1} \label{eq:P_t}
\end{align}
which satisfies
\begin{equation}
P_\tau = U_\tau Q^\dag_\tau[Q,Q^\dag]^{-1}U_\tau^{-1}.
\end{equation}
We have
\begin{equation}
 [Q, P_\tau] = \mr{id} - \Pi
\end{equation}
for all $\tau$.

Hence, the exact same proof as the proof of Theorem \ref{ThmGI}, with the $\eps$-regulator inserted and then taken to zero, shows that Wilson loop expectations with respect to the propagators $P_\tau$ are independent of $\tau$. Indeed, we define the operator 
\begin{align*}
\tilde D_\tau &= [Q,Q^\dag_\tau] + [d_{dR},Q_\tau^\dag]\\
& = [Q,Q^\dag] + [d_{dR},Q_\tau^\dag]
\end{align*}
as before, which is a Laplace-type operator plus a nilpotent operator. From this, we obtain the $\eps$-regulated families propagator $\tilde P_{\tau,\eps}$ as in the proof for Theorem \ref{ThmGI}, and we proceed in exactly the same way.

Since $P_0$ is the Coulomb gauge propagator, it remains to check that $P_1^{[0]}$ is the holomorphic gauge propagator and that $P_1^{[\pm 1]}$ have their $1$-form components belonging to $\Omega^{1,0}$. Indeed, if we do this, then
\begin{equation}
e^{\lambda_0\pd_{P_1}}Oe^I\bigg|_{conn,0} = e^{\lambda_0\pd_{P_{hol}}}O\bigg|_0. \label{eq:P1->Phol}
\end{equation}
for a gauge-invariant observable $O$, since $I$ vanishes whenever all the entries belong to $\Omega^{1,0}$.

First, we check that $P_1^{[0]}$ inverts $\bar\pd*\bar\pd: \Omega^{0,1} \to \Omega^{0,1}$. So let $\al = \pd f \in \Omega^{1,0}$ for some arbitrary function $f$. We have
\begin{align*}
 U Q^{[0],\dag} U^{-1}[Q,Q^\dag]^{-1}\bar\pd*\bar\pd \al &= U Q^{[0],\dag} U^{-1}[Q,Q^\dag]^{-1}\bar\pd*\bar\pd\pd f \\
 &= U Q^{[0],\dag} U^{-1}[Q,Q^\dag]^{-1}\bar\pd \left(-\frac{i}{2}\Delta f\right)\\
 &= -\frac{i}{2}U Q^{[0],\dag} U^{-1}\bar\pd f\\
 &= -\frac{i}{2}U Q^{[0],\dag} U^{-1}\left(\frac{1}{2}(1-i*)df\right)\\
 &= -\frac{i}{2}U Q^{[0],\dag} \left(df - \frac{1}{2}i*df\right)\\
 &= \frac{i}{2}U (*df)\\
 &= \frac{i}{2}(1+i*)(*df)\\
 &= i*\pd f\\
 &= \al.
\end{align*}
So $P^{[0]}_1 = P_{hol}$. Next, it follows from $U_1: \im *d \to \Omega^{1,0}_r$ that $P^{[1]}_1 = U_1P^{[1]}U_1^{-1}$ maps into $\Omega^{1,0}$ and $P^{[-1]}_1 = U_1P^{[-1]}U_1^{-1}$ annihilates $\Omega^{1,0}$. Thus, the $\cE^{[1]}_c$-components of $P^{[\pm 1]}(x,y)$ belong to $\Omega^{1,0}$.\End

We present the following lemma which makes
the $Q_\tau^\dag$ (and hence $P_\tau$) more explicit and which will be useful in the next section:

\begin{Lemma}
 We have
 \begin{align}
  Q^{\dag,[-1]}_\tau &= -\big((1 - \tau)\pd^* + (1+\tau)\bar\pd^*\big) \\
  Q^{\dag,[0]}_\tau &= (1 + \tau i*)(-\Pi_{\im *d}\,*)(1 - \tau i*) \label{eq:Q0dag}\\
  Q^{\dag,[1]}_\tau &= -\big((1 - \tau)\pd^* + (1+\tau)\bar\pd^*\big)
 \end{align}
\end{Lemma}

\Proof On $1$-forms, we can write $U_\tau$ as the matrix
$$U_\tau = \begin{pmatrix}
         1 & \tau i* \\
         0 & 1
        \end{pmatrix}$$
with respect to the decomposition $\Omega^1_c = \im d \oplus \im *d$. Thus, we have
$$U_\tau^{-1} = \begin{pmatrix}
         1 & -\tau i* \\
         0 & 1
        \end{pmatrix}$$
From this, we have
\begin{align}
Q^{\dag,[-1]}_\tau &= -d^*U_\tau^{-1}\\
&= -d^*(1-\tau i*) \\
&= -(d^* - \tau i*d) \label{eq:Qdag0-line3}\\
&= -(\pd^* + \bar\pd^*) + \tau i*(\pd + \bar\pd) \\
&= -(\pd^* + \bar\pd^*) - \tau(-*\pd* + *\bar\pd*) \\
&= -(\pd^* + \bar\pd^*) - \tau(\bar\pd^* - \pd^*) \\
&= -\big((1-\tau)\pd^* + (1+\tau)\bar\pd^*\big).
\end{align}
In the fourth line above, we used that $\pd$ and $\bar\pd$, acting on $\Omega^1_c$, are nontrivial only on $\Omega^{0,1}$ and $\Omega^{1,0}$, respectively, and these latter spaces have eigenvalues $\pm i$ with respect to $*$, respectively.

Likewise, we have
\begin{align}
Q^{\dag,[1]}_\tau &= -U_\tau d^* \\
&= -(1+ \tau i*)d^*\\
&= -d^* - \tau id* \label{eq:Qdag2line3}\\
&= -(\pd^* + \bar\pd^*) - \tau i(\pd + \bar\pd)* \\
&= -(\pd^* + \bar\pd^*) - \tau(-*\pd* + *\bar\pd*) \\
&= -\big((1-\tau)\pd^* + (1+\tau)\bar\pd^*\big).
\end{align}
The expression for $Q_\tau^{\dag,[0]}$ is straightforward.\End

\begin{Proposition}\label{Prop:well_defined}
 Definition \ref{Def:C_R2} is well-defined.
\end{Proposition}

\Proof The area form on the round sphere of radius $r$ is given by
$$d\sigma_r = \frac{4dxdy}{(1 + (x^2 + y^2)/r^2)^2}.$$
Let $S_r^2$ denote the sphere with the area forms $c_rd\sigma_r$ with $c_r \to \frac{1}{4}$. Then letting $r \to \infty$ constitutes a decompactification limit. Under this limit, we have
\begin{align*}
\lim_{r \to \infty} P_{hol, S_r^2} &= \left(\frac{c_r}{\pi}\frac{1}{(1+|z|^2/r^2)(1+|w|^2/r^2)}dz \frac{\bar z - \bar w}{z - w}dw\right)e_a\otimes e_a \\
&= P_{hol,\R^2}(z,w)
\end{align*}
Since the decompactification limit of the holomorphic gauge propagator exists, then a fortiori, the decompactification limit of holomorphic gauge Wilson loop expectations exist. Since Coloumb gauge expectations coincide witih holomorphic gauge expectations via Theorem \ref{Thm:C=Hol}, the result follows.\End

\subsection{2D Yang-Mills is finite}\label{Sec:Finite}

Fix a compatible Riemmanian metric and write $P = P_g$, where $P_g$ is the BV-propagator. We can
regulate the family of propagators $P_\tau$ interpolating between Coulomb gauge and holomorphic gauge (\ref{eq:P_t}) following (\ref{eq:BVprop_eps})--(\ref{eq:BVprop_eps2}), to obtain the family of regulated propagators \begin{equation}
P_{\tau,\eps} = Q_\tau^\dag \int_\eps^\infty e^{-t\Delta_g}dt, \qquad\eps > 0,\, 0 \leq \tau \leq 1.                                                                                                                  \end{equation}

Write $G_{BV,\eps} = \int_\eps^\infty e^{-t\Delta_g }dt$, which is a regulated Green's operator for $\Delta_g$ acting on the BV-complex $\cE$. We have
\begin{align*}
G_{BV,\eps} &= \oplus G_{BV,\eps}^{[i]}, \qquad G_{BV,\eps}^{[i]}: \cE^{[i]} \to \cE^{[i]}.
\end{align*}
We have
$$P_{\tau,\eps}^{[i]} = Q_\tau^{\dag, [i]}G^{[i+1]}_{BV,\eps}.$$

In this section, we prove the following result:

\begin{Theorem}\label{ThmFinite}
 For a closed surface $\Sigma$, we have
 \begin{equation}
   \lim_{\eps \to 0} e^{\lambda_0\pd_{P_{\tau,\eps}}}\left(W_{f,\gamma}e^{I}\right)\Big|_{conn, 0} \label{epsto0}
 \end{equation}
 exists, as a formal power series in $\lambda = \lam_0|\Sigma|$, for every $0 \leq \tau \leq 1$.
\end{Theorem}

In two-dimensions, Yang-Mills theory is superrenormalizable, which means that there are only finitely many one-particle irreducible\footnote{A Feynman diagram is one-particle irreducible if it cannot be disconnected by cutting an internal edge.} diagrams which are potentially divergent. This is a simple inspection: using the Fadeev-Popov procedure (which we will show to be equivalent to the BV procedure) the singularities of the Coulomb gauge propagator are logarithmic, so that the type of propagator insertions which lead to divergent integrals is highly constrained. Moreover, we only need to consider those one-particle irreducible diagrams that have external edges, since these ultimately need to be  connected to a Wilson loop observable.

It is easy to see that the only one-particle irreducible diagrams with external edges that are potentially divergent as $\eps \to 0$ are the ones in Figures I-III. Here we have omitted powers of $\lambda_0$ appearing in $\lambda_0\pd_{P_\eps}$ and in $I$, since for these diagrams they cancel to give an overall factor of $\lambda_0^0$. Individually, these diagrams are divergent as $\eps\to 0$, but we are only interested in sums over Feynman diagrams as occurring in (\ref{epsto0}). What we will show is that the sum of these diagrams remains finite as $\eps \to 0$.  The consequence is that the sum over \textit{all} Feynman diagrams in (\ref{epsto0}) become finite as $\eps \to 0$, since the only possible divergences come from a subgraph consisting of the sum of diagrams I--III.


\begin{figure}[h]
\centering
\includegraphics[scale=0.4]{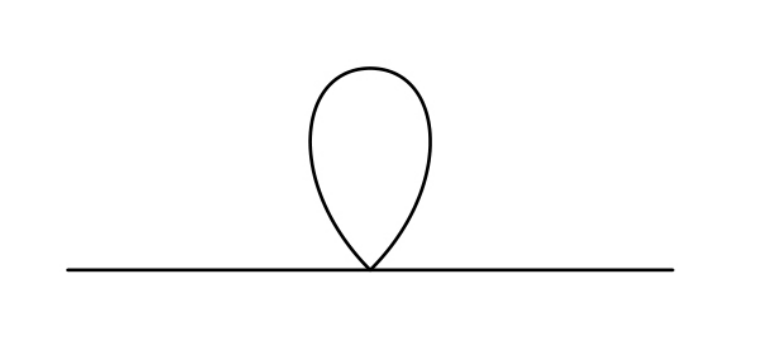}
\caption{Tadpole from $4$-point vertex arising from $\pd_{P_{\tau,\eps}} \left(-\frac{1}{8}\int \left<[A,A] \wedge *[A,A]\right>\right)$}
\end{figure}
\vspace{.25in}

\begin{figure}[h]
\centering
\includegraphics[scale=0.4]{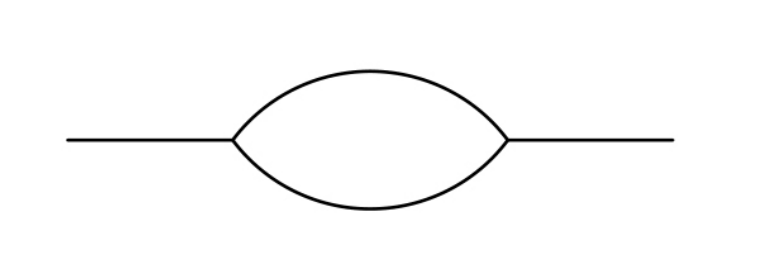}
\caption{Loop from $3$-point vertex arising from $\frac{1}{2}\pd_{P_{\tau,\eps}}^2 \frac{1}{2}\left(-\frac{1}{2}\int \left<[A,A] \wedge *dA\right>\right)^2$}
\end{figure}
\vspace{.25in}

\begin{figure}[h]
\vspace{.25in}
\centering
\includegraphics[scale=0.4]{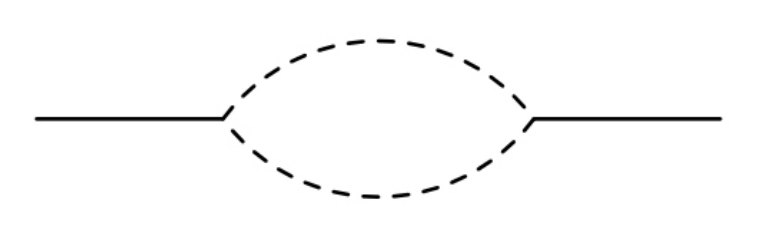}
\caption{Fermionic loop arising from $\frac{1}{2}\pd_{P_{\tau,\eps}}^2 \frac{1}{2}\left(\int \left<A, [A^\dag,X]\right>\right)^2$}
\end{figure}

Observe that each Feynman diagram has two pieces: a Lie-algebraic part and an analytic part. This is because the propagator $P_{\tau,\eps}$ factors into a differential-form part and the tensor $\mr{id}_\g \in \g \otimes \g$. Hence, the integrals arising from Feynman diagrams consist of Lie-algebraic contractions and differential-form contractions. For each of the above diagrams, we will compute each of these factors separately. (The individual Lie and analytic factors have an overall sign that depends on how one expresses Feynman diagrams as Wick contractions, but the product of these factors is always well-defined.) Here, a proper understanding of the signs and combinatorial factors attached to the different Feynman diagrams is absolutely essential, as they affect the sum leading to the cancellation of divergences.\\

In order to compute the Feynman diagrams I-III,
we need to perform some preliminary analysis of the precise form of the singularities of $P_{\tau,\eps}$ near the diagonal as $\eps \to 0$. The remainder of this section is divided into three subsections. First, we discuss the notational setup for our analysis near the diagonal, in particular, the singularity of the Green's function $G$ for the Laplace-Beltrami operator on $1$-forms. Second, we analyze the singularity of the Hodge projection $\Pi_{\im *d}$ and the composition $\Pi_{\im *d}G$ which occurs in our BV propagator. This allows us to understand the singularities of $P_{\tau,\eps}$ and its derivatives. Finally, in the third section, we piece together all these estimates to analyze the $\eps \to 0$ behavior of Feynman diagrams formed out of the $P_{\tau,\eps}$.\\

\subsubsection{Analysis near the diagonal}

Let $x$, $y$ be nearby points lying in a small coordinate chart of our surface $\Sigma$. Near the diagonal, the Green's function for the Laplacian on the space of $1$-forms takes the form
\begin{equation}
 G(x,y) = -\frac{\log \mr{d}(x,y)}{2\pi}\Phi_{x \leftarrow y} + l.o. \label{eq:Gsing}
\end{equation}
where $\Phi_{x\leftarrow y}$ denotes parallel transport (with respect to the Levi-Civita connection acting on forms) from $y$ to $x$, $\mr{d}(x,y)$ denotes the Riemannian distance between $x$ and $y$, and $l.o.$ denotes lower order terms (of order $O(d(x,y)\log d(x,y))$). In other words,
\begin{equation}
 \Delta^{-1}(\al) = \int_\Sigma G(x,y)\alpha(y) dV(y)
\end{equation}
where $dV$ is the Riemannian volume form. Using the $L^2$-pairing to identify $\Phi_{x \leftarrow y}$ with an element of $\Omega^1(\Sigma) \otimes \Omega^1(\Sigma)$, this is equivalent to $G(x,y)$ being the $L^2$-integral kernel of $\Delta^{-1}$.

In the above, we parametrized $G(x,y)$ in terms of two separate variables $x$ and $y$, one for each copy of $\Sigma$. Since the singularity of $G(x,y)$ and its derivatives occur along the diagonal, it will be convenient to parametrize $\Sigma \times \Sigma$ near the diagonal so that one coordinate parametrizes the diagonal submanifold of $\Sigma \times \Sigma$ and the other (local) coordinate measures the deviation from the diagonal via use of the exponential map. More precisely, we switch to the parametrization $(x,\exp_x(\mathbf{w}))$, where $\mb{w} \in T_x\Sigma$, so that $\mb{w} = 0$ corresponds to the diagonal submanifold. Thus, we consider the diffeomorphism
\begin{align}
\Psi: \mc{U} \subset T\Sigma & \to \Sigma \times \Sigma \nonumber \\
(x,\mb{w}) & \mapsto (x,\exp_x(\mb{w})), \label{eq:Psi}
\end{align}
where $U$ is a neighborhood of the zero section. The convenient feature about this coordinate system is that
the Riemannian distance $\mr{d}(x,\exp_x(\mb{w}))$ between two nearby points $x$ and $\exp_x(\mb{w})$ is simply
\begin{equation}
 \mr{d}(x,\exp_x(\mb{w})) = \|\mb{w}\|_x,
\end{equation}
where $\|\mb{w}\|_x$ is computed with respect to the Riemannian inner product on $T_x\Sigma$.
Thus, for $y = \exp_x(\mb{w})$ near $x$, we can write
\begin{align*}
 G(x,y) &= G(x,\exp_x(\mb{w}))\\
 &= -\frac{\log \|\mb{w}\|_x}{2\pi}\Phi_{x \leftarrow \exp_x(\mb{w})} + l.o.
\end{align*}
We will need to compute expressions involving exterior derivatives of the propagator and hence of $G(x,y)$. Using the local diffeomorphism (\ref{eq:Psi}), we translate derivatives in each of the component directions of $\Sigma \times \Sigma$ into derivatives with respect to the $(x, \mb{w})$--coordinate system. The exterior derivative on $\Sigma \times \Sigma$ splits into components along the first and second factor of $\Sigma$, which we denote by $d_x$ and $d_y$, respectively. (\textit{Warning:} The operation $d_x$ is ambiguous since it could also mean exterior derivative in the $x$-direction in the $(x,\mb{w})$-coordinate system. The meaning of $d_x$ will always be inferred from whether the function being differentiated is regarded as a globally defined function on $\Sigma \times \Sigma$ or as a function of $(x,\mb{w})$.) We are interested in computing the pull-backs $\Psi^*(d_x)$ and $\Psi^*(d_y)$ explicitly.

Pick a local coordinate system $x^\mu$ on $\Sigma$. This induces a product local coordinate system $(x^\mu, y^\nu)$ on $\Sigma \times \Sigma$, where the $x^\mu$ and $y^\mu$ can be identified using the identity map from the first factor of $\Sigma$ to the second factor. Consider a local orthonormal frame\footnote{We distingiush between an orthonormal basis $e_a$ in $\g$ and an orthonormal frame $e^\mu$ for $T^*\Sigma$ via roman versus latin indices, respectively.} $e^\mu = e^\mu(x)$ for $T_x^*\Sigma$, which induces a dual local frame $\pd_{e^\mu}$ for $T_x\Sigma$. In other words, $e^\mu$ is a coordinate system on $T_x\Sigma$, and thus we can write
$$\mb{w} = \mb{w}^\mu \pd_{e^\mu}, \qquad \mb{w} \in T_x\Sigma.$$
Performing a Taylor expansion of $\Psi(x,\mb{w})$ along $\mb{w} = 0$, we have
$$\Psi(x^\mu, \mb{w}^\nu) = (x^\mu, x^\mu + \mb{w}^\nu + O(\mb{w}^2)),$$
where $O(\mb{w}^2)$ denotes a ($x$-dependent) term which is a quadratic polynomial in $\mb{w}$. This follows from the fact that the derivative of the exponential map at the origin $\mb{w} = 0$ is the identity map. Since $d_x = dx^\mu \wedge \pd_{x^\mu} = e^\mu \wedge \pd_{e^\mu}$ and similarly for $d_y$, it follows that
\begin{align}
\Psi^*(d_x) &= e^\mu \wedge (\pd_{e^\mu} - \pd_{\mb{w}^\mu}) + O(\mb{w}) \label{eq:Psidx0}\\
\Psi^*(d_y) &= (e^\mu + d\mb{w}^\mu) \wedge \pd_{\mb{w}^\mu} + O(\mb{w}). \label{eq:Psidy0}
\end{align}
For the purposes of analyzing the singularities of Feynman diagrams (and thus of the $G(x,\exp_x(\mb{w})$), we can simplify (\ref{eq:Psidy0}) and (\ref{eq:Psidx0}) further.
Namely, introduce the somewhat ambiguous\footnote{
The criterion in (\ref{def:equiv}) is a priori ambiguous because it will be checked at a much later stage (the computation of Feynman diagrams) than when it is invoked. Fortunately, in two dimensions, the singularity structure is such that only leading order terms in $\mb{w}$ contribute.} notation ``$\equiv$'' to mean the following
\begin{equation}\equiv \;:\; \textrm{up to terms that will not contribute to divergences in Feynman diagrams} \label{def:equiv}
\end{equation}
Thus, we can write
\begin{align}
\Psi^*(d_x) &\equiv -e^\mu \wedge \pd_{\mb{w}^\mu} \label{eq:Psidx}\\
\Psi^*(d_y) &\equiv d\mb{w}^\mu \wedge \pd_{\mb{w}^\mu}. \label{eq:Psidy}
\end{align}
since the terms involving derivatives with respect to $\mb{w}$ will lead to the most singular terms when differentiating $G(x,\exp_x(\mb{w}))$ (and the lower order terms will not contribute to the singluarities in Feynman integrals.)

Likewise, we can simplify the expression for $G(x,\exp_x(\mb{w}))$ acting on $1$-forms. Use the $L^2$-pairing to identify the mapping $\Phi_{x \leftarrow \exp_x(\mb{w})}$ in (\ref{eq:Gsing}) as an element of $\Omega^1(\Sigma) \otimes \Omega^1(\Sigma)$. We have
\begin{equation}
\Phi_{x\leftarrow\exp_x(\mb{w})} = e^\mu(x) \otimes e^\mu(x) + O(\mb{w}) \label{eq:Phi_expansion}
\end{equation}
for $y = \exp_x(\mb{w})$ near $x$, where we can identify the left-most copy of $e^\mu(x)$ in the above as living in $T^*_y\Sigma$ using the local trivialization provided by our coordinate system $x^\mu$. Thus, we may write
\begin{equation}
G(x,\exp_x(\mb{w})) \equiv -\frac{\log \|\mb{w}\|}{2\pi}e^\mu \otimes e^\mu.
\end{equation}
More generally, letting $G_\eps(x,y)$ being the $L^2$-integral kernel of $\int_\eps^\infty e^{-t\Delta}dt$, we have
\begin{align}
G_\eps(x,\exp_x(\mb{w})) \equiv G_\eps^{\R^2}(\mb{w})e^\mu \otimes e^\mu,
\end{align}
since the heat-kernel on $\Sigma$ near the diagonal is to leading order equal to the heat-kernel on $\R^2$.

\subsubsection{Singularity analysis of $P_{\tau,\eps}^{[0]}(x,y)$}

The bosonic part of the propagator $P^{[0]}_{\tau,\eps}$ occurs in all the Feynman diagrams I-III, with III also containing the fermionic components $P^{[-1]}_{\tau,\eps}$ and $P^{[1]}_{\tau,\eps}$. We shall analyze the singularities of the $P^{[0]}_{\tau,\eps}$ in the present section; the singularities of the fermionic components will be handled in the next section.

We have
\begin{equation}
P_{\tau,\eps}^{[0]}(x,y) = \Big((1 +\tau i*)\circ \Pi_{\im *d}\circ (1 - \tau i*)\circ G_\eps\Big)(x,y), \label{eq:Pt0}
\end{equation}
where the integral kernel on the right-hand side is with respect to the $L^2$-inner product determined by the metric $g_\tau$. (Unless otherwise stated, BV operators with a superscript $[i]$ have their integrals defined with respect to the  BV pairing, whereas all other integral kernels are defined with respect to the $L^2$-pairing. We also suppress the Lie algebra dependence of our propagators in this section.). The case $\tau=0$ was established in (\ref{eq:P_FP=P_BV}) using (\ref{eq:PFPregulated}). The case $\tau > 0$ follows similarly using the expression (\ref{eq:Q0dag}).

It follows from (\ref{eq:Pt0}) that the singularity of $P_{\tau,\eps}^{[0]}(x,y)$ is determined by the singularity of $(\Pi_{\im *d} G_\eps)(x,y)$. This singularity is given by the following lemma:

\begin{Lemma}\label{Lemma:LemmaPi*dG}
 We have
 \begin{equation}\left(\Pi_{\im *d} G_\eps\right)(x,\exp_x(\mb{w})) \equiv \frac{1}{2}G_\eps(x,\exp_x(\mb{w})) \label{eq:Pi*dG}
 \end{equation}
 Consequently, we have
 \begin{equation}
 P_{\tau,\eps}^{[0]}(x,x) \equiv \frac{\log \eps^{-1/2}}{4\pi}(1 + \tau i)e^\mu \otimes (1 + \tau i)e^\mu
 \end{equation}

\end{Lemma}

\Proof First, we need to compute
\begin{align}
\Pi_{\im *d} &= d^*d G \label{eq:Piim*d}
\end{align}
in the sense of distributions. That is, we have to compute
$$\Pi_{\im *d}(\al) = \int_{\Sigma_y} G(x,y) \wedge *d^*d\al(y), \qquad \al \in \Omega^1(\Sigma).$$
Write $\beta(\mb{w}) = \exp_x^*(\al)(\mb{w})$ for short.
Thus,
\begin{align*}\int_{\Sigma_y} G(x,y) \wedge * d^*_yd_y\alpha(y)
&\equiv \int_{W_0} G(x,\exp_x(\mb{w})) \wedge * d^*_\mb{w}d_\mb{w}\beta(\mathbf{w})
\end{align*}
where $W_0$ stands for a neighborhood of the origin in $T_x\Sigma$.

Next, note that near $\mb{w} = 0$ in $T_x\Sigma$, the metric is given by $g_{\mu\nu}(\exp_x(\mb{w})) = \delta_{\mu\nu} + O(|\mb{w}|)$, since we write the component of $\mb{w} = \mb{w}^\mu\pd_{e^\mu}$, with respect to an orthonormal frame $\pd_{e^\mu}$. Also, observe that the associated coframe $e^\mu$, viewed as constant $1$-forms on $T_x\Sigma$, coincides with the coordinate $1$-form $d\mb{w}^\mu$. In the below, let $\eps_{\mu\nu}$ be the usual antisymmetric tensor and write $\beta = \beta_\mu e^\mu$. We have
\begin{align}
 \int_{W_0} G(x,\exp_x(\mb{w})) \wedge * d_\mb{w}^*d_\mb{w} \beta(\mathbf{w})
 &\equiv -\int_{W_0} \frac{1}{2\pi}\log |\mb{w}|(e^\mu \otimes d\mb{w}^\mu) \wedge * d_\mb{w}^*d_\mb{w} \beta(\mathbf{w})
\\
& \equiv e^\mu\int_{W_0} \frac{1}{2\pi}\log |\mb{w}| \pd_\nu\pd_\nu\beta_\mu \eps_{\nu\mu}d^2\mb{w}\\
&= -\lim_{\eps \to 0}e^\mu \int_{W_0\setminus B_\eps(0)}\frac{\mb{w}^\nu}{2\pi|\mb{w}|^2}\pd_\nu \beta_\mu \eps_{\nu\mu}d^2\mb{w}\\
&\equiv \lim_{\eps \to 0} e^\mu \int_{\pd B_\eps(0)}\frac{\mb{w}^\nu\beta_\mu }{2\pi|\mb{w}|^2}\eps_{\nu\mu}d\mb{w}^\mu \nonumber\\
&\qquad + \lim_{\eps \to 0}e^\mu \int_{W_0\setminus B_\eps(0)}\frac{1}{2\pi}\left(\frac{1}{|\mb{w}|^2} \frac{-2(\mb{w}^\nu)^2}{|\mb{w}|^4}\right)\beta_\mu\eps_{\nu\mu}d^2\mb{w}. \label{eq:twoterms}
\end{align}
Although we took $\beta$ to be smooth, observe that the second term of (\ref{eq:twoterms}) is finite as $\eps \to 0$ if the singular part of $\beta$ is radially symmetric. Indeed, writing $\beta_\mu(\mb{w}) = (\beta_{sing})_\mu(|\mb{w}|) + ({\beta_{smooth}})_\mu(0) + O(\mb{w})$, the first two terms yield integrals evaluating to zero, while the  $O(\mb{w})$ term yields a finite integral. In particular, we can let $\beta_\mu(\mb{w}) = (G_\eps)_{\mu\cdot}(\exp_y(\mb{w}),y)$. For the first term of (\ref{eq:twoterms}), using Stoke's Theorem, we get
\begin{align*}
\lim_{\eps \to 0} e^\mu \int_{B_\eps(0)}\frac{\mb{w}^\nu\beta_\mu d\mb{w}^\mu}{2\pi|\mb{w}|^2}\eps_{\nu\mu} &= e^\mu \lim_{\eps \to 0} \frac{1}{2\pi\eps^2}\int_{B_\eps(0)} \mb{w}^\nu\beta_\mu d\mb{w}^\mu\eps_{\nu\mu}\\
&= \frac{1}{2}\e^\mu \beta^\mu(0)\\
&= \frac{1}{2}\al(x).
\end{align*}
The above analysis implies (\ref{eq:Pi*dG}).

For the last equation, using (\ref{eq:Pt0}), we have
\begin{align*}
 P^{[0]}_{\tau,\eps}(x,x) &= \Big((1 +\tau i*)\circ \Pi_{\im *d}\circ (1 - \tau i*)\circ G_\eps\Big)(x,x)\\
 &= \Big((1 +\tau i*)\circ \Pi_{\im *d}\circ G_\eps \circ (1 - \tau i*)\Big)(x,x) \\
 &= ((1 +\tau i*) \otimes (1 + \tau i*))\Pi_{\im *d}G_\eps(x,x)\\
 &\equiv ((1 +\tau i*) \otimes (1 + \tau i*))\left(\frac{1}{2}\frac{\log \eps^{-1/2}}{2\pi}e^\mu \otimes e^\mu\right)\\
 &= \frac{\log \eps^{-1/2}}{4\pi}(1 + \tau i*)e^\mu \otimes (1+\tau i*)e^\mu.
\end{align*}
In the third line, we used that the $L^2$-adjoint of $(1 - \tau i*)$ on $\Omega^1(\Sigma)$ is $(1 + \tau i*)$.\End

Next, we compute the derivatives of $P^{[0]}_{\tau,\eps}(x,y)$.

\begin{Lemma}\label{Lemma:Pt1}
 We have
 \begin{align}
  (d_x \otimes 1) \circ P^{[0]}_{\tau,\eps}(x,y) &= (d_x \otimes (1 + \tau i*))G_\eps(x,y)  \label{eq:dxG}\\
  (1 \otimes d_y) P^{[0]}_{\tau,\eps}(x,y) &= \left((1 + \tau i*) \otimes d_y)G_\eps(x,y)\right) \label{eq:dyG}\\
  (d_x \otimes d_y) P^{[0]}_{\tau,\eps}(x,y) &= (d_x \otimes d_y)G_\eps(x,y) \label{eq:dxdyG}
 \end{align}
In the above, no sign-rule is used in evaluating $(1 \otimes d_y)$ and $(d_x \otimes d_y)$.
 \end{Lemma}

\Proof Using $d*\Pi_{\im *d} = 0$, $d\Pi_{\im *d} = d$, and $*G_\eps = G_\eps*$, then from (\ref{eq:Pt0}), we have
\begin{align*}
(d_x \otimes 1)P_{\tau,\eps}^{[0]}(x,y) &= \Big(d \circ (1 +\tau i*)\circ \Pi_{\im *d}\circ (1 - \tau i*)\circ G_\eps\Big)(x,y) \\
&= \Big(d \circ (1 - \tau i*)\circ G_\eps\Big)(x,y) \\
&= \Big(d \circ G_\eps \circ (1 - \tau i*)\Big)(x,y)\\
& = (d_x \otimes (1 + \tau i*))G_\eps(x,y).
\end{align*}
In the last line, we once again used that the $L^2$-adjoint of $(1 - \tau i*)$ on $1$-forms is $(1 + \tau i*)$. This proves (\ref{eq:dxG}).

Likewise, since the $L^2$-adjoint of $d$ is $d^*$, we have the following equality of integral kernels (derived from operators that map $\Omega^2(\Sigma; \g)$ to $\Omega^1(\Sigma; \g)$):
\begin{align*}
 (1 \otimes d_y)P_{\tau,\eps}^{[0]}(x,y) &= \Big((1 +\tau i*)\circ \Pi_{\im *d}\circ (1 - \tau i*)\circ G_\eps \circ d^*\Big)(x,y) \\
 &= \Big((1 +\tau i*)\circ \Pi_{\im *d}\circ (1 - \tau i*)\circ d^* \circ G_\eps^{(2)}\Big)(x,y) \\
 &= \Big((1 +\tau i*)\circ d^* \circ G_\eps^{(2)}\Big)(x,y)\\
 &= \Big((1 +\tau i*)\circ G_\eps \circ d^*\Big)(x,y)\\
 &= \Big((1 + \tau i*) \otimes d_y\Big)G_\eps(x,y)
\end{align*}
Here $G_\eps^{(2)}$ is the psuedo-inverse for $\Delta$ on $\Omega^2(\Sigma; \g)$. 
Likewise,
\begin{align*}
 (d_x \otimes d_y)P_{\tau,\eps}^{[0]}(x,y) &= (d\circ G_\eps \circ d^*)(x,y) \\
 &= (d_x \otimes d_y)G_\eps(x,y).
\end{align*}

\subsubsection{Computation of Feynman diagrams}

We now evaluate the Feynman diagrams given by Figures I-III.\\\

\noindent\textbf{Figure I.} (tadpole on 4 point function):\\

The only Wick contractions $\pd_{P_\eps} \left(-\frac{1}{8}\int \left<[A,A] \wedge *[A,A]\right>\right)$ which are nonzero are those which pair $A$'s from different Lie bracket terms (since $P$ is proportional to $\mr{id}_\g$). There are four such possible contractions.

Without loss of generality, regard the Wick contraction as contracting the second and third copy of $A$. In what follows, pick an orthonormal basis $e_a$ for $\g$ with respect to $\left<\cdot,\cdot\right>$. We have the Lie algebra factor $\left<[e_a, e_c], [e_c, e_b]\right>$, where $e_a$ and $e_b$ are Lie algebra factors coming from the external legs and we sum over repeated indices. Letting $C = \mr{Ad}(e_c)\mr{Ad}(e_c)$, we have
\begin{align*}
  \left<[e_a, e_c], [e_c, e_b]\right> &= \left<[e_c,[e_c,e_a]],e_b\right> \\
  &= C_{ab}
\end{align*}
where $C_{ab}$ is the $ab$ matrix element of $C$. Thus we have
\begin{align}
  (I)_{Lie} = C_{ab}
\end{align}
Note that for simple gauge group, $C_{ab} = -c_2(\mr{Ad}(\g))\delta_{ab}$, where $c_2(\mr{Ad}(\g))$ is the value of the quadratic Casimir in the adjoint-representation.

To compute the analytic factor of (I), write the external leg variable as $A = A^a e_a$, where $A^a$ is a scalar-valued $1$-form. We can treat the propagator as a real-valued differential form on $\Sigma \times \Sigma$, since the Lie algebra factor has been accounted for. By abuse of notation, we use the same notation for scalar-valued counterparts in this setting. We have
\begin{align*}
  (I)_{\mr{an}} &= 4\cdot -\frac{1}{8}\int_\Sigma \Big(A^a(x) \wedge \overbracket{\bullet\Big) \wedge *\Big(  \bullet}^{P_{\tau,\eps}^{[0]}(x,x)} \wedge A^b(x)\Big),
\end{align*}
where $4$ is the combinatorial factor arising from the number of Wick contractions. Using Lemma \ref{Lemma:LemmaPi*dG}, we have
\begin{align*}
  (I)_{\mr{an}} &\equiv -\frac{1}{2}\frac{\log \eps^{-1/2}}{4\pi} \int_\Sigma A^a(x) \wedge (1+i\tau*)e^\mu(x) \wedge *( (1+i\tau*)e^\mu(x) \wedge A^b(x) ) \\
  &= \frac{\log\eps^{-1/2}}{8\pi} \int_\Sigma (1-i\tau*)A^a(x) \wedge *(1-i\tau*)A^b(x)\\
&= \frac{(1-\tau^2)\log\eps^{-1/2}}{8\pi} \int_\Sigma A^a(x) \wedge *A^b(x).
  \end{align*}
Altogether, the Feynman integral corresponding to Figure I has a singular part equal to
\begin{align}
  (I) &= (I)_{Lie}(I)_{an} \\
&\equiv \frac{(1-\tau^2)\log\eps^{-1/2}}{8\pi} \int_\Sigma C_{ab} A^a \wedge *A^b.
\end{align}

\noindent\textbf{Figure II.} (one loop graph using 3 point vertices)\\

There are two possible types of Wick contractions that lead to divergences. For the first type, a propagator joins both copies of $dA$. For the second, $dA$ belongs to separate Wick contractions. The number of possible Wick contractions in each case is $4$. Call the resulting Feynman diagrams $(II_1)$ and $(II_2)$, respectively.

For $(II_2)$, regard each $dA$ of $\left<[A,A] \wedge *dA\right>$ as being contracted with the rightmost copy of $A$ in the $[A,A]$ term of the other copy of $\left<[A,A], *dA\right>$. So we get the Lie-algebraic factor
\begin{align*}
  (II_2)_{Lie} &= \left<[e_a, e_c],e_d\right>\left<[e_b,e_d],e_c\right>\\
  &= -\left<[e_a,e_c], e_d\right>\left<e_d, [e_b,e_c]\right> \\
  &= -\left<[e_a,e_c],[e_b,e_c]\right> \\
  &= C_{ab}.
\end{align*}

For the analytic factor, we have
\begin{align}
(II_2)_{\mr{an}} &= 4\cdot \frac{1}{4}\cdot \left(\frac{1}{2}\right)^2 \int_{\Sigma_x \times \Sigma_y}\big(A^a(x) \wedge \overbracket{\bullet \wedge *d_x \bullet\big) \big(A^b(y) \wedge \bullet\wedge * d_y \bullet}^{P^{[0]}_{\tau,\eps}(x,y)}\big)\\[-4.5ex]
&= \hphantom{4\cdot \frac{1}{4}\cdot \left(\frac{1}{2}\right)^2 \int_{\Sigma_x \times \Sigma_y}\big(A^a(x) \wedge \bullet \wedge *d_x} \underbracket{\hphantom{\bullet\big) \big(A^b(y) \wedge \bullet\,}}_{P^{[0]}_{\tau,\eps}(x,y)} \hphantom{\wedge * d_y \bullet\big)}\nonumber
\end{align}

Using Lemma \ref{Lemma:Pt1} to evaluate the derivatives of the $P_{\tau,\eps}^{[0]}(x,y)$ in terms of the Green's functions $G_\eps(x,y)$, and then using Lemma \ref{Lemma:SingInt} to simplify the resulting integral of derivatives of Green's functions, we have

\begin{align}
 (II_2)_{\mr{an}} &= 4\cdot \frac{1}{4}\cdot \left(\frac{1}{2}\right)^2 \int_{\Sigma_x \times \Sigma_y}\big(A^a(x) \wedge \overbracket{\bullet \wedge * \hphantom{\underbracket{\bullet\big) \big(A^b(y) \wedge \bullet}_{\big(d_xG_\eps \circ (1 - \tau i*)\big)(x,y)}}\wedge * \bullet}^{\big((1 + \tau i*) \otimes d_y\big)G_{\eps}(x,y)}\big)\label{eq:II_2_an_Wick}\\[-5.5ex] 
&= \hphantom{4\cdot \frac{1}{4}\cdot \left(\frac{1}{2}\right)^2 \int_{\Sigma_x \times \Sigma_y}\big(A^a(x) \wedge \bullet \wedge *} \underbracket{\bullet\big) \big(A^b(y) \wedge \bullet}_{\big(d_x \otimes (1 + \tau i*)\big)G_\eps(x,y)} \hphantom{\wedge * \bullet\big)}\nonumber\\
&\equiv -\frac{1}{16\pi^2}\int_{\Sigma_x}\int_{\eps^{1/2} < |\mb{w}| < 1}\Big(A^a(x) \wedge (1+\tau i*)e^{\mu_1}(x)\Big)\eps_{\nu_2\mu_2} \times \nonumber \\
&\qquad\qquad \Big(A^b(\exp_x(\mb{w})) \wedge (1+ \tau i*)e^{\mu_2}(\exp_x(\mb{w})\Big)\eps_{\nu_1\mu_1}\frac{\mb{w}^{\nu_1}\mb{w}^{\nu_2}}{|\mb{w}|^4} \label{eq:II_2_an_ps}
\end{align}
Only the terms even in $\mb{w}$ survive in (\ref{eq:II_2_an_ps}). Hence, stipulate $\nu_1 = \nu_2 =: \nu$ in the above, in which case, the factor $\eps_{\nu_1\mu_1}\eps_{\nu_2\mu_2}$ becomes $\eps_{\nu\mu}^2$ where $\mu_1=\mu_2 =: \mu$.
Thus, we get
\begin{align*}
(II_2)_{\mr{an}} &= -\frac{1}{16\pi^2}\int_{\Sigma_x}\int_{\eps^{1/2} < |\mb{w}| < 1}\Big((1-\tau i*)A^a(x) \wedge e^{\mu}(x)\Big)\\
& \hspace{1.5in} \times *\Big((1- \tau i*)A^b(\exp_x(\mb{w})) \wedge e^{\mu}(\exp_x(\mb{w})\Big)\eps_{\nu\mu}^2\frac{(\mb{w}^{\nu})^2}{|\mb{w}|^4}d^2\mb{w}
\\
& \equiv -\frac{1}{16\pi^2}\int_{\Sigma_x} (1 - \tau i*)A^a(x) \wedge *\Big((1 - \tau i*)A^b(x)\Big) \int_{\eps^{1/2} < |\mb{w}| < 1}\frac{1}{2}\frac{d^2\mb{w}}{|\mb{w}|^2}\\
&= \frac{(1-\tau^2)\log \eps^{-1/2}}{16\pi}\int_{\Sigma_x} A^a(x) \wedge *A^b(x)
\end{align*}
Here, we used that
$$\int_{\eps^{1/2} < |\mb{w}| < 1} \frac{(\mb{w}^\nu)^2}{|\mb{w}|^4}d^2\mb{w} = \frac{1}{2}\int_{\eps^{1/2} < |\mb{w}| < 1}  \frac{|\mb{w}|^2}{|\mb{w}|^4}d^2\mb{w}$$
by radial symmetry. So altogether, we obtain
\begin{equation}
 (II_2) \equiv \frac{(1-\tau^2)\log \eps^{-1/2}}{16\pi}\int_\Sigma C_{ab}A^a \wedge *A^b.
\end{equation}

For $(II_1)_{Lie}$, we have a Wick contraction of the two $dA$'s; let the other Wick contraction be on the right-most $A$'s. We get the Lie-algebraic factor:
\begin{align*}
  (II_1)_{Lie} &= \left<[e_a,e_c], e_d\right>\left<[e_b,e_c],e_d\right> \\
  &=  -C_{ab}
\end{align*}

For the analytic factor, we have
\begin{align}
(II_1)_{\mr{an}} &= 4\cdot \frac{1}{4}\cdot \left(\frac{1}{2}\right)^2 \int_{\Sigma_x \times \Sigma_y}\big(A^a(x) \wedge \bullet \wedge *d_x \overbracket{\bullet\big) \big(A^b(y) \wedge \bullet)\wedge * d_y \bullet}^{P^{[0]}_{\tau,\eps}(x,y)}\big)\\[-4.5ex]
&= \hphantom{4\cdot \frac{1}{4}\cdot \left(\frac{1}{2}\right)^2 \int_{\Sigma_x \times \Sigma_y}\big(A^a(x) \wedge\,}\underbracket{\hphantom{\bullet \wedge *d_x \bullet\big) \big(A^b(y) \wedge \bullet}\!}_{P^{[0]}_{\tau,\eps}(x,y)} \hphantom{\wedge * d_y \bullet\big)} \nonumber
\end{align}
As with $(II_2)_{an}$, using Lemma \ref{Lemma:Pt1} to evaluate the derivatives of the $P_{\tau,\eps}^{[0]}(x,y)$ in terms of the Green's functions $G_\eps(x,y)$, and then using Lemma \ref{Lemma:SingInt} to simplify the resulting integral of derivatives of Green's functions, we have
\begin{align}
 (II_1)_{an} &= 4\cdot \frac{1}{4}\cdot \left(\frac{1}{2}\right)^2 \int_{\Sigma_x \times \Sigma_y}\big(A^a(x) \wedge \bullet \wedge * \overbracket{\bullet\big) \big(A^b(y) \wedge \bullet\wedge * \bullet}^{(d_x \otimes d_y)G_\eps(x,y)}\big)\\[-4.5ex]
&= \hphantom{4\cdot \frac{1}{4}\cdot \left(\frac{1}{2}\right)^2 \int_{\Sigma_x \times \Sigma_y}\big(A^a(x) \wedge\,}\underbracket{\hphantom{\bullet \wedge *\bullet\big) \big(A^b(y) \wedge \bullet}\!}_{P^{[0]}_{\tau,\eps}(x,y)} \hphantom{\wedge * \bullet\big)} \nonumber\\
&\equiv \frac{1}{4}\int_{\Sigma_x}\int_{|\mb{w}| < 1}\Big(A^a(x) \wedge \underbracket{\bullet \Big)\Big(A^b(\exp_x(\mb{w}))\wedge \bullet}_{P^{[0]}_{\tau,\eps}(x,\exp_x(\mb{w}))}\Big) (-\pd_{\mb{w}^\nu}\pd_{\mb{w}^\nu}G^{\R^2}_\eps(x,\exp_x(\mb{w})))d^2\mb{w}\\
&\equiv \frac{1}{4}\int_{\Sigma_x}\int_{|\mb{w}| < 1}\Big(A^a(x) \wedge (1+\tau i*)e^\mu(x) \Big)\left(A^b(\exp_x(\mb{w}))\wedge (1 + \tau i*)e^\mu(\exp_x(\mb{w}))\right) \times \nonumber \\
&\hspace{1.5in} \frac{1}{2}G^{\R^2}_\eps(x,\exp_x(\mb{w}))(-\pd_{\mb{w}^\nu}\pd_{\mb{w}^\nu}G^{\R^2}_\eps(x,\exp_x\mb{w}))d^2\mb{w} \label{eq:useLemmaPi*dG}\\
&\equiv \frac{1}{8}\frac{\log \eps^{-1/2}}{2\pi}\int_{\Sigma_x}(1 - \tau i*)A^a(x) \wedge *(1 - \tau i*)A^b(x)\\
&= \frac{(1 - \tau^2)\log \eps^{-1/2}}{16\pi}\int_{\Sigma_x}A^a(x) \wedge *A^b(x).
\end{align}
Here, we used Lemma \ref{Lemma:LemmaPi*dG} in (\ref{eq:useLemmaPi*dG}). So altogether, we get
\begin{equation}(II_1) \equiv -\frac{(1-\tau^2)\log \eps^{-1/2}}{16\pi} \int_\Sigma C_{ab}A^a \wedge *A^b.\end{equation}\\
Altogether, the Feynman integral corresponding to Figure II satisfies
\begin{align*}
  (II) &= (II_1) + (II_2) \\
&\equiv 0.
\end{align*}

\noindent\textbf{Figure III.} (ghost loop) \\

Here we have two fermionic propagators, $P^{[1]}_{\tau,\eps} = Q^{\dag,[1]}_tG^{[2]}_{BV,\eps}$ and $P^{[-1]}_{\tau,\eps} = Q^{\dag,[-1]}G_{BV,\eps}^{[0]}$, corresponding to contracting $A^\dag$ with $X$ and vice versa. The Lie-algebraic factor we get is
\begin{align*}
(III)_{Lie} &=  \left<e_a, [e_c, e_d]\right>\left<e_b,[e_d,e_c]\right>\\
&= -\left<[e_a,e_c],e_d\right>\left<[e_b,e_c],e_d\right> \\
&= -\left<[e_a,e_c],[e_b,e_c]\right> \\
&= C_{ab}.
\end{align*}
For the analytic factor, we have to compute
\begin{align}
 (III_3)_{an} &= 2\cdot\frac{1}{4}\int_{\Sigma \times \Sigma} (A^a(x) \wedge \bullet \wedge \overbracket{\bullet)(A^b(y) \wedge \bullet}^{P^{[-1]}_{\tau,\eps}(x,y)} \wedge \bullet)\nonumber\\[-4.2ex]
 & \hphantom{\;= 2\cdot\frac{1}{4}\int_{\Sigma\times\Sigma} (A^a(x) \wedge\,\,}\underbracket{\hphantom{\bullet \wedge \bullet)(A^b(y) \wedge \bullet \wedge \bullet\!\!\!}}_{P^{[1]}_{\tau,\eps}(x,y)}  \label{eq:III_an}
\end{align}
where the $2$ is a symmetry factor arising from the fact that $P^{[1]}_{\tau,\eps}$ and $P^{[-1]}_{\tau,\eps}$ can arise from either of the two copies of the full propagator $P_{\tau,\eps}$).

For any given fixed $x$, let $\mb{z}$ and $\bar{\mb{z}}$ be local holomorphic-antiholomorphic coordinates in $\mb{w}$-space (with respect to the complex structure on $T_x\Sigma$). Thus if $\mb{w} = (\mb{w}^1, \mb{w}^2)$, where we make the identification $T_x\Sigma \cong \R^2$ equipped with the standard complex structure, then
\begin{align*}
\mb{z} &= \mb{w}^1 + i\mb{w}^2\\
\bar{\mb{z}} &= \mb{w}^1 - i\mb{w}^2.
\end{align*}
Let $e = \frac{1}{\sqrt{2}}d\mb{z}$ and $\bar e = \frac{1}{\sqrt{2}}d\bar{\mb{z}}$ be normalized holomorphic and anti-holomorphic $1$-forms, respectively. Likewise, let $\pd_e = \sqrt{2}\pd_z$ and $\pd_{\bar e} = \sqrt{2}\pd_{\bar z}$ be normalized holomorphic and anti-holomorphic tangent vectors, respectively. The holomorphic and anti-holomorphic part of the exterior derivative on $T_xX$, and their adjoints, can thus be expressed as
\begin{alignat*}{2}
\pd &= d\mb{z} \wedge \pd_{\mb{z}} = \sqrt{2}e \wedge \pd_{\mb{z}}  & \qquad\bar\pd &= d\bar{\mb{z}}  \wedge \pd_{\bar{\mb{z}}} = \sqrt{2}\bar e \wedge \pd_{\mb{\bar z}}\\
\pd^* &= -\sqrt{2}\pd_{\bar{\mb{z}}}\iota_{\pd_e} & \qquad \bar\pd^* &= -\sqrt{2}\pd_{\mb{z}}\iota_{\pd_{\bar{e}}}.
\end{alignat*}
(Note that $\pd$ subscripted by a variable denotes the partial derivative with respect to that variable whereas $\pd$ by itself denotes the holomorphic part of the exterior derivative in $\mb{w}$-space. Likewise for $\bar\pd$.)

On $\Omega^1(\Sigma; \g) = \cE^{[0]}$, the identity tensor with respect to the $BV$ convolution-pairing is
$$-i(\bar e \otimes e - e \otimes \bar e),$$
where the second factor is regarded as an element of $\cE^{[1]}$.
So
\begin{align}
G_{\eps=0,BV}^{[0]}(x, \exp_x(\mb{w})) &\equiv \frac{i}{4\pi}\log (\mb{z}\bar{\mb{z}})\left( \bar e \otimes e - e \otimes \bar e\right) \\
 Q^{\dag, [-1]}_\tau G^{[0]}_{\eps=0, BV}(x,\exp_x(\mb{w})) & = -\Psi^*\Big((1-\tau)\pd^* + (1+\tau)\bar\pd^* \Big)G^{[0]}_{\eps=0, BV}(x,\exp_x(\mb{w})) \\
 &\equiv -\frac{i\sqrt{2}}{4\pi}\left((1-\tau)\frac{1}{\bar{\mb{z}}}1 \otimes \bar e - (1+\tau)\frac{1}{\mb{z}}1\otimes e\right) \label{eq:Q+0G}
\end{align}
In the above, we are interested in $\eps = 0$ owing to Remark \ref{Rem:PS}.

On $\Omega^2(\Sigma; \g) = \cE^{[2]}$, the identity tensor with respect to the BV convolution pairing is given by
$$ie\wedge \bar e \otimes 1 \in \cE^{[2]} \otimes \cE^{[-1]}.$$
So
\begin{align}
G_{\eps=0,BV}^{[2]}(x, \exp_x(\mb{w})) &\equiv -\frac{i}{4\pi}\log (\mb{z}\bar{\mb{z}})\left(e \wedge \bar e \otimes 1\right)\\
 Q^{\dag, [1]}_\tau G_{\eps=0, BV}^{[2]}(x,\exp_x(\mb{w})) & = -\Psi^*\Big((1-\tau)\pd^* + (1+\tau)\bar\pd^* \Big) G_{\eps=0, BV}^{[2]}(x,\exp_x(\mb{w})) \\
 &\equiv \frac{i\sqrt{2}}{4\pi}\left((1-\tau)\frac{1}{\bar{\mb{z}}}\bar e \otimes 1 - (1+\tau)\frac{1}{\mb{z}}e\otimes 1\right)\label{eq:Q+2G}
\end{align}
Notice the difference in sign between (\ref{eq:Q+0G}) and (\ref{eq:Q+2G}); this implicitly accounts for the fermionic sign rule.

We can now compute $(III)_{an}$ via Lemma \ref{Lemma:SingInt}, Remark \ref{Rem:PS}, and formulas (\ref{eq:Q+0G}) and (\ref{eq:Q+2G}). Proceeding from (\ref{eq:III_an}), we have
\begin{align}
 (III)_{an} &\equiv \frac{1}{2}\cdot \frac{-(1-\tau^2)}{8\pi^2}\int_{\Sigma_x}\int_{\eps^{1/2} < |\mb{w}| < 1} \left(A^a(x) \wedge \bar e(x)\right)\nonumber \\
 & \hspace{1in}\left(A^b(\exp_x(\mb{w})) \wedge e(\exp_x(\mb{w})) \frac{d^2\mb{w}}{\mb{z}\bar{\mb{z}}}\right) + c.c. \label{eq:III_an_2}
 \end{align}
Here, only terms with both a $\mb{z}$ and $\bar{\mb{z}}$ contribute to the singularity in the product of $Q^{\dag, [1]}_\tau G_{\eps=0, BV}^{[2]}$ and $Q^{\dag, [-1]}_\tau G_{\eps=0, BV}^{[1]}$ (by symmetry), and $c.c.$ stands for complex conjugate (arising from switching $e$ and $\bar e$ in \ref{eq:III_an_2}). We have
\begin{multline*}
 \int_{\Sigma_x}\int_{\eps^{1/2} < |\mb{w}| < 1} \left(A^a(x) \wedge \bar e(x)\right)\left(A^b(\exp_x(\mb{w})) \wedge e(\exp_x(\mb{w})) \frac{1}{\mb{z}\bar{\mb{z}}}\right) + c.c.\\ \begin{split}&\equiv
 \int_{\Sigma_x} \left(A^a(x) \wedge \bar e(x)\right)*\left(A^b(x) \wedge e(x)\right)\int_{\eps^{1/2} < |\mb{w}| < 1} \frac{d^2\mb{w}}{|\mb{w}|^2} + c.c.\\
 &= 2\pi \log \eps^{-1/2}\int A^a(x) \wedge *A^b(x).
 \end{split}
\end{multline*}
It follows that
\begin{equation}
 (III) \equiv \frac{-(1-\tau^2)\log \eps^{-1/2}}{8\pi}\int C_{ab}A^a \wedge *A^b.
\end{equation}

\noindent\textbf{Proof of Theorem \ref{ThmFinite}: } From the above computations, we find that
$$(I) + (II) + (III) \equiv 0.$$
Thus, the limit (\ref{epsto0}) is finite since all divergences cancel.\End

\section{Exact Asymptotics vs. Perturbation Theory}\label{Sec:Asymp}

The relation (\ref{intro:expect}) for $\lambda\phi^4$ theories in dimensions two and three have been established, where it is known that $n$-point correlation functions have asymptotic expansions in small $\lambda$ equal to the formal series one obtains from perturbation theory (in fact, the latter can be Borel resummed to recover the $n$-point functions exactly) \cite[Ch 23.2]{GJ}. Such results do not directly address our line of inquiry however, since for scalar field theories, both perturbative and nonperturbative calculations involve the same scalar field and therefore use identical formulations. What makes the investigation of this paper notable is that it compares two different formulations: lattice (group-valued fields) versus continuum (Lie-algebra valued) fields.

For Yang-Mills theory, the central difficulty is that while $\lim_{\lambda \rightarrow 0}\left<W_{f,\gamma}\right>_\Sigma$ in two-dimensions can be evaluated by determining the asymptotics of heat kernels on Lie groups, the computation of $\left<W_{f,\gamma}\right>_{\Sigma, pert}$, to all orders in $\lambda$ is comparatively much harder to perform. Indeed, for Coulomb gauge, as the order in $\lambda$ increases, one has an increasingly complicated set of Feynman integrals to calculate. While holomorphic gauge is simpler since the theory becomes free (i.e. the interactions $I$ do not contribute to Feynman diagrams), the integrals one has to contribute are still highly nontrivial. This arises from $G$ being nonabelian (the nontrivial case), since then the combinatorics of Wick contractions arising from Lie-algebraic insertions into Wilson loop operators complicates the analysis.

Nevertheless, we are able to obtain the following results. To distinguish between limits of functions and the limits of asymptotic series we are about to perform, let
$\underset{\lam_0 \to 0}{\mr{a.s.}}\left<W_{f,\gm}\right>_{\Sigma}$
denote the asymptotic series in $\lam_0$ corresponding to the $\lam_0 \to 0$ limit of $\left<W_{f,\gm}\right>_{\Sigma}$, and similarly for
$\underset{\lam \to 0}{\mr{a.s.}}\left<W_{f,\gm}\right>_{\Sigma}$.
In what follows, we assume, as we always have, that the function $f$ is trace in an irreducible representation of $G$ so as to make use of the formula (\ref{eq:W_expansion}) for computing perturbative Wilson loop expectations.

\begin{Theorem}\label{Thm:SCC}
Let $\gm$ be a simple closed curve. Then
 \begin{equation}
  \underset{\lam \to 0}{\mr{a.s.}}\left<W_{f,\gm}\right>_{S^2} = \left<W_{f,\gm}\right>_{S^2,hol} \label{eq:ThmO3}
 \end{equation}
 In particular, the above equality holds in the decompactification limit:
 \begin{equation}
  \lim_{S^2\to\R^2}\underset{\lam_0 \to 0}{\mr{a.s.}}\left<W_{f,\gm}\right>_{S^2} = \left<W_{f,\gm}\right>_{\R^2,hol}.
 \end{equation}
\end{Theorem}

\Proof Theorem \ref{ThmExact1}  shows that the left-hand side of (\ref{eq:ThmO3}) is given by a Gaussian matrix integral. On the other hand, the work of \cite{GP}, shows that $\left<W_{f,\gm}\right>_{S^2, hol}$, for $\gm$ a circular contour, agrees with the same Gaussian matrix integral. For general $\gm$, we can find a diffeomorphism mapping $\gm$ to a circular contour and which preserves the area of the regions complementary to $\gm$ (following the lines of the proof of Theorem \ref{Thm:Area}). Thus, the result for general simple closed curves now follows from the invariance of holomorphic gauge under area-preserving diffeomorphisms via Theorem \ref{Thm:C=Hol}.\End

The next result indicates a subtlely in the various limits and gauges that arise in two-dimensional Yang-Mills theory. We have from \cite{NguSYM} that
\begin{equation}
 \left<W_{f,\gm}\right>_{\R^2} = \underset{\lam_0 \to 0}{\mr{a.s.}}\left<W_{f,\gm}\right>_{\R^2} = \left<W_{f,\gm}\right>_{\R^2, \mr{ax}}.
\end{equation}
On the other hand, we have the following

\begin{Theorem}\label{Thm:SwitchLimits}
 We have
 \begin{equation}
\lim_{S^2 \to \R^2}\underset{\lam_0 \to 0}{\mr{a.s.}}\left<W_{f,\gm}\right>_{S^2} \not= \underset{\lam_0 \to 0}{\mr{a.s.}}\left<W_{f,\gm}\right>_{\R^2},    \label{eq:distinct}
 \end{equation}
 i.e. decompactification and small coupling asymptotics do not commute. Consequently,
 \begin{equation}
  \underset{\lam_0 \to 0}{\mr{a.s.}}\left<W_{f,\gm}\right>_{\R^2} \neq \left<W_{f,\gm}\right>_{\R^2, hol}. \label{eq:distinct2}
 \end{equation}
 That is, the perturbative expansion in holomorphic gauge on $\R^2$ does not agree with the asymptotics of the exact expectation. Consequently, we have the following nonequivalence of gauges:
 \begin{equation}
  \left<W_{f,\gm}\right>_{\R^2,ax} \neq \left<W_{f,\gm}\right>_{\R^2,hol} = \left<W_{f,\gm}\right>_{\R^2,C}. \label{eq:distinct3}
 \end{equation}
 \end{Theorem}

\Proof Let $\gm$ be a simple closed curve. The left-hand side (\ref{eq:distinct}) is computed to all orders in (\ref{eq:ThmExact2}). The right-hand side is determined from
\begin{align}
 \left<W_{f,\gm}\right>_{\R^2}
 &= \int_G K_{\lambda_0|R|}(g)f(g)dg \nonumber\\
 &= \left(e^{-\lambda_0|R|\Delta/2}f\right)(1) \nonumber\\
 &= e^{-\lambda_0|R|c_2(\rho)/2}\dim \rho \label{eq:R2WLE}
\end{align}
where $f = \tr\, \rho$ and $c_2(\rho)$ is the quadratic Casimir for the irreducible representation $\rho$. The functions (\ref{eq:R2WLE}) and (\ref{eq:ThmExact2}) are not equal. In fact, they define entire functions in $\lambda_0$, and so their series expansions about $\lambda_0 = 0$ are not equal. In particular, Example 1 at the end of this section shows that (\ref{eq:R2WLE}) and (\ref{eq:ThmExact2}) can disagree at second order in $\lambda_0$. Hence, (\ref{eq:distinct}) follows, and from Theorem \ref{Thm:SCC} so does (\ref{eq:distinct2}). Since $\underset{\lam_0 \to 0}{\mr{a.s.}}\left<W_{f,\gm}\right>_{\R^2} = \left<W_{f,\gm}\right>_{\R^2,ax}$ via \cite{NguSYM}, we also have (\ref{eq:distinct3}).\End

We provide some explicit computations of the small coupling asymptotics of $\left<W_{f,\gm}\right>_{S^2}$ for $\gm$ a simple closed curve. The final answer is remarkably simple.

\begin{Theorem}\label{ThmExact1}
  Let $\gamma$ be a simple closed curve on $S^2$. Define
\begin{equation}
  \varrho = \lambda \frac{|R_1||R_2|}{|S^2|^2}
\end{equation}
where $R_1$ and $R_2$ are the two regions in the complement of $\gamma$. Then for any compact Lie group $G$, we have as $\lambda \to 0$
  \begin{equation}
    \underset{\lam \to 0}{\mr{a.s.}}\left<W_{f,\gamma}\right>_{S^2} = \frac{1}{(2\pi \varrho)^{d/2}}\int_\g f(\exp(X))e^{-|X|^2/2\varrho}dX. \label{eq:ThmExact1}
  \end{equation}
  In the decompactification limit (in which $|R_2|/|S^2| \to 1$, where $R_2$ is the unbounded region of $\gm$ viewed as a subset of $\R^2$), we obtain
  \begin{equation}
\lim_{S^2 \to \R^2}\underset{\lam_0 \to 0}{\mr{a.s.}}\left<W_{f,\gamma}\right>_{S^2} = \frac{1}{(2\pi |R_1|\lam_0)^{d/2}}\int_\g f(\exp(X))e^{-|X|^2/2|R_1|\lambda_0}dX. \label{eq:ThmExact2}
  \end{equation}
\end{Theorem}

\Proof Let $H$ be a maximal torus of $G$ with Lie algebra $\fh$. For the time being, we normalize our inner product $\left<\cdot,\cdot\right>$ on $\g$ so that the volume form it induces on $\g$ is the normalized Haar measure on $G$. This is so that the normalized Haar measure $dg$ occurring in the definition (\ref{def:heatkernel}) of the heat kernel $K_t$ coincides with the Riemannian volume induced from the bi-invariant metric determined by $\left<\cdot,\cdot\right>$. (Otherwise, we will pick up awkward scale factors in what follows.) Let $dY$ and $dX$ denote the volume forms on $\g$ and $\h$ induced from the Haar measure on $G$ and $H$, respectively.

Let $h$ be a regular element of $H$. We have the exact formula \cite[Section 7.2]{Cam90}
\begin{equation}
  K_t(h) = \sum_{Y \in \exp^{-1}(h)}\frac{1}{(2\pi t)^{d/2}}\left({\det}_Y(\exp_*)\right)^{-1/2}e^{-\frac{|Y|^2}{2t} + \frac{st}{12}}, \label{Arede}
\end{equation}
which expresses the heat kernel on a compact Lie group as a sum over geodesics. Here, $s$ is the scalar curvature of $G$, ${\det}_Y(\exp_*)$ is the determinant of the differential of the exponential map $\exp: \g \to G$ at $Y \in \h$, and $|\cdot|$ is the norm with respect to the inner product on $\g$. This determinant can be written as
\begin{equation}
  {\det}_Y(\exp_*) = \frac{j(\exp(Y))}{J(Y)},
\end{equation}
where $j$ and $J$ are ad-invariant functions on $G$ and $\g$ given by
\begin{align*}
  j(\exp(Y)) &= \prod_{\alpha \in R_+}|e^{\alpha(Y)/2} - e^{-\alpha(Y)/2}|^2 \\
  J(Y) &= \prod_{\alpha \in R_+}|\alpha(Y)|^2, \qquad Y \in \h.
\end{align*}
Here, $R_+$ is the set of positive roots of $G$. Thus the leading $t \to 0$ asymptotics of $K_t(h)$ for $h$ near $1$ is given by the term of  (\ref{Arede}) with $Y$ of smallest length, in which case $|Y| = \mr{dist}(h,1)$.

Let $t_i = \frac{\lambda|R_i|}{|S^2|}$, so that $t_1+t_2 = \lambda$. Let $W$ denote the Weyl group of $G$. Then by the Weyl integration formula
\begin{align}
  \frac{1}{K_{t_1+t_2}(1)}\int_G f(g)K_{t_1}(g)K_{t_2}(g)dg &=  \frac{1}{|W| K_{t_1+t_2}(1)}\int_H f(h)K_{t_1}(h)K_{t_2}(h)j(h)dh \label{eq:WIF}
\end{align}
where $dh$ is Haar measure on $H$. The asymptotics of (\ref{eq:WIF}) is determined entirely by the integral in a neighborhood of $1$. Using (\ref{Arede}) and keeping only the leading shortest geodesic term, (\ref{eq:WIF}) as $\lambda \to 0$ is asymptotically equal to
\begin{multline}
\frac{(t_1+t_2)^{d/2}}{|W|(2\pi)^{d/2}t_1^{d/2}t_2^{d/2}}\int_\h f(\exp(Y)) e^{-\frac{|Y|^2}{2t_1}}e^{-\frac{|Y|^2}{2t_2}}\left(\frac{j(\exp(Y))}{J(Y)}\right)^{-1}j(\exp(Y))dY \\
  = \frac{1}{|W|(2\pi \rho)^{d/2}}\int_\h f(\exp(Y))e^{-|Y|^2/2\varrho}J(Y)dY. \label{eq:Hasymp}
\end{multline}
On the other hand, this last integral is just
\begin{equation}
  \frac{1}{(2\pi \rho)^{d/2}}\int_\g f(\exp(X))e^{-|X|^2/2\varrho}dX \label{eq:GMM}
\end{equation}
by the Lie algebra version of the Weyl integration formula.

Suppose now our inner product on $\g$ is not the Haar inner product but $c^2$ times it. Observe that this scaling can be absorbed into the $*$ appearing in the Yang-Mills action, which has the effect of scaling areas by $c^{-2}$. In particular, we replace $\varrho \mapsto \varrho/c^2$ in (\ref{eq:GMM}). This is equivalent to replacing the Haar inner product with $c^2$ times it, i.e., our chosen inner product. Thus, (\ref{eq:GMM}) holds for the general case.\End

\begin{Remark}\label{Rem:Inst}
 In the proof of (\ref{eq:ThmExact1}), we neglected exponentially small contributions to (\ref{Arede}) arising from non-minimal length geodesics joining the identity element of $H$ to $h$. Such geodesics have nontrivial winding and we can regard their contribution to $K_t(h)$ as ``instanton contributions''.
\end{Remark}

In \cite{GP}, additional computations are done that provide an explicit check of Conjecture 1 to all orders in $\lambda$, namely the case of products of Wilson loops formed out of concentrically nested circles. The author has also done nontrivial partial checks in the case of a figure eight loop to second order in $\lambda$.\\

\noindent \textbf{Example 1.} Let $G = SU(2)$ equipped with the bi-invariant metric induced from trace in the standard representation on $\g$. We want to compute the exact asymptotics of $\left<W_{\chi_{m},\gamma}\right>_{S^2}$ where $\chi_{m}$ is trace in the $m$-dimensional representation. Define the function
\begin{equation}
  F(\varrho) = e^{-\varrho/4}(2 - \varrho).
\end{equation}
As $\lambda \to 0$, we have
  \begin{align}
    \left<W_{\chi_{m},\gamma}\right> \sim \begin{cases}
      F\big((m-1)^2\varrho\big) + F\big((m-3)^2\varrho\big) + \cdots + F(\varrho) & m \textrm{ even} \\
      F\big((m-1)^2\varrho\big) + F\big((m-3)^2\varrho\big) + \cdots + F(2^2\varrho) + 1 & m \textrm{ odd}.
    \end{cases} \label{SU(2)asymptotics}
  \end{align}
Indeed, we use (\ref{eq:GMM}) or (\ref{eq:Hasymp}) with
$$f = \chi_{m}(\exp(\theta I)) = e^{i(m-1)\theta} + e^{i(m-3)\theta} + \ldots + e^{-i(m-1)\theta}$$
where we have the generators
\begin{align*}
  I = \begin{pmatrix}
    i & 0 \\
    0 & -i
  \end{pmatrix}, \qquad
  J = \begin{pmatrix}
    0 & -1 \\
    1 & 0
  \end{pmatrix}, \qquad
  K = \begin{pmatrix}
    0 & -i \\
    -i & 0
  \end{pmatrix}
\end{align*}
for $\mathfrak{su}(2)$. So (\ref{eq:Hasymp}) becomes
\begin{align*}
  \int_{\frak{su}(2)}\chi_{m}(\exp(X))e^{-|X|^2/2\varrho}dX &= \int_{\R^3}\chi_{m}(\exp(x_1I+x_2J+x_3K))e^{-|x|^2/\varrho}\frac{d^3x}{(\pi \varrho)^{3/2}} \\
  &= \frac{4}{(\pi\varrho^3)^{1/2}}\int_0^\infty\chi_{m}(\exp(\theta I))e^{-|\theta|^2/\varrho}\theta^2d\theta.
\end{align*} This yields (\ref{SU(2)asymptotics}).

On the other hand, on $\R^2$ we have $\left<W_{f,\gm}\right>$ is given by
\begin{equation}
  \int_G K_{\lambda_0 |R|}(g) \chi_{m}(g)dg = e^{-\frac{(m^2-1)\lambda_0|R|}{4}}m \label{eq:R2asymp}
\end{equation}
since $\Delta \chi_{m} = \frac{m^2-1}{2}\chi_m$. In particular, the $S^2 \to \R^2$ limit of (\ref{SU(2)asymptotics}), in which $\varrho$ becomes $\lambda_0|R|$, does not equal (\ref{eq:R2asymp}), beginning at second order in $\lambda_0$.

\section{Discussion and Further Directions}\label{Sec:Discussion}

We conclude with some natural questions and directions for future research. \\

\renewcommand{\theenumi}{\arabic{enumi}.}

\noindent\textit{1. Why do we obtain asymptotic series that are entire functions?} The explicit asymptotic series we obtain Theorem \ref{ThmExact1} all define entire power series (for $f$ a polynomial function such as trace) in the coupling constant. Yang-Mills theory being free in axial and holomorphic gauges may make this consequence seem natural. On the other hand, the philosophy of resurgence theory \cite{DU} predicts that our series expansions should be non Borel-summable owing to the presence of instantons. At present, we have no way of reconciling this prediction with what is actually the case.\\

\noindent\textit{2. Compute asymptotics of more complicated Wilson loops}. The asymptotic formula (\ref{eq:ThmExact1}) and its generalization to products of Wilson loops obtained from nested, non-intersecting simple closed curves \cite{GPR} suggests that perhaps similar asymptotic formulae can be obtained for Wilson loops involving more complicated curves. What is remarkable about these formulas is that they are given by simple Gaussian integrals over the Lie algebra that are of the type appearing in random matrix theory. It would be of interest to investigate how complicated one may make a Wilson loop and still continue to obtain such formulas for exact asymptotics (to all orders).\\

\noindent\textit{3. Consider more general topologies}. In extending our results on the independence of the choice of gauge-fixing to surfaces of higher genus, the Batalin-Vilkovisky formalism would have to be carried out in the case when one has zero modes for the bosonic kinetic operator (corresponding to having a continuous moduli of flat connections). Since this difficulty is a finite-dimensional complication orthogonal to the infinite-dimensional nature of quantum field theory, we do not anticipate any fundamental obstacles albeit the technical details may be quite involved. Perhaps a very interesting scenario to consider would be to consider compact surfaces with boundary. Here, boundary conditions come into play and we have not considered how the two formulations, exact and perturbative, line up in this regard.

On the other hand, in higher genus, relating the asymptotics of the exact expection to perturbation theory may prove to be difficult, since now the set of flat connections about which to do perturbation theory becomes nontrivial. For each flat connection, we obtain a corresponding propagator formed out of the Green's operators for the Laplacian twisted by the such a flat connection. It is unclear to what extent explicit computations can be done for nontrivial flat connections.\\

\noindent\textit{4. Find a priori analytic relationships between the lattice and continuum formulation of quantum gauge theories.} This is of course not a new question but a very difficult one, since the basic variables in the two formulations are of different natures (Lie group elements versus Lie-algebra valued $1$-forms). So while an infinitesimal group element being approximated by an element in the Lie algebra provides the basis for how one discretizes the continuum theory, without precise estimates, one cannot establish any clear relation between the two. A resolution of this question would presumably shed light on some of the above questions and observations we have made.

\appendix

\section{Graded Vector Spaces}\label{Sec:App1}

A (real) \textit{graded vector space} $V$ is a vector space together with a decomposition $V = \oplus_{i \in \Z} V_i$ into vector spaces $V_i$ in degree $i$. If $v \in V_i$, then $|v| = i$ denotes its degree. An element is even or odd according to whether its degree is even or odd, so that we have a corresponding decomposition of $V = V_{ev} \oplus V_{odd}$ by parity. An ordinary vector space yields a graded vector space concentrated in degree zero.

For ordinary vector spaces $V$, one has the familiar notion of $\Sym^n(V)$ and $\Lambda^n(V)$, the symmetric and exterior powers of $V$. For graded vector spaces, one defines symmetric powers in the graded sense. Namely, let $\otimes^n V$ be the graded vector space whose graded components are
$$(\otimes^n V)_i = \bigoplus_{i_1 + \cdots + i_n = i}V_{i_1}\otimes \cdots \otimes V_{i_n}.$$
We have an action of $\Sym_n$ such that a transposition of adjacent elements acts via
$$u \otimes v \mapsto (-1)^{|u|\cdot|v|}v \otimes u.$$
Then $\Sym^n(V)$ is the $\Sym_n$-invariant subspace of $\otimes^n V$ with respect to the above action. We write
$$\Sym(V) = \bigoplus_{n \geq 0} \Sym^n(V)$$
to denote the total symmetric algebra on $V$. That is, $\Sym(V) \cong \Sym(V_{ev}) \otimes \Sym(V_{odd})$ as an algebra, with $\Sym(V_{ev})$ the symmetric (in the ordinary sense) algebra on $V_{ev}$ and $\Sym(V_{odd})$ the exterior algebra on $V_{odd}$.

If $V$ is a graded vector space, then its dual space $V^*$ is the graded vector space given by
$$V^*_i = (V_{-i})^*,$$
that is, the degree $i$ component of $V^*$ is the dual space of $V_{-i}$. In this way, the evaluation pairing
$$V^* \otimes V \to \R$$
is a degree zero map.

\begin{Definition}\label{DefFun}
  Given a graded vector space $V$, a \textit{function} on $V$ is an element of $\Sym(V^*)$. A \textit{vector field} on $V$ is an element of $\Sym(V^*) \otimes V$.
\end{Definition}

For $V$ an ordinary vector space, the above coincides with the ordinary notion of a polynomial function or polynomial vector field. 

A (linear) map $f: V \to V'$ of graded vector spaces has degree $p$ if $f(V_i) \subset V'_{i+p}$. If the degree is not specified, it is understood to be degree zero. The commutator of two maps $f,g: V \to V$ of degrees $|f|$ and $|g|$ is defined using the appropriate sign rule:
$$[f,g] = fg - (-1)^{|f||g|}gf.$$

\subsection{Directional derivatives} \label{Sec:DD}

The space $(V^*)^{\otimes n}$ denote the space of $n$-multilinear maps from $V^{\otimes n}$ to $\R$. It has a natural action of $\Sym_n$ induced from the one on $V^{\otimes n}$. For $v \in V$, define the contraction operator
\begin{align}
  \pd_v: (V^*)^{\otimes n} & \to (V^*)^{\otimes (n-1)} \nonumber \\
v_1^* \otimes \cdots \otimes v_n^* & \mapsto \sum_i (-1)^{|v|( |v_1|+\cdots+|v_{i-1}|)}v_i^*(v) \Big(v_1^*\otimes\cdots \otimes v_{i-1}^* \otimes v_{i+1}^*\otimes\cdots \otimes v_n^*\Big). \label{eq:pd}
\end{align}
In other words, $\pd_v$ is the directional derivative with respect to $v$, where in the graded setting, it is a derivation of degree $|v|$ using the above signed Leibniz rule.

More generally, given an element $K = u \otimes v \in V^{\otimes 2}$, we can define the operation
\begin{equation}
\pd_K = \frac{1}{2}\pd_v\pd_u. \label{eq:contractK}
\end{equation}
This operation extends bilinearly to any $K \in V^{\otimes 2}$ and depends only on the component of $K$ in $\Sym^2(V)$.

One can consider contractions using ``non-constant coefficient" vector fields, i.e., elements of $\Sym(V^*) \otimes V$. If $fv \in \Sym(V^*) \otimes V$, with $f \in \Sym(V^*)$, then
$$\pd_{fv} = f\pd_v$$
Likewise, we have
\begin{equation}
  \pd_{vf} = (-1)^{|f||v|}f\pd_v \label{eq:pdvf}
\end{equation}
using the usual sign rules.

\subsection{Infinite-dimensional case}

The above considerations generalize to the infinite-dimensional setting needed for quantum field theory. Instead of a finite-dimensional graded vector space, we instead have the space of sections $\cE$ of a graded vector bundle $E$ over a smooth manifold $M$. We leave the regularity of the elements of $\cE$ unspecified (i.e. whether they are smooth or distributional), it usually being clear from the context. The dual space $\cE^*$ of the smooth elements of $\cE$ consists of distributions on $M$ valued in the dual bundle of $E$. Multilinear functionals on $\cE$ are elements of $\cE^* \otimes \cdots \otimes \cE^*$, where the tensor product is completed\footnote{\label{fnote:complete} We have $C^\infty(M_1 \times M_2) = C^\infty(M_1) \otimes C^\infty(M_2)$, where the right-hand side is completed in the sense of nuclear Frechet spaces. This readily extends to the case of smooth sections of a vector bundle, so that $\Gamma(E_1 \boxtimes E_2) = \Gamma(E_1) \otimes \Gamma(E_2)$.}. Given $v \in \cE$, one defines $\pd_v: (\cE^*)^{\otimes n} \to (\cE^*)^{\otimes n-1}$ as above. Likewise for $K \in \cE^{\otimes 2}$, which we regard as an integral kernel of an operator (defined with respect to a pairing on $\cE$, see Section \ref{SecGI}), we can define $\pd_K$ as above. If $K$ is not smooth, $\pd_K$ may be ill-defined when evaluated on a multilinear functional. In particular, when $K$ arises as the integral kernel of a differential operator, one can obtain divergent Feynman integrals from the contractions $\pd_K$ applied to local functionals. An element of multilinear map $(\cE^*)^{\otimes n}$ is said to be \textit{local} if is given by the integral of a polydifferential function of $\cE$ over $M$.

\section{Wick's Theorem}\label{Sec:Wick}

Wick's Theorem comes in two cases: bosonic and fermionic. The first case gives us a combinatorial formula for the integration of monomials against Gaussian measures. We assume the reader is familiar with this result and only record it here for notational purposes. Consider $\R^d$ with the standard inner product $(\cdot,\cdot)$ and let $A = A_{ij}$ be a symmetric nondegenerate $d\times d$ matrix. It determines a bilinear form $(\cdot,A\cdot)$ and a normalized Gaussian measure $d\mu_A = \left(\frac{\det A}{(2\pi)^{d}}\right)^{1/2}e^{-(x,Ax)/2}d^dx$.

\begin{Lemma}(Bosonic Wick's Theorem)
Consider the monomial $f(x) = x^{i_1}\cdots x^{i_{2n}}$. Then
\begin{equation}
  \int d\mu_A f(x) = \frac{1}{2^{n}n!}\sum_{\sigma \in S_{2n}}A^{i_{\sigma(1)}i_{\sigma(2)}}\cdots A^{i_{\sigma(2n-1)}i_{\sigma(2n)}}.
\end{equation}
Here $A^{ij}$ denotes the inverse matrix of $A_{ij}$.
\end{Lemma}
As is well-known, the sum on the right-hand-side has an elegant description in terms of Feynman diagrams. For further details, see e.g. \cite{Cos, PS}.

Next, consider the fermionic case of Wick's Theorem. For this we have to introduce the notion of integration over fermionic (odd) variables. Let $V$ be an odd vector space spanned by $\xi_1,\ldots, \xi_d$. Then we can define the partial integration operator $\int d\xi_i$ by
$$\int d\xi_i = \pd_{\xi_i},$$
where $\pd_{\xi_i}$ is defined as in (\ref{eq:pd}). Thus,
$$\int d\xi_i \xi_i = 1$$
and more generally, if $f(\xi)$ does not depend on $\xi_i$, then
\begin{align*}
  \int d\xi_i \xi_i f(\xi) &= f(\xi)\\
\int d\xi_i f(\xi) &= 0.
\end{align*}
We write $\int d\xi_d\ldots d\xi_1$ as shorthand for $\int d\xi_d\cdots \int d\xi_1$. Thus,
$$\int d\xi_d\ldots d\xi_1\, \xi_1\ldots \xi_d = 1$$
and
$$\int d\xi_d\ldots d\xi_1 \,\pd_{\xi_i} f(\xi) = 0$$
for all $i$.

Suppose we have an even number of Grassman variables $d = 2m$. Given a nondegenerate skew-symmetric matrix $A_{ij}$, we get the bilinear expression $\xi_i A_{ij}\xi_j/2$, which we abbreviate as $(\xi,A\xi)/2$. Letting $d\mu = d\xi_1\ldots d\xi_{d}$, we have the following:

\begin{Lemma}\label{LemmaPfaff}
  We have
    $$\int d\mu e^{-(\xi,A\xi)/2} = \mr{Pfaf}(A).$$
\end{Lemma}

Let
$$d\mu_A = \mr{Pfaf}(A)^{-1}d\mu e^{-(\xi,A\xi)/2}$$
denote the normalized ``Gaussian measure" on odd variables. We can integrate (polynomial) functions in the $\xi_i$ against this density. We have the identity
$$A^{ik}\pd_{\xi_k}(e^{-(\xi,A\xi)/2}f(\xi)) = e^{(-\xi,A\xi)/2}\big(-\xi_i f(\xi)+A^{ik}\pd_{\xi_k}f(\xi)\big) $$
Letting $f(\xi) = \xi_j$ and integrating both sides, we conclude that
\begin{align}
  \int d\mu_A \xi_i\xi_j &= A^{ij}. \label{2pt}
\end{align}
Iteration of this yields the following formula:

\begin{Lemma}\label{Lem:FWT} (Fermionic Wick's Theorem)
  Consider the monomial $f(\xi) = \xi_{i_1}\cdots \xi_{i_{2n}}$. Then
  \begin{equation}\int d\mu_A f(\xi) = \frac{1}{2^nn!}\sum_{\sigma \in S_{2n}}(-1)^\sigma A^{i_{\sigma(1)}i_{\sigma(2)}}\cdots A^{i_{\sigma(2n-1)}i_{\sigma(2n)}}. \label{eq:Fwick}
  \end{equation}
\end{Lemma}

It is often the case that the set of odd variables is partitioned into two separate sets, $\omega_1\ldots, \omega_m$ and their corresponding ``conjugate" variables $\omega_1^*, \ldots, \omega_m^*$. Moreover, we are given the bilinear expression $\omega_i^*B_{ij}\omega_j$ with $B_{ij}$ is an arbitrary matrix, which we abbreviate by $\omega^*B\omega$. One can think of this as letting  $\xi_k = \omega_k^*$ and $\xi_{m+k} = \omega_k$ for $1 \leq k \leq m$ and
\begin{equation}
  A = \begin{pmatrix}0 & -B \\ B& 0 \end{pmatrix} \label{AB}
\end{equation}
in the above. Let $d\mu = d\omega_1^*d\omega_1\cdots d\omega_m^*d\omega_m$.

\begin{Lemma}\label{LemmaDet}
  $$\int d\mu e^{-\omega^* B \omega} = \det B.$$
\end{Lemma}

Since
\begin{align*}
  \mr{Pfaf}(A) &= (-1)^{\binom{m}{2}}\mr{det}(B) \\
  d\xi_1\cdots d\xi_{2m} &= (-1)^{\binom{m}{2}}d\omega_1^*d\omega_1\cdots d\omega_m^*d\omega_m,
\end{align*}
the formulas in Lemmas \ref{LemmaPfaff} and \ref{LemmaDet} agree.

We suppose $B_{ij}$ is invertible. Let $d\mu_B = \frac{1}{\det B}d\mu e^{-(\xi^*,B\xi)}$. From
$$B^{ik}\pd_{\omega_k^*}(e^{-(\omega^*,B\omega)}f) = e^{(-\omega^*,B\omega)}\Big(-\omega_i f + B^{ik}\pd_{\omega_k^*}f\Big),$$
we have
\begin{equation}
 \int d\mu_B \omega_i\omega_j^* = B^{ij}, \label{eq:fermloop0}
\end{equation}
or equivalently
\begin{align}
  \int d\mu_B \omega_i^*\omega_j = -B^{ij}. \label{eq:fermloop}
\end{align}

One can attribute this minus sign to the minus sign that occurs in
\begin{align}
  A^{-1} = \begin{pmatrix}
    0 & B^{-1} \\ -B^{-1} & 0
  \end{pmatrix}
\end{align}
so that (\ref{2pt}) and (\ref{eq:fermloop0}--\ref{eq:fermloop}) agree.

\begin{Remark}
The above considerations explain the convention that fermionic loops in quantum field theoretic computations are weighted with a minus sign. This is a basis dependent statement, however. One should regard (\ref{Lem:FWT}) as fundamental, with (\ref{eq:fermloop}) a consequence of a particular (albeit common) parametrization.
\end{Remark}

One can unify the bosonic and fermionic cases of Wick's Theorem using the framework of graded vector spaces. Let $A(\cdot,\cdot)$ be a symmetric pairing on a graded vector space $V$, i.e., an element of $\Sym^2(V^*)$, which is nondegenerate. Such a pairing determines a dual pairing $P$ on $V^*$ which is an element of $\Sym^2(V)$. We call $P$ the \textit{propagator}.

If we pick a basis $v_i$ (of homogeneous degree elements) for $V$, then if $A_{ij} = A(v_i,v_j)$, we have $P = A^{ij}v_iv_j$. Define
$$d\mu_A = c_A\prod dv_i e^{-A_{ij}v_iv_j/2}$$
with $c_A$ chosen so that so that $\int d\mu_A = 1$. We have the following:

\begin{Lemma}\label{LemmaWickP}(Wick's Theorem, unified version)
 For $f$ a (polynomial) function on $V$,
  \begin{equation}
      \int d\mu_A f(x) = (e^{\pd_P}f)(0). \label{eq:unifiedWick}
  \end{equation}
\end{Lemma}
Here, $\pd_P: \Sym^n(V^*) \to \Sym^{n-2}(V^*)$ is the \textit{Wick contraction} operator, defined as in Section \ref{Sec:DD}, and
$$e^{\pd_P} = \sum_{n=0}^\infty \frac{(\pd_P)^n}{n!}$$
is the sum over all Wick contractions weighted with the appropriate symmetry factor. The right-hand side of (\ref{eq:unifiedWick}) is evaluated at zero so that the maximal number of Wick contractions have been made (in the Feynman-diagrammatic picture, we only sum over vacuum diagrams, i.e. those without external legs).

\section{Explicit BV computations}\label{Sec:AppBV}

Here we elucidate some of the abstract details of Section \ref{SecGI} and perform explicit computations. Let us start by making a general remark about functionals, following the presentation in Section \ref{Sec:App1}, in order to clarify bothersome sign issues. We have our space of fields $\cE$. It is the space of sections of a graded vector bundle over a base manifold $\Sigma$. Functionals are elements of $\Sym(\cE^*)$. This means that if we are given a functional $S \in \Sym(\cE^*)$, a permutation of its arguments induces an appriopriate sign change in the value of $S$. \textit{This is not to be confused with the differentio-geometric manipulation of the expression that defines $S$.}

\begin{Example}
 In the 2d Yang-Mills setting, consider the functional $I \in \Sym^3(\cE^*)$ given by
 $$I(A,A,X) = \int_{\Sigma} \left<A \wedge *[A,X]\right>.$$
 Here we have expressed $I$ as a polynomial (the first and second arguments are evaluated on the same element $A$), rather than as a symmetric multilinear map evaluated on distinct inputs. Note how $[A,X] = -[X,A]$, so that the pairing $[\cdot,\cdot]$ does not obey the proper sign rule when interchanging $A$ with $X$ (since $A$ is even and $X$ is odd). However, the symmetry property of $S$ means that we define, e.g.
 $$I(X,A,A) = I(A,X,A) = I(A,A,X).$$
\end{Example}

Thus, when we perform computations such as $\{F,G\}$, \textit{it is important to note not just the expression that defines $F$ but also the manner in which $F$ was defined as an element of $\Sym(\cE^*)$, i.e., the ordering of the inputs.} In the above example, this distinguishes the order of elements in the functional $S$ (which obeys graded vector space sign rules) from the order of elements in the operation $[\cdot,\cdot]$, which obeys the usual rule for the Lie bracket of different forms.

\subsection{Functional Derivatives}
Our goal is to express the BV functional derivative and BV bracket explicitly using functional derivative notation. This functional derivative notation, while formal-looking, is a very convenient (and standard) computational ansatz for carrying out the evaluation of rigorously defined, though abstract, operations. Namely, it is a way of expressing the directional derivative (see \ref{Sec:DD}) in the infinite-dimensional setting in such a way that an explicit coordinate system is used (although all results are independent of which choice).

First we start in the finite-dimensional setting \cite{Sch}, in which we have a finite-dimensional graded vector space with a BV-pairing $\left<\cdot,\cdot\right>$ (i.e., nondegenerate, skew-symmetric, degree $-1$). Suppose we are given a BV-symplectic coordinate basis $x_i,\xi_j$, i.e.
$$\left<\pd_{x^i},\pd_{\xi^j}\right> = -\left<\pd_{\xi^j},\pd_{x^i}\right> = \delta_{ij},$$
with $x^i$ and $\xi^j$ even and odd, respectively. For any function $F$, we have
\begin{align*}
\delta^{BV}F &= \pd_{\xi^i}F\pd_{x^i} + (-1)^{|F|}\pd_{x^i}F\pd_{\xi^i}\\
&= \pd_{x^i}[\pd_{\xi^i}F] + \pd_{\xi^i}[\pd_{x^i}F]
\end{align*}
 where in the last line, the $[\cdot]$ means we regard the inside expression as a coefficient of the vector field on the left and not an expression to be differentiated. This is consistent with the definition
 $$\pd_vF = (-1)^{|v|}\left<v,\delta^{BV}F\right>.$$
Moreover, it follows from the definition $\{F,G\} = \pd_{\delta^{BV}F}G$
that
\begin{align*}
\{F,G\} = \pd_{\xi^i}F\pd_{x^i}G + (-1)^{|F|}\pd_{x^i}F\pd_{\xi^i}G.
\end{align*}
The integral kernel of the identity operator with respect to BV convolution, defined as in (\ref{BVconv}) but with the BV-pairing at hand, is equal to $K_0 = \pd_{x^i} \otimes \pd_{\xi^i} + \pd_{\xi^i} \otimes \pd_{x^i}$. So the BV Laplacian is given by
\begin{align*}
\Delta_{BV} &= \pd_{K_0}\\
&= \pd_{\xi^i}\pd_{x^i}
\end{align*}

\begin{Example}
In 2D Yang-Mills, we have the BV-pairing (\ref{eq:BVpairing_dens}). Pick local coordinates $x^\mu$ on $\Sigma$. This gives us the following local ``basis'' of functional derivatives
$$\frac{\delta}{\delta X^a(x)}, \frac{\delta}{\delta A^a_\mu(x)}, \frac{\delta}{\delta A^{\dag,a}_\mu(x)}, \frac{\delta}{\delta X^a_{\mu\nu}(x)}.$$
This is just another way of writing the restriction of an element of $\cE$ (which is a section of a bundle over $\Sigma$) to the point $x$, which yields an element dual to the coordinate evaluation monomials at $x$, i.e. the monomials $X^a(x)$, $A^a_\mu(x)$, $A^{\dag,a}_\mu(x)$, $X^a_{\mu\nu}(x)$. Thus,
$$\frac{\delta}{\delta A^a_\mu(x)}A^b_\nu(y) = \delta^{ab}\delta_{\mu\nu}\delta^{(2)}(x-y),\qquad \mathrm{etc.}$$

We have a corresponding local coordinate volume form $dV^{\mu\nu} := \frac{1}{2}\eps_{\mu\nu}dx^\mu \wedge dx^\nu$
Any local action functional $S$ can be (locally) written in the form
$$\int_\Sigma dV^{\mu\nu}  \mathcal{S}$$
where $\mathcal{S}$ is the action functional density associated to $S$ (and $dV^{\mu\nu}$).

We can write the BV-pairing (\ref{eq:BVpairing_dens}) formally as\footnote{Note the minus sign occurring on the right-hand side of (\ref{eq:AAdag}) has to do with the skew-symmetry of the wedge-product of $1$-forms. Had we switched $\nu$ and $\mu$, there would be no minus sign, as is consistent with the BV-pairing being the wedge pairing on $1$-forms.}
  \begin{align}
    \left<\frac{\delta}{\delta A_\mu(x)}, \frac{\delta}{\delta A^\dag_\nu(y)}\right>_{BV} &= \delta^{(2)}(x-y)dV^{\mu\nu} = -\left<\frac{\delta}{\delta A^\dag_\nu(x)},\frac{\delta}{\delta A_\mu(y)}\right>_{BV} \label{eq:AAdag}\\
    \left<\frac{\delta}{\delta X^\dag_{\mu\nu}(x)}, \frac{\delta}{\delta X(y)}\right>_{BV} &= \delta^{(2)}(x-y)dV^{\mu\nu} =
-\left<\frac{\delta}{\delta X(x)}, \frac{\delta}{\delta X^\dag_{\mu\nu}(y)}\right>_{BV} \label{eq:XXdag}
  \end{align}

We have
$$\delta^{BV}S = \int_\Sigma d^2x \bigg((-1)^{|S|}\frac{\delta}{\delta \psi(x)} \mathcal{S} \fd{\tilde\psi(x)} + \fd{\tilde\psi(x)} \mathcal{S} \fd{\psi(x)}\bigg),$$
where $\frac{\delta}{\delta \psi}$ varies over all even functional derivative basis elements $\frac{\delta}{\delta A^a_\mu(x)}, \frac{\delta}{\delta X_{\mu\nu}(x)}$
and $\fd{\tilde\psi(x)}$ varies over all corresponding odd dual basis elements $\eps_{\mu\nu}\fd{A^{\dag,a}_\nu(x)}, \fd{X^a(x)}$. The $\int d^2x$ is just notation for a formal sum over all $x$, i.e. $\delta^{BV}S$ acts at every $x \in \Sigma$. It is now easy to verify that
$$\pd_\phi S = (-1)^{|\phi|}\int_\Sigma dV^{\mu\nu}\left<\phi, \delta^{BV}\mathcal{S}\right>_{BV}$$
for any $\phi$.

Next, for $\eps > 0$, let $K_{\eps}(x,y)$ denote the integral kernel of $e^{-\eps\Delta}$, with $K_{\eps}^{\psi,\tilde\psi}(x,y)$ its components with respect to basis elements $\psi$ and $\tilde\psi$. Then the regulated BV-Laplacian is given by
$$\Delta_{BV,\eps} S =  K_\eps^{\psi,\tilde\psi}(x,y)\fd{\psi(x)}\fd{\tilde\psi(y)}S.$$
This is well-defined for $\eps > 0$ since then $K_\eps(x,y)$ is smooth. On the other hand, if $S$ is a local action functional, in which case, it is supported on the small diagonal inside $\cE^{\otimes n}$ if $S$ is of polynomial degree $n$, then $\Delta^{BV}_0S$ is ill-defined since $K_0(x,y)$ is the delta-function along the diagonal.

Now we consider $S$ arising from terms in the Yang-Mills action functional. The quadratic part is given by
 \begin{align*}
  S_{kin} &= S_{kin, bos}(A,A) + S_{kin, fer}(A^\dag, X)\\
  &= \frac{1}{2\lambda_0}\int \left<A, d*dA\right>_{BV}  - \frac{1}{\lambda_0}\int_\Sigma\left<A^\dag, dX\right>_{BV}\\
  &= \frac{1}{2\lambda_0}\int dV^{\mu\nu} A^a_\mu(x) (d*dA^a)_\nu(x) - \frac{1}{\lambda_0}\int_\Sigma dV^{\mu\nu} A^{\dag,a}_\mu(x) (dX^a)_\nu(x)
 \end{align*}
 So
 \begin{align*}
  Q &= \lambda_0\{S_{kin},\cdot\}\\
  &= \int d^2x\bigg( (d*dA^a)_\mu(x)\fd{A^{\dag,a}_\mu(x)} -(dX^a)_\mu(x)\fd{A^a_\mu(x)} - (dA^{\dag,a})_{\mu\nu}(x)\frac{\delta}{\delta X^{\dag,a}_{\mu\nu}(x)}\bigg)\\
  &= Q^{[1]} + Q^{[0]} + Q^{[2]}.
 \end{align*}
Note that 
\begin{equation}
\{S_{kin}, S_{kin}\} = 0 
\end{equation}
since $Q^2 = 0$.

We check the master equation (\ref{CME}), which decomposes into 3 parts (according to ghost number $1,2,3$ respectively). Since $S = S_{bos} + S_{gh} - I_{CE}$, we have
\begin{align}
\{S_{gh},S_{bos}\} = -Q^{[0]}\bar I_{bos} - Q^{[1]}\bar I_{gh} + \frac{1}{2}\{\bar I_{gh},\bar I_{bos}\} &= 0 \label{eq:CME1}\\
\frac{1}{2}\{S_{gh},S_{gh}\} - \{I_{CE}, S_{gh}\} = -Q^{[0]}\bar I_{gh} + \frac{1}{2}\{\bar I_{gh},\bar I_{gh}\} -Q^{[2]}\bar I_{CE} + \{\bar I_{CE}, \bar I_{gh}\} &= 0 \label{eq:CME2}\\
\{I_{CE},I_{CE}\} &= 0, \label{eq:CME3}
\end{align}
where $\bar{I}_{gh} = \lam_0 I_{gh}$, $\bar{I}_{CE} = \lam_0 I_{CE}$.
We verify each of the above equations.

We have (\ref{eq:CME1}) since $\{S_{gh},S_{bos}\}$ yields the change in $S_{bos}$ under an infinitesimal gauge transformation, and $S_{bos}$ is gauge-invariant. 

Next, for (\ref{eq:CME2}), with $I_{gh} = I_{gh}(A^\dag, A, X)$, we have
\begin{align*}
-Q^{[0]}\bar I_{gh}(A^\dag,X_1,X_2) &= \bigg(\int d^2x\, (dX_1)^a_\mu(x)\fd{A^a_\mu(x)}\bigg)\bar I_{gh}(A^\dag,A,X_2) + \leftrightarrow\\
&= -\int \left<A^\dag, [dX_1,X_2]\right> + \leftrightarrow  \\
&= -\int\left<A^\dag, d[X_1,X_2]\right>,
\end{align*}
where the operation $\leftrightarrow$ is symmetrization (in the graded sense) over $X_1$ and $X_2$. Here, we have a minus sign occurring in the above due to having to move $-Q^{[0]}$ past $A^\dag$ to functionally differentiate the second argument $A$. We have
\begin{align}
 \{\bar I_{gh},\cdot\} = \int d^2x\bigg( [A,X]^a_\mu(x)\fd{A^{a}_\mu(x)} + [A^\dag,X]^a_\mu(x)\fd{A^{\dag,a}_\mu(x)} - (A^\dag \wedge A)^a_{\mu\nu}(x)\fd{X^a_{\mu\nu}(x)}\bigg)
\end{align}
The minus sign in the last term occurs because it arises from
$$\fd{X^{a}(x)}I_{gh}(A^\dag,A,X)\fd{X^{\dag,a}_{\mu\nu}(x)},$$
and the odd derivation $\fd{X^a(x)}$ has to move past the fermionic $A^\dag$ in the first argument of $I_{gh}(A^\dag,A,X)$ to functionally differentiate the 3rd position. Thus,
\begin{align*}
\frac{1}{2}\{\bar I_{gh},\bar I_{gh}\}(A^\dag, A, X_1, X_2) &= \frac{1}{2}\left(\int \left<[A, X_1] \wedge [A^\dag, X_2]\right> + \int \left<[A^\dag \wedge X_1], [A \wedge X_2]\right>\right) + \leftrightarrow\\
&= -\frac{1}{2}\int\left<[A^\dag \wedge A] , [X_1,X_2]\right> + \leftrightarrow \\
&= -\int \left<A^\dag, [A,[X_1,X_2]]\right>.
\end{align*}
Finally, for $I_{CE} = I_{CE}(X^\dag, X_1, X_2)$, since
$$\{\bar I_{CE},\cdot\} = \int d^2x [X_1,X_2]^a(x)\fd{X^a(x)},$$
we have
$$-\{\bar I_{CE},S_{gh}\}(A^\dag, A, X_1,X_2) = \frac{1}{\lam_0}\int \left<A^\dag, d_A[X_1, X_2]]\right>$$
where as before, we pick up an additional minus sign for moving the the odd vector field $\{\bar I_{CE},\cdot\}$ past the first argument $A^\dag$ to the third argument of $I_{gh}$. Altogether, summing the terms on the left-hand side of (\ref{eq:CME2}), the equation holds.

Finally (\ref{eq:CME3}) follows from
\begin{align*}
\{\bar I_{CE}, \bar I_{CE}\}(X^\dag, X_1,X_2,X_3) &= \int \left<X^\dag, [[X_1,X_2],X_3]\right> - \int \left<X^\dag, [X_1, [X_2,X_3]]\right> + \leftrightarrow\\
&= -\int \left<X^\dag, [X_2,[X_1,X_3]]\right> + \leftrightarrow\\
& = 0,
\end{align*}
where we symmetrize over the all the $X_i$'s and the last line follows from the Jacobi identity.
\end{Example}

\section{Some Singular Integral Computations}

For computing the divergences of Feynman diagrams, we have to compute integrals formed out of (derivatives of) the Green's function. In dimension two, only the leading order singularity of the Green's function contributes to the singular part of Feynman diagrams. Hence, the singular parts of the integrals we consider can be recast explicitly as singular integrals on $\R^2$.

On $\R^2$, the Green's function for the Laplacian $\Delta$ is given by
\begin{equation}
G^{\R^2}(\mb{w}) = -\frac{1}{2\pi}\log|\mb{w}|. \label{eq:GR^2}
\end{equation}
We also have its heat-kernel regulated version
\begin{align}
G_\eps^{\R^2}(\mb{w}) &= \int_\eps^L e^{-t\Delta}dt \nonumber \\
&= \int_\eps^L \frac{1}{4\pi t}e^{-|\mb{w}|^2/4t}dt. \label{eq:GepsR^2}
\end{align}
where $L$ is a fixed, albeit arbitrary constant.  (Different choices of $L$ will only change $G_\eps^{\R^2}$ by a smooth term, but a finite $L$ must be chosen to ensure convergence of (\ref{eq:GepsR^2}), unlike the case of $\Sigma$ compact).

Let $\equiv$ denote equality up to terms which are finite as $\eps \to 0$. We have the following integral computations:

\begin{Lemma}\label{Lemma:SingInt}
 Let $f$ be any smooth function. Then
 \begin{align}
 \int_{|\mb{w}|<1} \pd_{\mb{w}^\mu}G_\eps^{\R^2}(\mb{w})\pd_{\mb{w}^\nu}G_\eps^{\R^2}(\mb{w})f(\mb{w})\,d^2\mb{w} & \equiv \frac{1}{4\pi^2}\int_{\eps^{1/2} < |\mb{w}| < 1} \frac{\mb{w}^\mu \mb{w}^\nu}{|\mb{w}|^4}f(\mb{w})\,d^2\mb{w}  \label{eq:SingInt1}
 \\[2ex]
 &= \frac{\delta_{\mu\nu}\log \eps^{-1/2}}{4\pi}f(0).
\end{align}
Consequently,
\begin{align}
-\int_{|\mb{w}|<1} \big(\pd_{\mb{w}^\mu}\pd_{\mb{w}^\mu}G_\eps^{\R^2}(\mb{w})\big)G_\eps^{\R^2}(\mb{w})f(\mb{w})\,d^2\mb{w} &\equiv \frac{\log \eps^{-1/2}}{2\pi}f(0). \label{eq:SingInt2}
\end{align}
 \end{Lemma}

The first equation will be used in the evaluation of the Feynman diagrams $(II_2)$ and $(III)$ from Section \ref{Sec:Finite}. The second of these will be used to estimate the Feynman diagram $(II_1)$.\\

\Proof We have
\begin{equation}\int_{|\mb{w}|<1} \pd_{\mb{w}^\mu}G_\eps^{\R^2}(\mb{w})\pd_{\mb{w}^\nu}G_\eps^{\R^2}(\mb{w})f(\mb{w})\,d^2\mb{w} = \frac{1}{64\pi^2}\int_{|\mb{w}| < 1} \frac{\mb{w}^\mu\mb{w}^\nu}{|\mb{w}|^4}\left(\int_{\eps/|\mb{w}|^2}^{L/|\mb{w}|^2}\frac{1}{t^2}e^{-1/4t}dt\right)^2f(\mb{w})d^2\mb{w}.\label{eq:II_2_flat}\end{equation}
We split this integral into two regions: $|\mb{w}| < \eps^{1/2}$ and $|\mb{w}| > \eps^{1/2}$. For the first region, the integral is uniformly bounded in $\eps$ since $\pd_{\mb{w}^\mu}G = O(\eps^{-1}|\mb{w}|)$ as $\mb{w} \to 0$. For the second region,
\begin{multline*}
\frac{1}{64\pi^2}\int_{\eps^{1/2} < |\mb{w}| < 1} \frac{\mb{w}^\mu\mb{w}^\nu}{|\mb{w}|^4}\left(\int_{\eps/|\mb{w}|^2}^{L/|\mb{w}|^2}\frac{1}{t^2}e^{-1/4t}dt\right)^2f(\mb{w})d^2\mb{w}\\
\begin{split}
&\equiv \frac{1}{64\pi^2}\int_{\eps^{1/2} < |\mb{w}| < 1} \frac{\mb{w}^\mu\mb{w}^\nu}{|\mb{w}|^4}\left(\int_{0}^{L/|\mb{w}|^2}\frac{1}{t^2}e^{-1/4t}dt\right)^2f(\mb{w})d^2\mb{w}\\
&= \frac{1}{4\pi^2}\int_{\eps^{1/2} < |\mb{w}| < 1} \frac{\mb{w}^\mu\mb{w}^\nu}{|\mb{w}|^4}e^{-|\mb{w}|^2/4L}f(\mb{w})d^2\mb{w}\\
&\equiv \frac{1}{4\pi^2}\int_{\eps^{1/2} < |\mb{w}| < 1} \frac{\mb{w}^\mu\mb{w}^\nu}{|\mb{w}|^4}f(\mb{w})d^2\mb{w}.
\end{split}
\end{multline*}
The first line follows from
$$\int_0^{\eps/|\mb{w}|^2}\frac{1}{t^2}e^{-1/t}dt = e^{-|\mb{w}|^2/\eps}$$
and
$$\int_{\eps^{1/2}<|\mb{w}|<1} \frac{e^{-|\mb{w}|^2/\eps}}{|\mb{w}|^2}d^2\mb{w}$$
being finite as $\eps \to 0$. Equation (\ref{eq:SingInt1}) now follows, with the right-hand side leading to a logarithmic singularity as $\eps \to 0$ for $\mu=\nu$. Equation (\ref{eq:SingInt2}) then follows from integration by parts.

\begin{Remark}\label{Rem:PS}
 One can regard Lemma \ref{Lemma:SingInt} as stating the heat-kernel regularization is equivalent to ``point-splitting regularization'' (for the kinds of integrals under consideration), where the latter involves letting $\eps = 0$ in the propagator but excising an $\eps^{1/2}$ tubular-neighborhood of the singular locus $\mb{w} = 0$ (thus we ``split'' apart singular terms). This is evident for (\ref{eq:SingInt1}), which involves two terms that have first derivatives of $G_\eps^{\R^2}$. For (\ref{eq:SingInt2}), one has to interpret the point-splitting appropriately. Here, when $\eps = 0$, two derivatives of the Green's function yields a delta-function at $\mb{w}=0$, which when separated from the other singular term by a distance of $\eps^{1/2}$, yields the right-hand side of (\ref{eq:SingInt2}).
\end{Remark}


\end{document}